\documentclass[twocolumn]{aastex631}
\usepackage{url}

\shorttitle{Radio Emission of Post-Merger Galaxies}
\shortauthors{Walsh \& Burke-Spolaor}

\begin{document}

%\title{Post-Merger Galaxies in the Radio: AGN, Supernova Remnant Populations, and Possible Jet-ISM Feedback}

\title{Prevalence of Compact Nuclear Radio Emission in Post-Merger Galaxies and its Origin}%: AGN, Supernova Remnant Populations and Hints of Jet-ISM Feedback}

\author[0000-0003-1551-1340]{Gregory Walsh}
\affiliation{Department of Physics and Astronomy, 
West Virginia University, 
Morgantown, WV 26506, USA}
\affiliation{Center for Gravitational Waves and Cosmology,
West Virginia University, 
Chestnut Ridge Research Building, 
Morgantown, WV 26506, USA}

\author[0000-0003-4052-7838]{Sarah Burke-Spolaor}
\affiliation{Department of Physics and Astronomy, 
West Virginia University, 
Morgantown, WV 26506, USA}
\affiliation{Center for Gravitational Waves and Cosmology,
West Virginia University, 
Chestnut Ridge Research Building, 
Morgantown, WV 26506, USA}

\correspondingauthor{Gregory Walsh}
\email{gvw0001@mix.wvu.edu}

\begin{abstract}
Post-merger galaxies are unique laboratories to study the triggering and interplay of star-formation and AGN activity. Combining new, high resolution, 10 GHz Jansky Very Large Array (VLA) observations with archival radio surveys, we have examined the radio properties of 28 spheroidal post-merger galaxies. We find a general lack of extended emission at (sub-)kiloparsec scales, indicating the prevalence of compact, nuclear radio emission in these post-merger galaxies, with the majority (16/18; 89\%) being radio-quiet at 10 GHz. Using multi-wavelength data, we determine the origin of the radio emission, discovering 14 new radio AGN and 4 post-mergers dominated by emission from a population of supernova remnants. Among the radio AGN, almost all are radio-quiet (12/14; 86\%). We discover a new dual AGN (DAGN) candidate, J1511+0417, and investigate the radio properties of the DAGN candidate J0843+3549. 4 of these radio AGN are hosted by SF emission-line galaxies, suggesting that radio AGN activity may be present during periods of SF activity in post-mergers. The low jet powers and compact morphologies of these radio AGN also point to a scenario in which AGN feedback may be efficient in this sample of post-mergers. Lastly, we present simulated, multi-frequency observations of the 14 radio AGN with the Very Long Baseline Array (VLBA) and the VLBI capabilities of the Next Generation Very Large Array (ngVLA) to assess the feasibility of these instruments in searches for supermassive black hole binaries (SMBHBs). 
\end{abstract}


\section{Introduction} \label{sec:intro}
Theoretical studies predict that galaxy mergers are the main contributor to the buildup of stellar mass in galaxies and to the formation of bulges and massive elliptical galaxies \citep{Springel_00,Cox+08,DiMatteo+08,Torrey+12}. Integral to this evolution is what role galaxy mergers have in the triggering of an active galactic nucleus (AGN) and/or intense starburst activity. Early studies of ultra-luminous infrared galaxies (ULIRGs), which are at least partially powered by a heavily obscured AGN \citep{Lonsdale+06}, found a nearly ubiquitous fraction hosted by interacting galaxy systems \citep{Murphy+96,Veilleux+02}. These early works suggested that ULIRGs are triggered by galaxy merger-induced processes. Likewise, intense starburst activity has been observed in merging systems \citep{Tacconi+08}. For both cases, the triggering of these phenomenon are caused by nuclear inflows of gas produced by gravitational torques during the merger process \citep{Hopkins+06}, linking the growth of supermassive black holes (SMBHs) and their host galaxies. Indeed, observed correlations between the SMBH and host galaxy properties confirm their co-evolution \citep{Kormendy&Richstone_95,Magorrian+98,Ferrarese&Merrit_00,Gebhardt+00,Tremaine+02,Gultekin+09,McConnell&Ma_13,Sahu+19}. Thus, detailed studies of galaxy mergers at various stages of evolution are needed to fully realize the astrophysical processes governing these phenomena.

Observational studies are in conflict with one another over the role mergers play in triggering an AGN. While many have found either an increased AGN fraction in merging systems \citep{Ellison+11,Satypal+14,Donley+18,Goulding+18} or an increased merger fraction in AGN hosts \citep{Chiaberge+15,Fan+16,Gao+20,Marian+19,Breiding+23}, others have found no such connection between AGN and mergers \citep{Grogin+05,Cisternas+11,Bohm+13,Villforth+17,Lambrides+21}. Selection biases almost certainly contribute to this dissonance. Different AGN selection criterion, e.\,g., mid-infrared \citep[e.\,g.,][]{Satyapal+14,Donley+18,Goulding+18}, X-ray \citep[e.\,g.,][]{Grogin+05,Villforth+17}{}{}, optical \citep[e.\,g.,][]{Bohm+13}{}{}, and radio \citep[e.\,g.,][]{Chiaberge+15,Breiding+23}{}{}, and selection of mergers at various stages of their evolution also necessarily select different astrophysical scenarios \citep[e.\,g.,][]{Sanders+88}. 

Among the evolutionary stages of merger systems, post-merger galaxies, those in which the stellar nuclei have coalesced, perhaps present the most unique laboratories to study these triggering mechanisms and the effects of AGN feedback on star formation as a result of the advanced stage of the merger. Small samples of post-merger systems have found hints at enhancement in the star-formation rate \citep{Ellison+13_pm} and AGN incidence over galaxies in close pairs \citep{Carpineti+12,Bickley+23}. The effectiveness of AGN feedback, however, is questioned when examining post-merger galaxies. Post-mergers appear to host a significantly higher fraction of post-starburst galaxies \citep{Ellison+22,Li+23}, characterized as having recently experienced intense star-formation activity that was rapidly truncated \citep{C&Sh_87}, although this itself does not imply that an AGN is the main driver of this quenching. Indeed, this low efficiency scenario is corroborated by the works of \citet{Kaviraj+15} and \citet{Shabala+17}. Both of these studies determined that the onset of merger-triggered AGN activity is delayed with respect to the peak of starburst activity, significantly limiting its ability to impact the star-formation rate of the host galaxy. Expanding the overall population of post-merger galaxies for which we can study these evolutionary effects is important towards understanding the general trends in their behavior.

Post-merger galaxies are also ideal targets in which to search for supermassive black hole binaries (SMBHBs). As all massive galaxies are believed to harbor a SMBH \citep{Kormendy&Ho_13}, a major galaxy merger should lead to the formation of a SMBHB. At the initial stages of SMBHB evolution, dynamical friction is the dominant mechanism through which the SMBHs lose energy and momentum \citep{Begelman+80}, eventually settling into the gravitational center of the merger remnant. Simulations of galaxy mergers have found that this phase of the evolution may be as short as 1 Gyr \citep{D&A_17,Kelley+17}, shorter than the timescale over which the bulges will centralize. Thus, by the time the stellar nuclei have merged, the resident SMBHs are likely to already reside in the gravitational center of the merger remnant \citep[Cf.][]{D&B_17,Kelley+17}{}{}. At parsec-scale separations, the SMBHs will form a gravitationally-bound SMBHB. Here, several different processes, of varying efficiency, are hypothesized to contribute to the shrinking of the binary's orbit; the so-called `last parsec' problem. If the SMBHB is able to overcome the last parsec, it will reach sub-pc orbital separation, where the emission of low-frequency gravitational waves will efficiently bring the binary to merger. Thus, establishing a population of observed SMBHBs at various orbital separations is key towards our understanding of the nHz gravitational wave population, which will soon be probed by pulsar timing arrays \citep{Agazie+23_gwb,Agazie+23_ind}. Critical to this aspect, however, is the poorly understood evolution of SMBHBs themselves. Low efficiency at pc-scale can create a scenario in which binaries at these orbital separations are still present in the post-merger host galaxy. Observations of galaxy mergers at this post-merger phase must be taken to better understand SMBHB evolution.

In this paper we present a multi-wavelength analysis of 30 galaxies identified as post-mergers in Galaxy Zoo to study their emission mechanisms. The paper is organized as follows. Section~\ref{sec:sample} presents the post-merger sample. Section~\ref{sec:obs_cal} describes new 10 GHz observations taken with the Karl G. Jansky Very Large Array (VLA) of this post-merger sample, while in Section~\ref{sec:surveys}, we describe the multi-frequency data obtained via archival radio surveys. The optical emission-line classifications of each post-merger galaxy, and the radio luminosities, morphologies and spectra of these sources are presented in Section~\ref{sec:source_properties}. In Section~\ref{sec:org_of_re}, we present analyses to determine the origin of the radio emission in our 10 GHz-detected post-merger galaxies. We then discuss the prevalence of radio-quiet emission in these post-mergers, the impact of AGN feedback in radio AGN hosts, and the properties of the SF-dominated radio sources in Section~\ref{sec:discussion}. Lastly, in Section~\ref{sec:ngvla}, we present simulated, multi-frequency observations of the radio AGN with the Very Long Baseline Array (VLBA) and Next Generation VLA (ngVLA) to assess the feasibility of SMBHB searches for these post-merger galaxies. Our results are summarized in Section~\ref{sec:summary}.

Throughout this paper, we have adopted a $\Lambda$CDM cosmology with $H_0=67.4$ km s$^{-1}$ Mpc$^{-1}$ and $\Omega_m=0.315$ \citep{Planck_20}. We use the radio spectral index convention $S_\nu \propto \nu^\alpha$.

\section{Sample} \label{sec:sample}
Our sample consists of the 30 spheroidal post-merger (SPM) galaxies presented by \citet[][hereafter C12]{Carpineti+12}. The C12 sample was selected from the larger sample of \citet[][]{Darg+10}. These authors constructed a catalog of 3003 local (0.005 $< z <$ 0.1) galaxy merger systems identified through visual inspection via the Galaxy Zoo Project \citep[][]{Lintott+08}. Of these 3003, 370 of these merging systems were considered strongly perturbed, e.\,g., the presence of strong tidal tails, but could not be clearly divided into a pair of two interacting galaxies. From the parent sample of 370 late-stage merger systems, C12 selected their sample of 30 SPM galaxies via the distinct visual characteristics of SPMs: SPM galaxies are defined as a single galaxy that displays morphological disturbances associated with a recent merger event, e.\,g., tidal tails, and contain only a single dominant bulge, making them the likely progenitors of early-type galaxies. As a consistency check, after visual inspection of each SPM candidate, C12 utilized the SDSS parameter \verb|fracdev| in the optical $r$-band for a quantitative representation of the bulge-dominant nature of each system. \verb|fracdev| represents the likelihood of the surface brightness distribution to be fit by a de Vaucoulerus profile: pure bulge systems have a value of 1; pure exponential or disc-like distributions have a value of 0. All of the 30 SPM galaxies have \verb|fracdev| $>0.5$, signifying they are bulge dominated, with most having \verb|fracdev| $\geq 0.8$. Further, the 30 SPM systems are all of high stellar mass (10.3 $\leq \mathrm{log}(M_*) \leq$ 11.76), typical of early-type galaxies. Additionally, C12 found that these 30 SPM systems are diverse in their large-scale environments. Using the environment parameter $\rho_g$ defined by \citet[][]{Schawinski+07_env}, C12 found two SPM systems which inhabit a cluster environment, 19 in a group environment, and 9 in a field environment.

\section{Observations and Data Calibration}\label{sec:obs_cal}
High resolution, Karl G. Jansky Very Large Array (VLA) observations of the 30 SPM galaxies were taken from 2016 to 2022. \object[WISEA J101833.64+361326.7]{J1018+3613} was observed on May 26, 2016 (PI: S. Burke-Spolaor) using the S- (2-4 GHz) and X-band (8-12 GHz) receivers of the VLA while in B configuration. 3C 186 was used for bandpass calibration and J1018+3542 was used to perform phase referencing. \object[Mrk 1018]{J0206$-$0017}, \object[VV 713]{J1445+5134}, \object[IC 1102]{J1511+0417}, \object[CGCG 135-026]{J1511+2309} and \object[IC 4630]{J1655+2639} were observed in two observing programs on November 15, 2020 and December 12, 2020 (PI: P. Breiding) using the S- and X-band receivers of the VLA while in BnA and A configuration. 3C 286 and 3C 138 were used for bandpass calibration and a nearby phase reference calibrator for phase calibration of each target. We obtained observations for the remaining 24 SPM galaxies on May 19, and May 28, 2022 using the X-band receiver of the VLA with 4 GHz bandwidth while in A configuration. We observed 3C 147 and 3C 286 for bandpass calibration and a nearby phase reference calibrator for phase calibration of each target. Our observations were designed to reach a nominal sensitivity of $15\,\mu$Jy for each target, with a $3\sigma$ detection threshold of $45\,\mu$Jy. Two of the SPM galaxies, J0908+1407 and J0933+1048, were removed from the analysis due to technical issues during the observing session. For that reason, only the remaining 28 SPM galaxies are included in this analysis.  

The data sets were calibrated either using the VLA calibration pipeline\footnote{\url{https://science.nrao.edu/facilities/vla/data-processing/pipeline}} or following standard data calibration techniques. The VLA calibration pipeline could not be used for the observing session that contained technical issues. To check for calibration consistency, we calibrated the data set without technical errors using both the VLA calibration pipeline and our manual calibration routine. We achieved the same results using both calibration methods for this data set.

The data were inspected, flagged, and imaged in the Common Astronomy Software Applications (CASA; \citealt{CASA_22}) package. To account for the large fractional bandwidths at each band, $\sim$ 60\% and $\sim$ 40\% respectively for S- and X-band, we used multi-Taylor, multi-frequency synthesis deconvolution (MTMFS; \citealp{MTMFS_11}) when cleaning our images. Because of the limited $uv$-coverage from our observations, we utilized a Briggs weighting scheme \citep[][]{Briggs_95} with a robust parameter of 0.7 to suppress the sidelobes present in images of moderately strong sources ($S_\nu>1$ mJy). Otherwise, we used a natural weighting scheme when imaging. Where applicable, we performed standard self-calibration techniques on the target data to improve the image quality.

\section{Radio Surveys}\label{sec:surveys}
We wish to construct a broadband radio SED for each of the 28 SPM galaxies in our survey. In this section, we describe the surveys used to construct each SED. 

It is important to note that each survey, observed at a different frequency and with different angular resolutions, is, by nature, sensitive to different forms of radio emission and may or may not suffer from source confusion. Low frequency ($<1$ GHz) surveys are more sensitive to diffuse emission, likely associated with star formation, and are more likely to suffer from source confusion due to their larger resolution elements. High frequency surveys, in contrast, generally resolve out this same extended, diffuse emission if it is of low surface brightness, making them good identifiers of compact radio jets and cores, features associated with an AGN. This can create an artificial steepening of the radio spectrum due to the high frequency surveys missing flux density information recovered at lower frequency for diffuse emission, or confusion from background source blending.

We attempted to mitigate these issues by visually inspecting all of the survey images used in our analysis. In particular, we are interested in the spectral index measurement of the nuclear radio emission, which will be obtained through the 3 and 10 GHz flux density measurements, or their limits (see Section~\ref{sec:radio_spec}). Background source blending is not an issue because of the high angular resolution at these frequencies and the low redshift ($z<0.1$) of these SPMs. For lower frequency observations, visual inspection showed no complex structure in the vast majority of the radio maps. For those select few that do have complex structure, we describe our procedures for their flux density measurements in Section~\ref{sec:radio_morph}. This is likewise true for any high frequency image that showed complex structure.

For each radio survey, we considered a source to be detected if it was found in an available source catalog, e.\,g., LoTSS, RACS, and FIRST, or the signal-to-noise (S/N) ratio in the respective survey image is $>5\sigma$, where $\sigma$ is the local image RMS noise. In some cases, a source was identified at only a $3\sigma$ significance. We considered the $3\sigma$ source a true detection if it was spatially coincident with a source of $\geq 5\sigma$ detection in any of the other radio surveys. Each image was inspected visually to assure that no sources were missed. It is important to distinguish this to properly account for the difference in sensitivities between the various radio surveys used. If we only classified sources at the $5\sigma$ level and greater, our multi-frequency analyses would be incomplete and not truly representative of the SPM sample radio population within the limits of each survey. The 10 GHz map of \object[KUG 1013+394]{J1015+3914}, presented in Figure~\ref{fig:1015}, illustrates this point. The diffuse, $3\sigma$ emission at 10 GHz would not by itself be substantial for a detection. However, J1015+3514 is detected at high signal-to-noise ratio in all other surveys that observed it, and this diffuse 10 GHz emission is spatially coincident with those detections. For this reason, we consider the 10 GHz source a true detection and include it as part of our analyses.

For $3\sigma$ sources, we determined the flux density by performing a 2D Gaussian fit to the observed radio emission using the task \verb|IMFIT| in CASA.

\subsection{LoTSS}\label{sec:lotss}
The LOw Frequency ARray (LOFAR; \citealp{van_Haarlem+13}) Two Meter Sky Survey (LoTSS; \citealp{Shimwell+22}) is an on-going survey covering the northern sky above $+34\degr$ conducted at a central observing frequency of 144 MHz. For our analysis, we use the second data release\footnote{\url{https://lofar-surveys.org/dr2\_release.html}}, which covers 27\% of the northern sky with a resolution of 6$\arcsec$ and median RMS sensitivity of 83 $\mu$Jy beam$^{-1}$. 16 of the 28 SPM galaxies in our survey fall within the LoTSS DR2 sky coverage, of which 12 were detected. 

\subsection{RACS}\label{sec:racs}
The Rapid ASKAP Continuum Survey (RACS; \citealp{McConnell+20,Hale+21}) is the first large-area survey completed using the Australian Square Kilometer Array Pathfinder (ASKAP; \citealp{Hotan+21}). RACS covered the entire southern sky up to a declination $+41\degr$ with a median field RMS sensitivity of 250 $\mu$Jy beam$^{-1}$. RACS-low, as part of the RACS DR1\footnote{\url{https://research.csiro.au/casda/the-rapid-askap-continuum-survey-stokes-i-source-catalogue-data-release-1/}}, was observed at a central frequency of 887.5 MHz with a resolution of $15\arcsec$. 14 of the 28 SPM galaxies in our survey fall within the RACS sky coverage, of which 7 were detected.

\subsection{FIRST}\label{sec:first}
The Faint Images of the Radio Sky at Twenty Centimeters (FIRST; \citealp{Helfand+15}) survey was a VLA survey conducted at 1.4 GHz and observed the entire sky north of $+10\degr$ and south of $+65\degr$, covering 10,575 deg$^2$. The survey resolution is given at 5$\arcsec$ with a typical RMS sensitivity of 150 $\mu$Jy beam$^{-1}$. 27 of the 28 SPM galaxies in our survey fall within the FIRST sky coverage, of which 14 were detected. We used the flux density and RMS values listed for each source from the catalog of \citet[][]{Helfand+15}, except for 3$\sigma$ sources, which were not included in this catalog.

\subsection{NVSS}\label{sec:nvss}
The NRAO VLA Sky Survey (NVSS; \citealp{Condon+98}) was a VLA survey conducted at 1.4 GHz and observed the entire sky north of $-40\degr$. The nominal resolution of the survey is 45$\arcsec$ with a typical RMS sensitivity of 450 $\mu$Jy beam$^{-1}$. Because of the large angular resolution, we used the NVSS catalog data for only one source, \object[WISEA J130414.50+652044.2]{J1304+6520}, which was not part of the FIRST sky coverage, but was observed and detected by NVSS.

\subsection{VLASS}\label{sec:vlass}
The VLA Sky Survey (VLASS; \citealp{Lacy+20}) is an on-going VLA survey conducted at S-band, covering the frequency range 2-4 GHz, which will cover the whole sky observable by the VLA ($\delta>-40\degr$) over three observing epochs. Each observing epoch is designed to reach a nominal RMS sensitivity of 120 $\mu$Jy beam$^{-1}$ with a resolution of 2.5$\arcsec$. VLASS has currently completed two observing epochs, with raw and calibrated data sets and Quick Look images available for both epochs 1 and 2. The flux density accuracy of Quick Look sources in the first campaign of the first epoch of VLASS (VLASS1.1) were affected by antenna pointing errors, giving systematically lower flux density measurements of 10\% with a scatter of $\pm 8$\% for flux densities below $\approx 1$ Jy (see VLASS Memo 13\footnote{\url{https://library.nrao.edu/public/memos/vla/vlass/VLASS_013.pdf}} for more detail). For this reason, we used only the campaigns from the second epoch of VLASS (VLASS2.1 and VLASS2.2) for the S-band flux density of sources of interest. As mentioned in Section~\ref{sec:obs_cal}, we observed 6 of the SPM galaxies in separate VLA observing campaigns at S-band. We used these 3 GHz VLA observations for these sources to derive source parameters instead of any corresponding VLASS detections for them. The remaining 22 SPM galaxies have all been observed in the second VLASS campaign, of which 7 were detected. To extract the flux density of the detected sources, we used the CASA task \verb|IMFIT| to fit a two-dimensional Gaussian to the source in each Quick Look image.

\section{Source Properties}
\label{sec:source_properties}
Following the detection criterion of Section~\ref{sec:surveys}, 75\% (12/16) of the sources with available LoTSS data were detected; 50\% (7/14) with available RACS data were detected; 54\% (15/28) with available 1.4 GHz data, from either FIRST or NVSS, were detected; 32\% (7/22) with available VLASS data were detected, with a 100\% detection rate for the remaining 6 with separate 3 GHz VLA observations; and 67\% (18/28) were detected by our 10 GHz VLA observations.

\subsection{Emission-Line Activity}
\label{sec:em_line}
\begin{figure*}[t]
    \centering
    \includegraphics[width=0.95\textwidth,trim=28mm 0mm 30mm 10mm,clip]{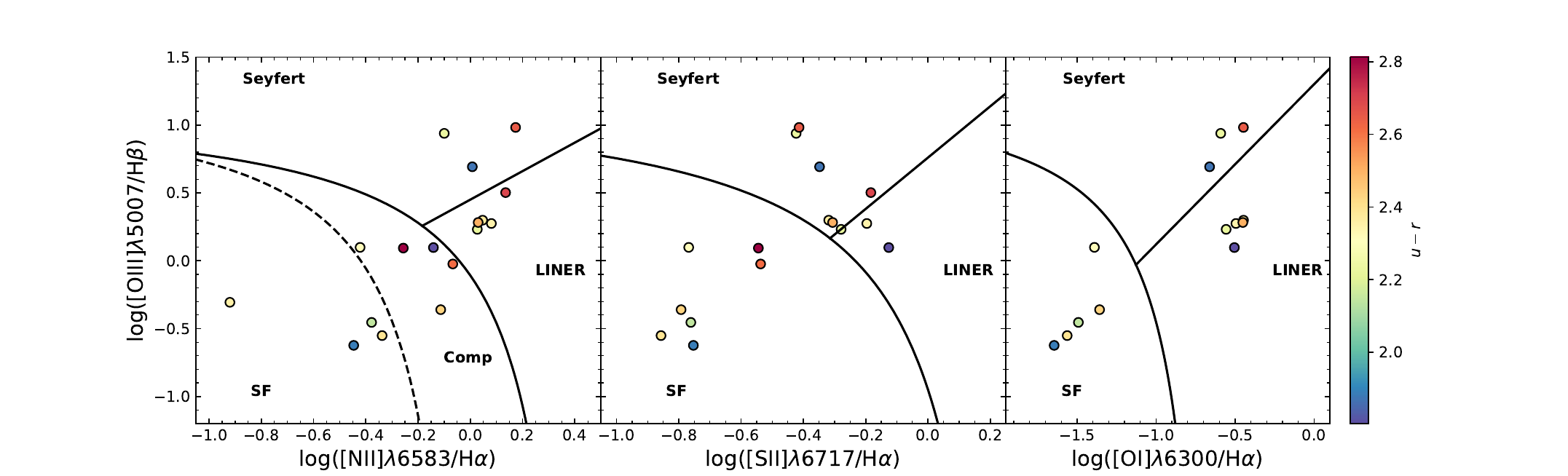}
    \caption{The BPT diagnostic diagrams for the emission-line galaxies in the C12 SPM sample. Each point is colored by its $u-r$ color, with the colorbar indicating the range of values in the scale. Even for the most blue SPMs, the SPMs are still predominantly red in color. The dashed line in the [NII]$\lambda6583$/H$\alpha$ diagram (left panel) is the empirical SF line of \citet{Kauffmann+03} and the straight line that divides Seyferts and LINERs is from \citet{Schawinski+07_bpt}. Seyferts and LINERs are divided in the [SII]$\lambda6717$/H$\alpha$ (middle panel) and [OI]$\lambda6300$/H$\alpha$ (right panel) diagrams from the line of \citet{Kewley+06}. In all diagrams, the solid line is the theoretical maximum from the starburst models of \citet{Kewley+01}. Galaxies that fall between the lines of \citet{Kauffmann+03} and \citet{Kewley+01} are SF-AGN composite, while those below \citet{Kauffmann+03} are purely SF.
    }
    \label{fig:bpt}
\end{figure*}

Emission-line diagnostics are a powerful tool to probe the dominant ionization mechanism in a galaxy. To examine the emission-line behavior of the 30 SPM galaxies, we have used the OSSY catalog \citep{Oh+11} to obtain the intrinsic fluxes of the H$\beta$, [OIII]$\lambda5007$, H$\alpha$, [NII]$\lambda6583$, [SII]$\lambda6717$, and [OI]$\lambda6300$ emission lines. \citet{Oh+11} determined these values by performing a spectral fitting routine to the SDSS DR7 spectrum of each source. If the signal-to-noise (S/N) ratio of the H$\beta$, [OIII]$\lambda5007$, H$\alpha$ or [NII]$\lambda6583$ lines was $<3$, we classified the galaxy as quiescent. For the remaining galaxies, we followed the standard BPT diagram analysis \citep{BPT_81}. For the [NII]/H$\alpha$ diagnostic, we used the demarcation of \citet{Kauffmann+03} to distinguish between pure star-forming (SF) and SF-AGN composite galaxies. Composite galaxies and AGN are divided using the theoretical maximum starburst model from \citet{Kewley+01}. AGN are then subdivided between Seyferts and LINERs (Low-Ionization Nuclear Emission-line Regions) by the division of \citet{Schawinski+07_bpt}. The best indication of Seyfert or LINER behavior is achieved by using the [OI]$\lambda6300$ emission line \citep{Schawinski+07_bpt}. However, the [OI]$\lambda6300$ line is typically weaker than any of the other lines used and we only employed this diagnostic if the [OI]$\lambda6300$ line was detected with a S/N ratio $\geq3$. Otherwise, we employed the [SII]$\lambda6717$ diagnostic to distinguish between Seyfert AGN and LINERs. For these two diagnostics, we used the Seyfert-LINER demarcation lines of \citet{Kewley+01}. If neither line was detected, we used the [NII] diagnostic to distinguish between Seyferts and LINERs.

The results of our BPT analysis are presented in Figure~\ref{fig:bpt}, where each data point colored by its $u-r$ color. The emission-line classification of each SPM galaxy is listed in Table~\ref{table:source_info}. It should be noted that for even the bluest of the SPM galaxies in the C12 sample, their overall $u-r$ color is still predominantly red. This is expected, as C12 found that the $u-r$ colors of this SPM sample is indicative of a recent star-formation episode, e.\,g,. bluer than an early-type control sample, but one that peaked prior to the merger coalescence, e.\,g., redder than a sample of ongoing mergers (see Figure 5 of C12). 

The BPT diagnostic for J0206$-$0017 deserves special attention. The middle panel of Figure~\ref{fig:bpt} shows only 16 of the 17 identified active galaxies. This is because the data point for J0206$-$0017 has log([SII]$\lambda6717$/H$\alpha$)=-1.27. In comparison to the much larger sample of active SDSS galaxies used by \citet{Kewley+06}, there are no galaxies which approach this value of J0206$-$0017. This is most likely attributable to the fact that J0206$-$0017 is a known changing-look AGN with asymmetric broad-line emission \citep{Cohen+86,McElroy+16}. The prescription used by OSSY to determine the line fluxes would not have accounted for the extremely broad nature of the H$\alpha$ and H$\beta$ lines for this source, and we would most likely need to perform our own spectral fitting routine to extract a reliable flux value for the narrow emission-line components to these broad lines.
Because of this, we have classified J0206$-$0017 as a Seyfert AGN instead of as a star-forming galaxy as would be determined by its BPT diagnostics.

We also note that the spectra of J0908+1407, J1511+2309, and J1655+2639 all contain H$\beta$ absorption. In all of these cases, the H$\beta$ absorption appears to be of stellar origin. Through visual inspection, there does not appear to be a significant blueshift in the H$\beta$ absorption, which would be representative of an AGN-related outflow (e.\,g., \citealt{Williams+17}). For any H$\beta$ emission present in these sources, the S/N ratio of the emission line was $<3$. Although the emission lines of [OIII]$\lambda5007$, H$\alpha$, and [NII]$\lambda6583$ are all detected with S/N ratio $\geq3$ in these spectra, for consistency, we classified them as quiescent because of the weak H$\beta$ emission.

In total, we found that 43\% (13/30) of the C12 SPM galaxies were classified as quiescent. The remaining $\approx$57\% (17/30) were classified as either purely SF (10\%; 3/30), SF-AGN composite ($\approx$13\%; 4/30), Seyfert AGN ($\approx$13\%; 4/30), or LINER (20\%; 6/30) from their BPT diagnostics. In comparison to the emission-line diagnostics performed by C12, our analysis finds a higher percentage of quiescent galaxies ($16\%\pm6\%$ to 43\%), a lower percentage of Seyfert AGN ($42\%\pm6\%$ to 13\%), and a similar percentage of LINERs ($26\%\pm6\%$ to 20\%) and star-forming galaxies ($16\%\pm6\%$ to 10\%). Direct comparison is somewhat ambiguous though, since C12 did not use the SF-AGN composite classification for their BPT analysis. It is unclear where the composite systems we identified would fall in the analysis of C12. It is interesting, however, that we arrive at different conclusions for the number of quiescent galaxies considering both the OSSY catalog and C12 used the \verb|gandalf| code \citep{Sarzi+06} to perform emission-line fitting of the spectra. We would expect, then, that the S/N ratio of the requisite emission lines would not change between the two analyses. Even if the 3 H$\beta$ absorption spectra are considered as active galaxies by C12, this only marginally reduces the percentage of quiescent galaxies we have identified from 43\% to 30\%, which is still a factor of 2 greater than what was found by C12.

\begin{deluxetable*}{ccccc}[ht]
%\tabletypesize{\scriptsize}
\tablewidth{0pt}
\tablecaption{Spheroidal Post-Merger Sample \& BPT Classification}
\label{table:source_info}
\tablecolumns{5}
\tablehead{
\colhead{Source} & \colhead{RA} & \colhead{Dec} & \colhead{$z$} & \colhead{BPT}\\
\colhead{(1)} & \colhead{(2)} & \colhead{(3)} & \colhead{(4)} & \colhead{(5)}
}
\startdata
J0206$-$0017 & 31.567 & -0.291 & 0.043 & AGN \\ 
J0759+2750 & 119.952 & 27.839 & 0.067 & Composite \\ 
\object[WISEA J083309.47+152353.4]{J0833+1523} & 128.289 & 15.398 & 0.076 & Quiescent \\ 
J0843+3549 & 130.937 & 35.828 & 0.054 & AGN \\ 
J0851+4050 & 132.978 & 40.836 & 0.029 & LINER \\ 
J0908+1407 & 137.156 & 14.122 & 0.088 & Quiescent \\ 
J0916+4542 & 139.212 & 45.700 & 0.026 & Composite \\ 
J0933+1048 & 143.447 & 10.811 & 0.085 & Quiescent \\
J1015+3914 & 153.992 & 39.243 & 0.063 & Starforming \\ 
J1018+3613 & 154.640 & 36.224 & 0.054 & AGN \\ 
J1041+1105 & 160.266 & 11.096 & 0.053 & LINER \\ 
\object[WISEA J105646.98+124544.9]{J1056+1245} & 164.196 & 12.762 & 0.092 & Quiescent \\
J1113+2714 & 168.419 & 27.241 & 0.037 & Starforming \\ 
\object[WISEA J111732.56+375746.9]{J1117+3757} & 169.385 & 37.963 & 0.096 & LINER \\ 
J1124+3005 & 171.142 & 30.095 & 0.055 & LINER \\ 
J1135+2913 & 173.781 & 29.891 & 0.046 & Starforming \\ 
\object[SPRC 027]{J1144+2309} & 176.183 & 23.162 & 0.048 & Quiescent \\
\object[WISEA J123013.16+114613.3]{J1230+1146} & 187.554 & 11.770 & 0.089 & Quiescent \\ 
\object[WISEA J125350.03+394416.9]{J1253+3944} & 193.458 & 39.738 & 0.092 & Quiescent \\
J1304+6520 & 196.060 & 65.346 & 0.083 & AGN \\ 
\object[ARP 196 NED01]{J1314+2607} & 198.656 & 26.123 & 0.074 & Quiescent \\
\object[WISEA J132654.49+565320.7]{J1326+5653} & 201.726 & 56.889 & 0.090 & Quiescent \\
\object[2MASX J14053947+4001556]{J1405+4001} & 211.414 & 40.032 & 0.084 & Quiescent \\
J1433+3444 & 218.327 & 34.735 & 0.034 & LINER \\ 
J1445+5134 & 221.438 & 51.581 & 0.030 & Composite \\ 
J1511+0417 & 227.771 & 4.294 & 0.042 & LINER \\ 
J1511+2309 & 227.964 & 23.151 & 0.052 & Quiescent \\ 
J1517+0409 & 229.454 & 4.162 & 0.037 & Quiescent \\
J1617+2512 & 244.426 & 25.206 & 0.031 & Composite \\ 
J1655+2639 & 253.790 & 26.663 & 0.035 & Quiescent \\ 
\enddata
\tablecomments{Column 1: Source name.
Column 2: Right Ascension.
Column 3: Declination.
Column 4: Spectroscopic redshift from SDSS.
Column 5: BPT classification.
}
\end{deluxetable*}
\subsection{Radio Flux Densities and Luminosities}
\label{sec:flux_lum}
Flux density measurements were obtained either from survey catalog entries or from the CASA task \verb|IMFIT| when reported values were not available. The integrated flux density measurements and their associated errors for each source are summarized in Table~\ref{table:flux_values}. For sources identified in the LoTSS, RACS, and FIRST/NVSS catalogs, the measurement error in Table~\ref{table:flux_values} is the error quoted by each catalog summed in quadrature with a 5\% uncertainty in the absolute flux scale. For VLASS and $3\sigma$ detections in any of the archival radio surveys, the error is the RMS image noise and 5\% uncertainty in the absolute flux scale. For sources detected by the 3 and 10 GHz VLA observations, the errors are the image RMS and a 3\% uncertainty in the absolute flux scale \citep{Perley&Butler_17}.

\begin{deluxetable*}{lccccc}[t!]
\tablewidth{0pt}
\tablecaption{Integrated Flux Density Values}
\label{table:flux_values}
\tablecolumns{6}
\tablehead{
\colhead{Source} & \colhead{$S_\mathrm{LoTSS}$ (mJy)} & \colhead{$S_\mathrm{RACS}$ (mJy)} & \colhead{$S_\mathrm{1.4GHz}$ (mJy)} &
\colhead{$S_\mathrm{3GHz}$ (mJy)} & \colhead{$S_\mathrm{10GHz}$ (mJy)} \\
\colhead{(1)} & \colhead{(2)} & \colhead{(3)} & \colhead{(4)} & \colhead{(5)} & \colhead{(6)}
}
\startdata
J0206$-$0017 & -- & $5.38 \pm 0.81$ & $3.36 \pm 0.18$ & $1.91 \pm 0.07$ & $1.64 \pm 0.06$ \\
J0759+2750 & $11.3 \pm 0.6$ & $4.23 \pm 0.52$ & $3.45 \pm 0.21$ & $1.93 \pm 0.15$ & $0.786 \pm 0.016$ \\
J0833+1523 & -- & $<1.65$ & $<0.41$ & $<0.48$ & $<0.06$ \\
J0843+3549* & $29.7 \pm 0.8$ & -- & $3.77 \pm 0.17$ & $2.79 \pm 0.13$ & $0.539 \pm 0.017$ \\
J0851+4050 & $1.0 \pm 0.2$ & -- & $<0.43$ & $<0.37$ & $0.226 \pm 0.015$ \\
J0916+4542 & $2.6 \pm 0.2$ & -- & $0.782 \pm 0.13$ & $<0.33$ & $0.11 \pm 0.013$ \\
J1015+3914** & $11.5 \pm 0.4$ & -- & $1.36 \pm 0.19$ & $1.13 \pm 0.13$ & $0.526 \pm 0.024$ \\
J1018+3613 & $23.4 \pm 0.9$ & -- & $16.2 \pm 0.8$ & $8.75 \pm 0.25$ & $2.72 \pm 0.08$ \\
J1041+1105 & -- & $3.3 \pm 0.5$ & $<0.39$ & $<0.45$ & $0.08 \pm 0.012$ \\
J1056+1245 & -- & $<1.2$ & $<0.41$ & $<0.47$ & $<0.04$ \\
J1113+2714 & -- & $1.08 \pm 0.30$ & $1.32 \pm 0.14$ & $<0.35$ & $0.063 \pm 0.012$ \\
J1117+3757 & $2.0 \pm 0.2$ & -- & $<0.43$ & $<0.33$ & $<0.04$ \\
J1124+3005 & $0.9 \pm 0.2$ & -- & $<0.44$ & $<0.33$ & $<0.03$ \\
J1135+2913 & $15.2 \pm 0.6$ & -- & $3.38 \pm 0.22$ & $2.89 \pm 0.14$ & $0.355 \pm 0.013$ \\
J1144+2309 & -- & $<1.38$ & $<0.45$ & $<0.34$ & $<0.04$ \\
J1230+1146 & -- & $<16$ & $<0.65$ & $<0.65$ & $<0.05$ \\
J1253+3944 & $<0.30$ & -- & $<0.39$ & $<0.41$ & $<0.04$ \\
J1304+6520 & $12.3 \pm 0.5$ & -- & $2.41 \pm 0.24$ & $2.95 \pm 0.22$ & $0.99 \pm 0.29$ \\
J1314+2607 & $<0.30$ & $<0.78$ & $<0.41$ & $<0.40$ & $<0.04$ \\
J1326+5653 & $<0.30$ & -- & $<0.46$ & $<0.56$ & $<0.04$ \\
J1405+4001 & $<0.30$ & -- & $<0.41$ & $<0.50$ & $<0.04$ \\
J1433+3444 & $15.6 \pm 0.7$ & -- & $2.69 \pm 0.19$ & $2.02 \pm 0.16$ & $1.15 \pm 0.036$ \\
J1445+5134 & $28.2 \pm 1.0$ & -- & $11.9 \pm 0.58$ & $6.06 \pm 0.14$ & $1.05 \pm 0.049$ \\
J1511+0417* & -- & $<2.3$ & $1.55 \pm 0.17$ & $1.068 \pm 0.032$ & $1.342 \pm 0.039$ \\
J1511+2309 & -- & $11.8 \pm 1.7$ & $1.32 \pm 0.19$ & $0.495 \pm 0.031$ & $0.249 \pm 0.017$ \\
J1517+0409 & -- & $<1.8$ & $<0.45$ & $<0.41$ & $0.12 \pm 0.013$ \\
J1617+2512 & -- & $1.47 \pm 0.46$ & $1.41 \pm 0.15$ & $1.06 \pm 0.13$ & $0.22 \pm 0.012$ \\
J1655+2639 & -- & $7.8 \pm 0.59$ & $4.70 \pm 0.23$ & $2.50 \pm 0.07$ & $0.77 \pm 0.02$ \\
\enddata
\tablecomments{Column 1: Source name. 
Column 2: LoTSS (144 MHz) flux density and error \citep{Shimwell+22}. Upper limits indicate a $3\sigma$ non-detection, whereas no entry means the source was not included in the survey field.
Column 3: RACS (888 MHz) flux density and error \citep{McConnell+20,Hale+21}.
Column 4: 1.4 GHz flux density and error, reported from either FIRST (27/28; \citealp{Helfand+15}) or NVSS (1/28; \citealp{Condon+98}).
Column 5: 3 GHz flux density and error, reported from either VLASS (22/28; \citealp{Lacy+20}) or for the first time from VLA observations (6/28).
Column 6: 10 GHZ flux density and error, reported from our VLA observations.
*: These flux density measurements are reported for the dominant component when the source is resolved into a multi-component morphology.
**: The flux density at 10 GHz was found after applying a \textit{uv}-taper to the image plane.}
\end{deluxetable*}
The observed radio flux densities span $0.90-30$ mJy, with a median of 12 mJy, for the 12 LoTSS detections; $1.1-12$ mJy, with a median of 4.2 mJy, for the 7 RACS detections; $0.78-16$ mJy, with a median of 2.7 mJy, for the 15 FIRST/NVSS detections; $0.50-8.8$ mJy, with a median of 2.0 mJy, for the 13 VLASS/3 GHz VLA detections; and $0.06-2.7$ mJy, with a median of 0.50 mJy, for the 18 with a 10 GHz VLA detection.

\begin{figure*}[t]
    \centering
    \includegraphics[width=\textwidth, trim=10mm 5mm 10mm 10 mm, clip]{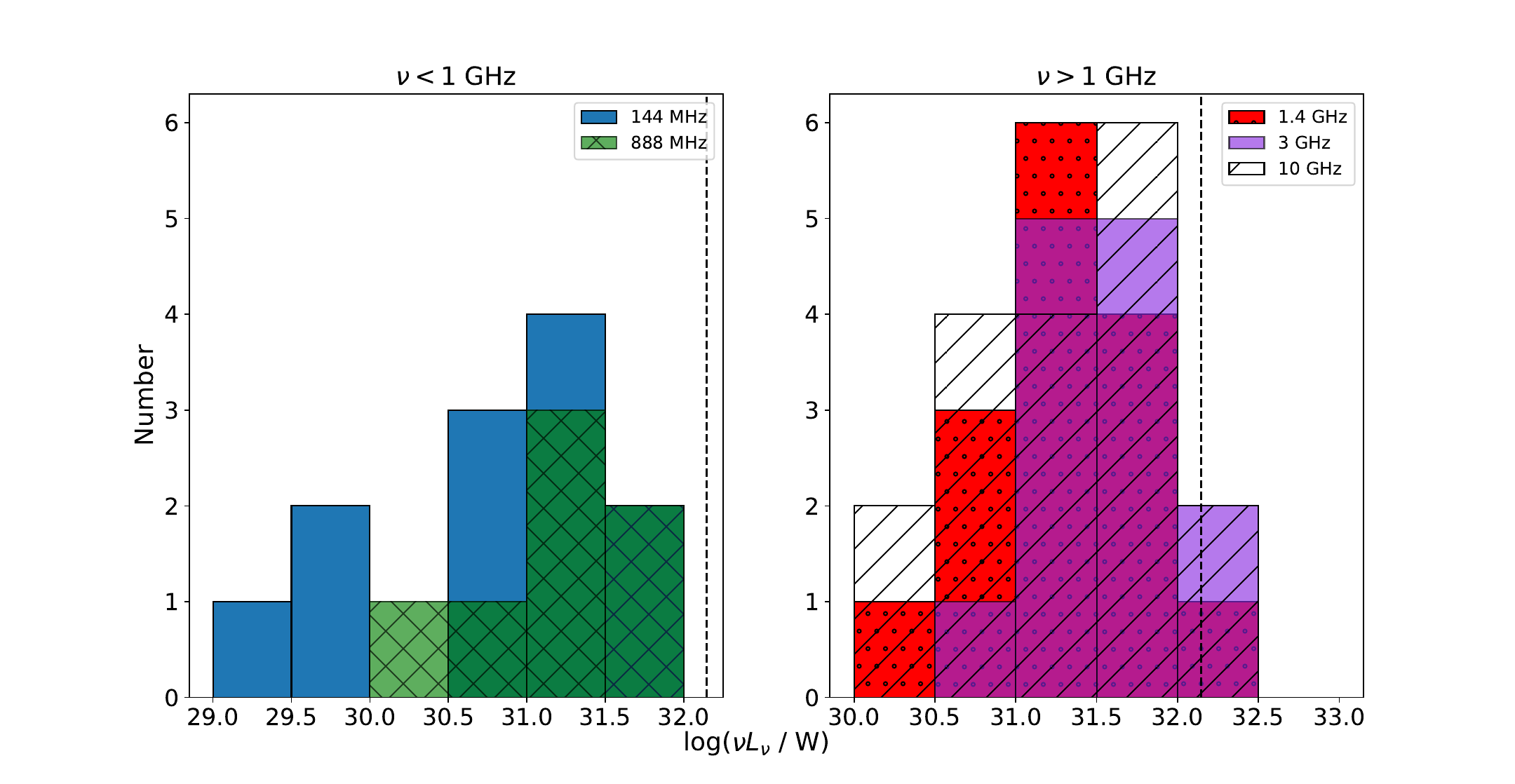}
    \caption{\textit{Left}: Luminosity ($\nu L_\nu$) distribution of the radio sources associated with each spheroidal post-merger (SPM) galaxy from the radio surveys observed in the MHz regime: LoTSS (solid blue) and RACS (hatched green). Some of the SPM galaxies were not observed in each survey. The dashed vertical line represents the demarcation between radio-loud and radio-quiet objects at 1.4 GHz, or $\nu L_\nu = 1.4\times10^{32}$ W. \textit{Right}: Same as left but for radio surveys observed in the GHz regime: FIRST or NVSS (dotted red), VLASS or 3 GHz VLA (solid purple), and 10 GHz VLA (hatched).} 
    \label{fig:det_hist}
\end{figure*}

Each SPM galaxy has an associated redshift measurement from SDSS. We used this and the flux density measurement to calculate a luminosity, in units of Watts (W), at each observing frequency for all of the detected radio sources. We show the luminosity distributions for the $\nu < 1$ GHz (LoTSS, RACS) and $\nu > 1$ GHz (FIRST/NVSS, VLASS, VLA) surveys in Figure~\ref{fig:det_hist}. After calculating the luminosities for each source, we compared these to the standard radio-loud demarcation spectral luminosity $\nu L_\nu\approx10^{32}$ W. A luminosity value above this demarcation is considered radio-loud; radio AGN dominate the radio luminosity function above this demarcation \citep{Condon+02, Kimball+11}. We find that J1018+3613 is radio-loud at GHz frequencies, and J1304+6520 is radio-loud at 3 and 10 GHz. The remaining radio sources are all radio-quiet objects. 

\subsection{Radio Morphology}\label{sec:radio_morph}
We describe the bulk radio morphology properties of the detected sources in each of the radio surveys. 
Each individual source and its intensity maps are discussed and presented in Appendix~\ref{sec:app_a}.

For each source, we categorize the morphology into one of the following classifications: 
\begin{enumerate}
    \item \textbf{Unresolved}: The peak-to-integrated flux density ratio is unity within $3\sigma$ uncertainty and the source is a single Gaussian component that does not exhibit any flux beyond the synthesized beam. 0/12 sources are unresolved by LoTSS; 3/7 sources by RACS; 12/15 sources by FIRST/NVSS; 5/13 sources by VLASS or 3 GHz VLA; and 6/18 sources by our 10 GHz VLA observations.
    \item \textbf{Marginally Resolved}: The peak-to-integrated flux density ratio is unity within $3\sigma$ uncertainty and the source is marginally extended along one axis of the synthesized beam. 1/12 sources are marginally resolved by LoTSS; 2/7 sources by RACS; 2/15 sources by FIRST/NVSS; 2/13 sources by VLASS or 3 GHz VLA; and 1/18 sources by our 10 GHz VLA observations.
    \item \textbf{Resolved}: The peak-to-integrated flux density ratio of the source is significantly less than unity, and the deconvolved major and minor axes have non-zero size. 11/12 sources are resolved by LoTSS; 2/7 sources by RACS; 1/15 sources by FIRST/NVSS; 6/13 sources by VLASS or 3 GHz VLA; and 9/18 sources by our 10 GHz VLA observations.
    \item \textbf{Multi-component}: The intensity map of the radio source shows two or more distinct radio components common to one central engine. 0/12 sources are multi-component in LoTSS; 0/7 in RACS; 0/15 in FIRST/NVSS; 1/13 in VLASS or 3 GHz VLA; and 2/18 in our 10 GHz VLA observations.
\end{enumerate}

Visual inspection of the FIRST intensity map of J1511+2309 source (Figure~\ref{fig:1511+2309}) shows two distinct radio components separated by $\sim 10\arcsec$, or 10 kpc at the redshift of the source. The central component is compact and spatially coincident with the optical nucleus of the host galaxy. The second component is extended in morphology and has no optical counterpart. However, there is no clearly definable jet axis which extends from the primary component in any of the 1.4, 3, or 10 GHz maps that may link the two radio components to a common central engine. The VLASS map, which is closer in resolution to FIRST than the 3 GHz VLA observations of this source, further resolves this second component into two more distinct components, neither of which lie along a preferential axis and appear to be unassociated with one another or the primary component. This is similarly true upon examination of the 3 GHz, higher resolution map produced from VLA observations of J1511+2309. For these reasons, we concluded that the second component detected by FIRST is unassociated with this SPM and did not classify the radio emission associated with this galaxy as multi-component. Similarly, the 144 MHz LoTSS map of \object[WISEA J084344.98+354942.3]{J0843+3549} shows two distinct radio components separated by $27.3\arcsec$. The second radio source, observed to the southwest of the primary source, is associated with the galaxy cluster GMBCG J130.93151+35.82210 \citep{Hao+10} at a redshift of $z=0.475$ \citep{Rozo+15}. We did not classify the 144 MHz morphology of J0843+3549 as multi-component because of this. 

\begin{figure*}[t]
    \centering
    \includegraphics[width=\textwidth, trim=10mm 2mm 10mm 10 mm, clip]{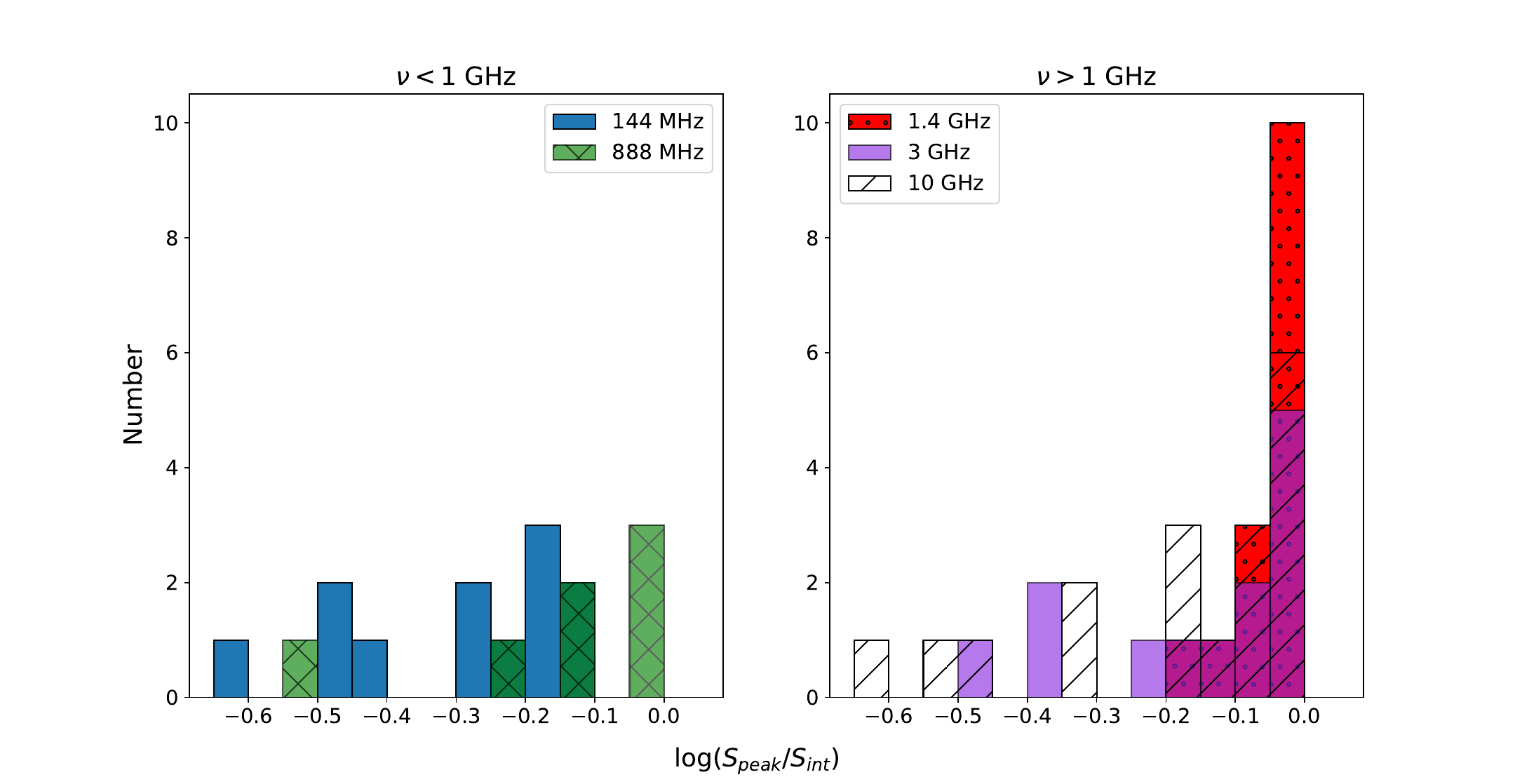}
    \caption{\textit{Left}: Distribution of peak-to-integrated flux density ratios of the 10 GHz-detected SPM galaxies in the LoTSS (solid blue) and RACS (hatched green). Some of the SPM galaxies were not observed in each survey. More negative values indicate resolved structure of the radio source. \textit{Right}: Same as left but for FIRST or NVSS (dotted red), VLASS or 3 GHZ VLA (solid purple), and 10 GHz VLA (hatched).}
    \label{fig:ratios}
\end{figure*}

Figure~\ref{fig:ratios} shows the logarithm of the peak-to-integrated flux density ratios for each of the detected sources. It is clear that at the lowest frequency, 144 MHz, each of the radio sources has an extended emission component. For most of these sources, this extended emission is diffuse and non-axisymmetric, meaning it is unlikely from an AGN. Each of the LoTSS sources, however, still displays an unresolved component that is spatially coincident with the optical center of the host galaxy. \object[UGC 09367]{J1433+3444} is the only source with collimated emission. We discuss this in Section~\ref{sec:J1433+3444}. For the peak-to-integrated flux density ratio of this source, we determined the total integrated flux density of the unresolved, nuclear emission plus the diffuse component by applying a mask to the region of interest in the intensity map within the CASA task \verb|VIEWER|. However, the 144 MHz flux density reported in Table~\ref{table:flux_values} and used in Section~\ref{sec:radio_spec} is only that of the unresolved emission. We did this to mitigate the effects of artificial steepening of the radio spectrum of the nuclear emission, which is the main emission region of interest for our study. For the 7 sources with a detection by RACS, we find a mix of unresolved, resolved, and moderately resolved sources. 

At 1.4 GHz, the majority of sources become unresolved, with the unresolved emission being spatially coincident with the optical nucleus of the host galaxy. This shows that compact, nuclear emission is prevalent in the radio-detected SPM galaxies. However, higher resolution observations at 3 GHz reveal that extended emission is indeed prevalent in the majority of sources (8/13), and the unresolved emission of J1511+0417 resolves into two components. Likewise, our 10 GHz observations further reveal extended emission of these nuclear radio sources. With the high resolution of our observations, many of the formerly unresolved sources at lower frequencies now show a diffuse, non-axisymmetric component to the nuclear radio source (\object[WISEA J075948.40+275019.4]{J0759+2750}, J1015+3914, J1135+2953, J1304+6520 and \object[WISEA J161742.15+251220.5]{J1617+2512}). Like J1511+0417 at 3 GHz, J0843+3549 resolves into two components at 10 GHz.

\subsection{Radio Spectra}\label{sec:radio_spec}
The radio spectrum is a useful tool for interpreting the underlying physical characteristics of the radio source, including the dominant production mechanism for the observed emission. For AGN, the radio emission is dominated by synchrotron emission from a distribution of relativistic electrons, creating a distinctive non-thermal power law spectrum. For older jetted and lobe structures, the highest energy electrons in the distribution will radiate away the fastest, causing a break in the power law at higher frequencies and creating a steep radio spectrum with a power law slope $\alpha<-0.5$. Radio cores, which are associated with the region of emission closest to the active SMBH itself, are actively injected with fresh high energy electrons, creating a flat spectrum with a power law slope $\alpha>-0.5$. For HII regions, thermal emission is dominant at rest-frame frequencies of $\nu>10$ GHz, and is characterized by a power law slope $\alpha\approx-0.1$. At $\nu<10$ GHz, the non-thermal emission from supernova remnants (SNR) dominates, with varying power law slopes. Both mechanisms of emission can be self-absorbed at low frequency, causing a characteristic spectral turnover and inverted slope $\alpha>0$. Identifying and quantifying the power law slope $\alpha$, as well as the curvature and peak frequency, if present, can greatly aid in the interpretation of the radio source.

To explore this parameter space, we constructed a radio SED using the multi-frequency flux density measurements we have tabulated for each of the 18 radio sources we detected with our 10 GHz observations. We have chosen not to include J1117+3757 and \object[WISEA J112434.22+300544.7]{J1124+3005}, since these two SPMs were only detected by LoTSS. 12 of these 18 radio sources were observed and detected in 4 or more of the radio surveys we have used. We considered these 12 to be well-sampled radio spectra for SED analysis.

To constrain the overall shape of the radio SED for these 12 well-sampled radio sources, we have performed a two-fold fitting procedure. First, the spectrum is fit by a simple power law of the form
\begin{equation}
\label{eqn:power_law}
    S_\nu = A \nu^\alpha ,
\end{equation}
where $S_\nu$ is the flux density in mJy, $\nu$ is the observing frequency in GHz, and $\alpha$ is the spectral index value. The second fit describes a parabola in log space and accounts for curvature in the overall shape of the radio spectrum:
\begin{equation}
\label{eqn:curve_pl}
    S_\nu = A\nu^\alpha \mathrm{e}^{q(\ln\nu)^2}\,,
\end{equation}
where $S_\nu$, $\nu$, and $\alpha$ are identical to Eqn.~\ref{eqn:power_law}, and $q$ gives the spectral curvature. For cases of significant curvature, e.\,g., $|q|\geq0.2$ \citep{Duffy&Blundell_12}, $\alpha$ and $q$ lead to a peak frequency $\nu_{peak}$ of
\begin{equation}
\label{eqn:peak_freq}
    \nu_{peak} = \mathrm{e}^{-\alpha/2q}\,.
\end{equation}
Here, $q$ is strictly phenomenological. Physically-motivated synchrotron self-absorption or free-free absorption models, or models with multiple electron populations, would require the use of more free parameters, e.\,g., more flux density measurements at distinct frequencies, than were available for this analysis \citep{Tingay+15}. However, $q$ is still an important constraint to describe the overall shape of the radio spectrum and can hint at the underlying physical mechanism of the radio emission \citep{Callingham+17,Nyland+20,Patil+22}. 

For the remaining 6 sources without well-sampled spectra, the maximum number of detections for a single source across all the surveys used is 3. Then, we could not perform the curved power law fit to these spectra given the paucity of data. In addition, the spectral index values determined by our two-fold fitting procedure are often difficult to interpret for such a wide frequency range, spanning approximately 2 decades in frequency for some sources. To obtain a representative spectral index value, we performed a linear fit to the 3 and 10 GHz flux density values for each of the eighteen 10 GHz-detected radio sources. For those sources without a 3 GHz detection, this estimate provides a lower limit to the actual spectral index value. We chose to use the 3 and 10 GHz flux density values because our ultimate goal is to characterize the nuclear radio emission detected by our high resolution VLA observations. These observing frequencies have the highest angular resolution among the surveys used for our analysis, giving us the best approximation to the true spectral index value of the nuclear radio emission. 

The two-point spectral index $\alpha_3^{10}$ is given by
\begin{equation}\label{eqn:alpha}
    \alpha_3^{10} = \frac{\mathrm{log}(S_3/S_{10})}{\rm{log}(3/10)}\,.
\end{equation}
with an associated error of
\begin{equation}\label{eqn:alpha_error}
    \sigma_\alpha = \frac{1}{\rm{ln}(10/3)}\sqrt{\left(\frac{\sigma_{S_3}}{S_3}\right)^2 + \left(\frac{\sigma_{S_{10}}}{S_{10}}\right)^2}\, ,
\end{equation}
given by standard propagation of errors.

\begin{figure*}
    \centering
    \includegraphics[scale=0.85]{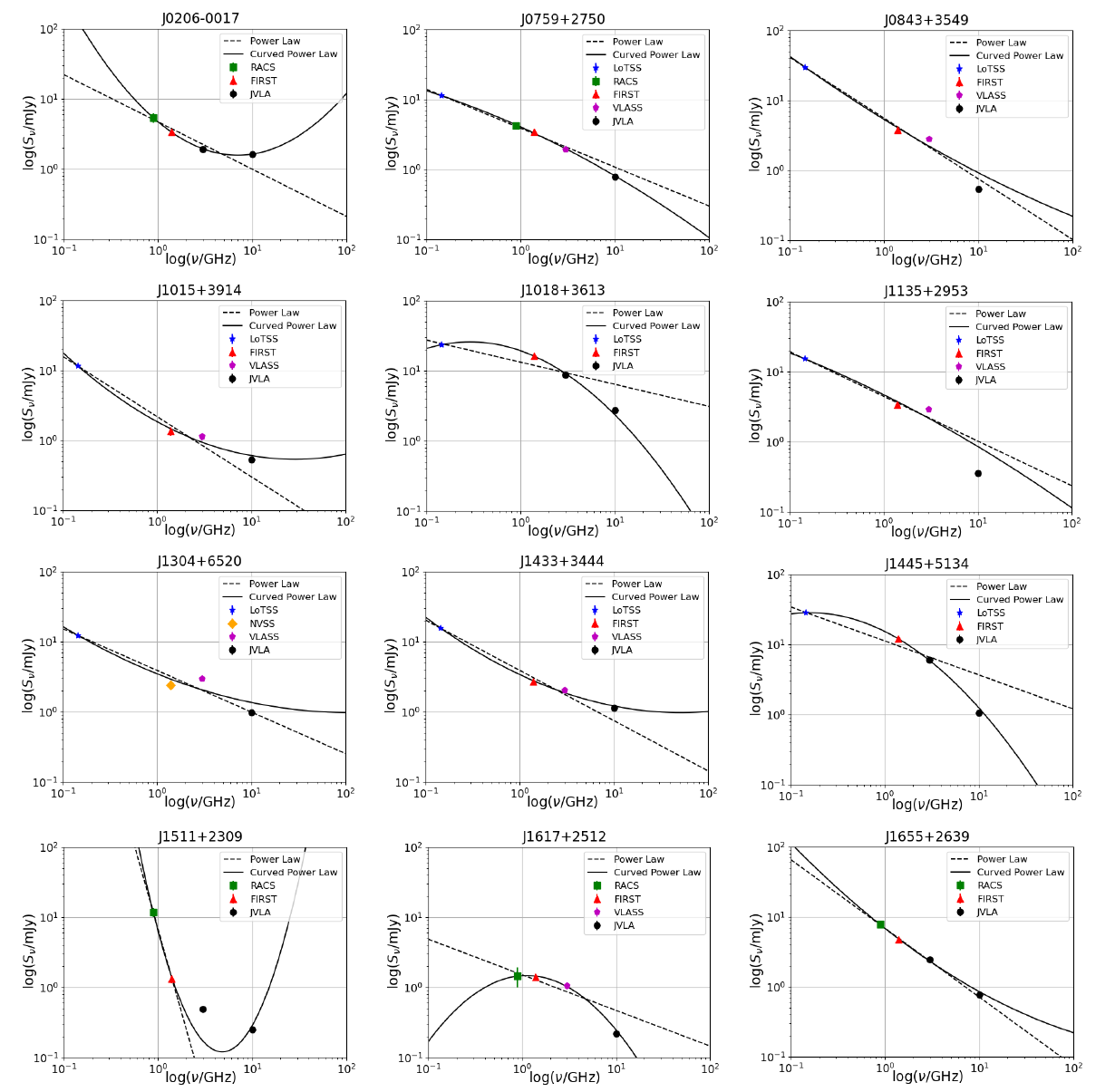}
    \caption{Broadband radio SED for each of the 12 well-sampled radio sources detected by our 10 GHz observations. Radio surveys used are labeled by different markers and colors, with the key in the upper right of each SED. 1$\sigma$ errors are plotted for each flux density measurement. The best-fit power law and curved power law are plotted as the dashed and solid black lines, respectively.}
    \label{fig:seds_alldetect}
\end{figure*}

\begin{figure*}
    \centering
    \includegraphics[scale=0.85]{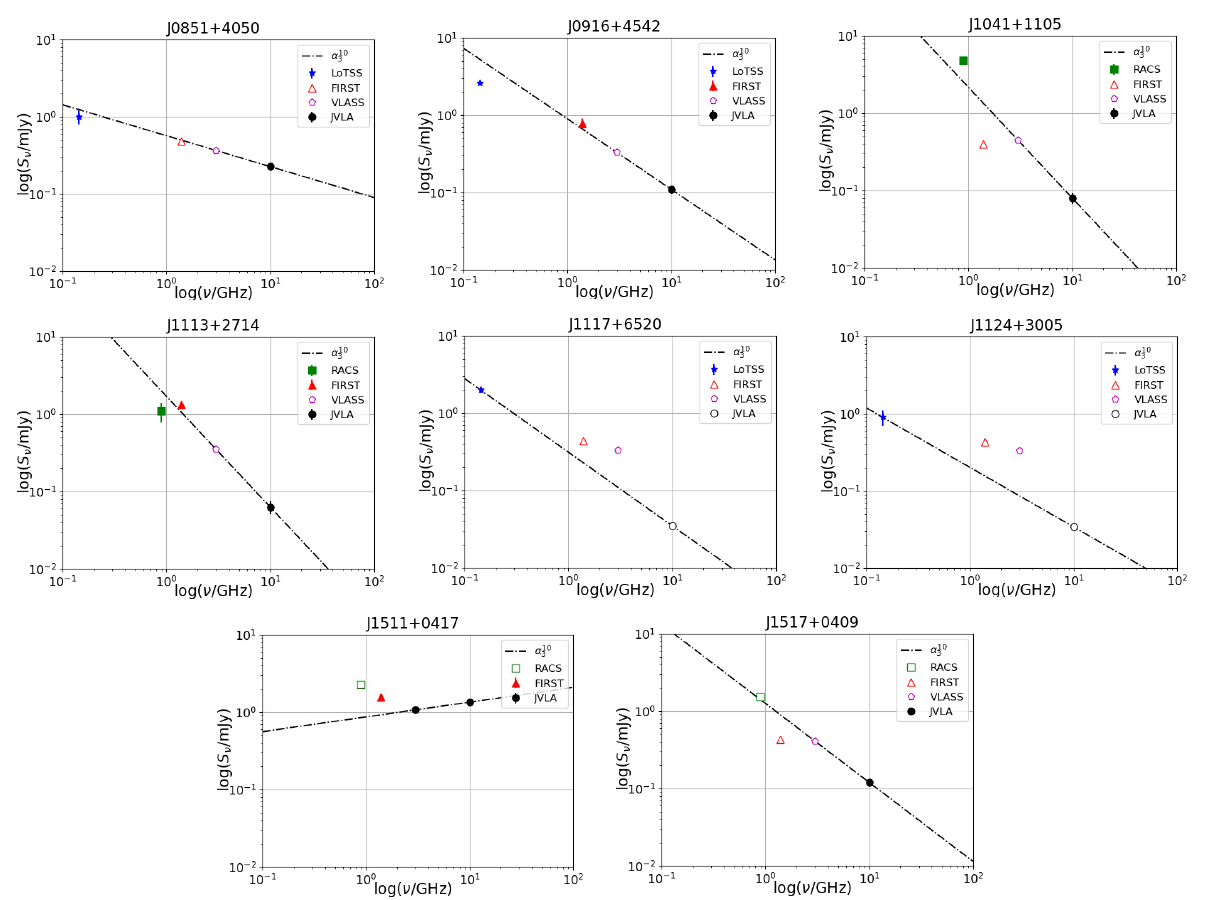}
    \caption{Broadband radio SED for each of the 6 radio sources detected by our high resolution, 10 GHz observations that were a non-detection by 1 or more of the radio surveys used. Radio surveys are labeled by different markers and colors, with the key in the upper right of each SED. Unfilled markers indicate a $3\sigma$ non-detection in that survey. The two-point spectral index $\alpha_3^{10}$ determined from a linear fit to the 3 and 10 GHz fluxes, or their limits, is plotted as the dot-dashed black line.}
    \label{fig:seds_somedetect}
\end{figure*}

Figure~\ref{fig:seds_alldetect} and Figure~\ref{fig:seds_somedetect} show each radio SED and the results of our fitting analyses for the eighteen 10 GHz-detected radio sources. Table~\ref{table:spec_fit} lists the reduced $\chi^2$ values of the power law and curved power law fits for the 12 well-sampled radio spectra. For each of these, the spectral curvature parameter $q$ is provided for those spectra that are better fit by a curved power law than a simple power law ($\chi^2_\mathrm{red,PL} > \chi^2_\mathrm{red,CPL})$. We also list the peak frequency $\nu_\mathrm{peak}$ for the 3 sources that have $q\leq-0.2$. Table~\ref{table:spec_fit} also lists the two-point spectal index value $\alpha_3^{10}$, or its lower limit, for all 18 sources.

We find that 5 of the 12 well-sampled radio SEDs show evidence of significant curvature: J0206$-$0017, J1018+3613, J1445+5134, J1511+2309 and J1617+2512. Of these, J1018+3613, J1445+5134 and J1617+2512 are all peaked spectrum objects, while J0206$-$0017 and J1511+2309 have $q>0.2$, indicative of an inverted spectrum. Visual inspection of the radio SEDs for these two sources show that their actual nature is most likely not inverted. Instead, it's probable that larger-scale emission components with steeper spectra are being resolved out by the higher frequency, higher resolution observations, leaving only the most compact features as the sole contributor to the recovered flux density and causing the spectrum to flatten. Indeed, the flat-spectrum nature of J0206$-$0017 was confirmed by \citet{Walsh+23}. We expect that higher frequency observations of J1511+2309 would confirm the presence of a flat-spectrum object in this source. We did not find evidence of significant curvature in the remaining 7 well-sampled radio SEDs.

The two-point spectral index values $\alpha_3^{10}$ span a range $0.19\geq\alpha_3^{10}\geq-1.74$. The majority of sources (10/13) have a steep spectral index, e.\,g., $\alpha<-0.5$; J0206$-$0017 and J1433+3444 have a flat spectral index, e.\,g., $-0.5\leq\alpha\leq0$; and J1511+0417 has an inverted spectral index value of $\alpha_3^{10}>0$. The spectrum of J1511+0417 is likely truly inverted, unlike the spectra of J0206$-$0017 and J1511+2309, though more flux density measurements are required to confirm this. Of those sources with lower limits on $\alpha_3^{10}$, \object[UGC 04635]{J0851+4050} likely has a flat spectral index given that its $\alpha_3^{10}>-0.41$. The limits for the remaining 4 radio sources leave their radio spectral class ambiguous.

\begin{deluxetable*}{lccccc}[t!]
\tablewidth{0pt}
\tablecaption{Radio Spectral Fitting Parameters}
\label{table:spec_fit}
\tablecolumns{6}
\tablehead{
\colhead{Source} & \colhead{$\chi^2_{\rm PL}$} & \colhead{$\chi^2_{\rm CPL}$} & \colhead{$\alpha_3^{10}$} &
\colhead{$q$} & \colhead{$\nu_{\rm peak}$ (MHz)} \\
\colhead{(1)} & \colhead{(2)} & \colhead{(3)} & \colhead{(4)} & \colhead{(5)} & \colhead{(6)}
}
\startdata
J0206$-$0017 & 76.0 & 0.23 & -0.13 $\pm$ 0.04 & 0.29 $\pm$ 0.01 & -- \\
J0759+2750 & 113 & 1.52 & -0.75 $\pm$ 0.07 & -0.04 $\pm$ 0.02 & -- \\
J0843+3549* & 95.0 & 531 & -1.37 $\pm$ 0.05 & -- & -- \\
J0851+4050 & -- & -- & $> -0.41$ & -- & -- \\
J0916+4542 & -- & -- & $> -0.91$ & -- & -- \\
J1015+3914 & 7164 & 5673 & -0.64 $\pm$ 0.10 & 0.10 $\pm$ 0.07 & -- \\
J1018+3613 & 1048 & 21.8 & -0.97 $\pm$ 0.03 & -0.19 $\pm$ 0.02 & 290 \\
J1041+1105 & -- & -- & $> -1.43$ & -- & -- \\
J1113+2714 & -- & -- & $> -1.42$ & -- & -- \\
J1135+2913 & 1348 & 1538 & -1.74 $\pm$ 0.05 & -- & -- \\
J1304+6520 & 13.9 & 186 & -0.91 $\pm$ 0.07 & -- & -- \\
J1433+3444 & 64.7 & 4.27 & -0.47 $\pm$ 0.07 & 0.08 $\pm$ 0.02 & -- \\
J1445+5134 & 1458 & 19.3 & -1.46 $\pm$ 0.04 & -0.18 $\pm$ 0.02 & 155 \\
J1511+0417* & -- & -- & 0.19 $\pm$ 0.03 & -- & -- \\
J1511+2309 & 217 & 114 & -0.57 $\pm$ 0.08 & 1.61 $\pm$ 0.36 & -- \\
J1517+0409 & -- & -- & $> -1.02$ & -- & -- \\
J1617+2512 & 216 & 4.92 & -1.31 $\pm$ 0.11 & -0.37 $\pm$ 0.08 & 1114 \\
J1655+2639 & 5.80 & 19.5 & -0.98 $\pm$ 0.03 & -- & -- \\
\enddata
\tablecomments{Column 1: Source name. 
Column 2: Reduced $\chi^2$ value for a power law fit to the radio SED. Only reported for sources with 4 or more survey detections.
Column 3: Reduced $\chi^2$ value for a curved power law fit to the radio SED. Only reported for souces with 4 or more survey detections.
Column 4: Spectral index value, and its error, for the optically-thin emission, or its lower limit. This was estimated by using the 3 and 10 GHz flux density values, or the upper limit on the 3 GHz flux for non-detections at this frequency.
Column 5: Spectral curvature parameter determined by the curved power law fit. Only reported if $\chi^2_{\rm red,\,PL} > \chi^2_{\rm red,\,CPL}$.
Column 6: Peak frequency of the radio SED. Only reported if $\chi^2_{\rm red,\,PL} > \chi^2_{\rm red,\,CPL}$ and $q < 0$. \\
Note: J0843+3549 and J1511+0417 are resolved into two distinct components at 10 GHz. For both, we used the dominant, southern component to calculate $\alpha_3^{10}$. See Figure~\ref{fig:0843} and Figure~\ref{fig:1511+0417} for the 10 GHz contour maps of J0843+3549 and J1511+0417, respectively. \\
}
\end{deluxetable*}

The radio SED of J0843+3549, J1015+3914, J1135+2953, and J1304+6520 each show a significant deviation from their best-fit simple and curved power laws at 1.4 GHz. This is reflected in the abnormally high reduced $\chi^2$ values for the fits of these sources. This can be interpreted in two ways. First, the significant deviation at 1.4 GHz may be explained by intrinsic variability of the radio source. The observations at 144 MHz (LoTSS), 888 MHz (RACS), 3 GHz (VLASS/VLA), and 10 GHz (VLA) are quasi-contemporaneous; at most, the observations were taken within 5 years of one another. Yet, the 1.4 GHz observations, by either FIRST or NVSS, were conducted well over a decade ago, at least, at the time of this analysis. If the source underwent significant variability over a years-long timescale prior to its most recent observation at 1.4 GHz, all of the flux density measurements at frequencies besides 1.4 GHz would reflect this. It is possible, then, that these sources have naturally varied over this intervening time span and are no longer well fit to the quasi-contemporaneous data points of the other surveys. The corresponding variability amplitudes at 1.4 GHz, assuming a power law fit to the spectrum without the 1.4 GHz flux density properly describes the spectral shape, range from 12\%-61\%. These amplitudes are certainly plausible, given that some sources have been found to reach variability amplitudes higher than 2400\% at this observing frequency \citep{Nyland+20}. Alternatively, the flux densities at lower frequencies are representative of a separate electron population. For example, LoTSS sources in the local universe will be dominated by diffuse emission associated with star-formation processes. Using higher frequency and higher spatial resolution observations, this diffuse emission will be resolved out. If there is a second, distinct population of electrons producing a radio source that is more compact and hosted by the same SPM, this will become apparent by a break in the broadband radio spectrum. Essentially, the two electron populations, both of which are located within the same host galaxy, are confused with each other at low frequency, and only the high frequency observations we have used are truly representative of the second, more compact, nuclear source. Follow-up observations at high angular resolution with better frequency sampling, including 1.4 GHz, are required to distinguish between the two methods we have outlined that may produce this observed break in the radio spectrum. The nature of these sources is further discussed in Section~\ref{sec:org_of_re}. 

\section{Origin of Radio Emission}
\label{sec:org_of_re}
\begin{figure}[t!]
    \centering
    \includegraphics[width=\linewidth, trim=5mm 1mm 12mm 11mm, clip]{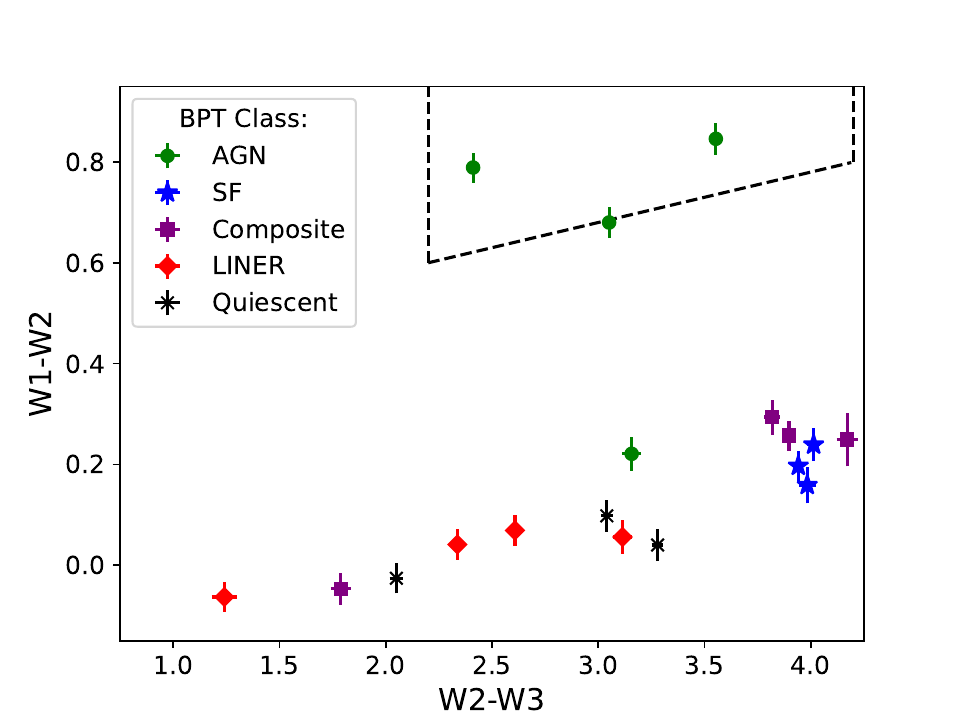}
    \caption{\textit{WISE} color-color diagram for the eighteen 10 GHz-detected SPM galaxies in our sample. The dashed black lines define the region of mid-IR AGN taken from the selection criterion of \citet{Jarrett+11}. Two of our sources, J0206$-$0017 and J0843+3549, are selected as mid-IR AGN, while a third, J1018+3613, falls in this region within $1\sigma$ error. Each of these has multi-wavelength evidence for an AGN. We conclude that their radio emission is associated with the AGN and remove these sources from further analyses to avoid possible introduction of systematics by the AGN to the SFR calculation of each host SPM.
    }
    \label{fig:wise_cc}
\end{figure}

In this section, we assess the origin of the radio emission in our SPM sample. Our analysis will make use of mid-IR fluxes available from the ALLWISE source catalog to perform a dust correction to the far-UV (FUV) flux for each SPM, for which we wish to calculate the star-formation rate (SFR). In this paradigm, active star formation heats dust grains in the surrounding interstellar medium that re-radiate this energy as thermal emission in the mid-IR. However, it is possible for an AGN to also assume this heat engine role and this will introduce systematic effects into our calculation of the SFR using the FUV fluxes. As such, we first identify if any of our sources are AGN by mid-IR selection criterion. To do this, we utilize a \textit{WISE} color-color diagram to search for mid-IR AGN using the selection criterion of \citet{Jarrett+11} for each of the eighteen 10 GHz-detected radio sources. These results are shown in Figure~\ref{fig:wise_cc}. We find that J0843+3549 and J0206$-$0017 are within the mid-IR AGN box of \citet{Jarrett+11}, and J1018+3613 is consistent with a mid-IR AGN within errors. The remaining 15 sources are well outside of the AGN region and occupy the region of star-forming galaxies, e.\,g., spiral galaxies, luminous infrared galaxies (LIRGS) and starburst/ultraluminous infrared galaxies (ULIRGS) \citep{Wright+10}. J0206$-$0017 \citep{Osterbrock_81,Cohen+86,McElroy+16, Walsh+23}, J0843+3549 \citep{VC&Veron_01,Stern&Laor_12,Koss+18}, and J1018+3613 (\citealt{Stern&Laor_12}, Walsh et al. in prep) are all known AGN, in addition to being identified as Seyfert AGN via their emission-line ratios as discussed in Section~\ref{sec:em_line}. To mitigate the impact of potential systematics introduced by the mid-IR AGN to our SFR calculation, we conclude that the radio emission in each of these sources is associated with the AGN and remove them from the analyses described in this section. For each of the remaining fifteen 10 GHz-detected radio sources, we consider the following origin scenarios for their radio emission: thermal emission from star-forming regions, synchrotron emission from individual radio supernova (RSN) or a population of supernova remnants (SNRs), or an AGN.

\subsection{Radio Excess}
\label{sec:radio_excess}
We begin by searching for excess radio emission in each of the remaining 15 SPM radio sources. To do this, we predict what the SFR for each of the SPMs would be from their 1.4 GHz luminosity and compare this radio-predicted SFR to that calculated from the FUV emission of the host galaxy. Sources that have an over-prediction of the SFR from their radio luminosity are called radio excess. These radio excess sources cannot be explained from SF processes alone, while those that do not show radio excess can be, though aren't necessarily.

We first predict the radio-based SFR using the 1.4 GHz luminosity. To do this, we use Equation 17 of \citet{Murphy+11}:
\begin{equation}
    \left(\frac{\mathrm{SFR}_\mathrm{1.4GHz}}{M_\odot\, \mathrm{yr}^{-1}}\right) = 6.35 \times 10^{-29} \left(\frac{L_\mathrm{1.4GHz}}{\mathrm{erg\, s^{-1}\, Hz^{-1}}} \right)\, .
\end{equation}
This SFR is based on the FIR-radio correlation, which relates the galactic FIR properties to the galactic radio continuum properties. \citet[][]{Murphy+11} note that the expected contribution to the total radio emission from non-thermal processes is negligible for some cases in which the emission is co-spatial with an active HII region. \citet[][]{Condon&Yin_90} argue that this is the case only for small HII regions, for which stars with $M > 8 M_\odot$ can escape before exceeding their lifetime of $<3\times10^7$ yr. For our sources that are identified as SF or SF-AGN composite galaxies via their optical emission-line ratios, the SDSS spectroscopic fiber has a diameter of $3\arcsec$. We only know that within the galactic region covered by the spectroscopic fiber, an active HII region is, at least partially for composite galaxies, contributing to the ionization. The area covered by the SDSS fiber is much larger than the synthesized beamwidth of our 10 GHz VLA observations ($\sim0.2\arcsec$). Because none of our sources show features comparable in angular size to the SDSS fiber, for the consistency of the analysis, we continued under the assumption that the HII region is large enough such that there could be significant non-thermal radio emission spatially coincident with it.

To estimate the radio-based SFR, we use the 1.4 GHz luminosity values derived from the FIRST or NVSS, for J1304+6520, catalog entry for each source. We do this instead of extrapolating to the 1.4 GHz luminosity using $\alpha_3^{10}$ because, as mentioned in Section~\ref{sec:radio_spec}, some sources exhibit clear breaks from their best-fit power law at 1.4 GHz (J0843+3549, J1015+3914, J1135+2953, and J1304+6520), and the synthesized beams of FIRST and NVSS, $5\arcsec$ and $45\arcsec$, respectively, are better matched to galaxy-scale properties than the synthesized beam of our 10 GHz observations. It is important to most accurately trace the galaxy-scale radio emission because the radio-FIR correlation has been shown to deviate from a linear correlation for regions of radio emission with low thermal fractions \citep[][]{Hughes+06}.  At 1.4 GHz, the expected thermal contribution to the total radio emission for any radio source is approximately 5 to 10\% \citep{Condon_92, Murphy_13}. This is true even for starburst systems, for which \citet{Murphy_13} found a thermal fraction of 5\% in a sample of 31 local starburst galaxies. We assume, then, that the radio sources we have detected are not extraordinary in this regard, and have low thermal fractions at 1.4 GHz. However, the spatial scale range for which \citet{Hughes+06} found the radio-FIR correlation to deviate from a linear correlation is from 50-250 pc for their low thermal fraction radio sources. At $5\arcsec$ resolution, the smallest spatial scale probed for our 18 source sample is approximately 3 kpc, or an order of magnitude larger than what was found by \citet{Hughes+06}. Because of this, we do not expect any deviations from the standard radio-FIR correlation using the 1.4 GHz luminosity for our sources.

We now calculate the host galaxy SFRs for 13 SPMs that had FUV measurements available from \textit{GALEX}. The remaining 2 did not have \textit{GALEX} measurements available and we describe the calculation of their SFRs later on in this section. We first correct the FUV luminosity for dust absorption by using the 25$\mu$m \textit{WISE} luminosity for each source, following \citet[][]{Hao+11}:
\begin{equation}
    L(\mathrm{FUV})_\mathrm{corr} = L(\mathrm{FUV})_\mathrm{obs} + 3.89 L(25\mu \mathrm{m})\, , 
\end{equation}
where all luminosity values are in units of erg s$^{-1}$. Here, we have used the available \textit{WISE} 22$\mu$m flux density as a proxy for the 25$\mu$m luminosity, since the flux density ratio between these two values is expected to be unity for early-type galaxies \citep[][]{Jarrett+13}. After calculating the \textit{WISE}-corrected FUV luminosity, we find the host galaxy SFR following Table 1 of \citet[][]{Kennicutt&Evans_12} for the FUV band:
\begin{equation}
    \left(\frac{\mathrm{SFR}}{M_\odot\, \mathrm{yr}^{-1}}\right) = 4.5 \times 10^{-44} L(\mathrm{FUV})_\mathrm{corr}\,  .
\end{equation}
Using this method, the 1$\sigma$ uncertainty on the SFR is 0.13 dex \citep[][]{Hao+11}. For the two SPM detections without available \textit{GALEX} FUV measurements, J1304+6520 and J1511+2309, we use the H$\alpha$ luminosity to calculate the host galaxy SFR. \citet[][]{Kennicutt+09} provide a dust-attenuated correction to the H$\alpha$ luminosity using the 25 $\mu$m luminosity:
\begin{equation}
    L(\mathrm{H}\alpha)_\mathrm{corr} = L(\mathrm{H}\alpha)_\mathrm{obs} + 0.020 L(25 \mu \mathrm{m})\, .
\end{equation}
As before, all luminosity values are in erg s$^{-1}$. For $L(\mathrm{H}\alpha)_\mathrm{obs}$, we use the values provided by the OSSY catalog \citep{Oh+11} for each of the two optical spectra. We again follow Table 1 of \citet[][]{Kennicutt&Evans_12} to calculate the host galaxy SFR using the dust-corrected H$\alpha$ luminosity:
\begin{equation}
    \mathrm{SFR} = 5.37 \times 10^{-42} \left(\frac{L(\mathrm{H}\alpha)_{corr}}{\mathrm{erg\, s^{-1}}}\right)\, .
\end{equation}
The 1$\sigma$ uncertainty for the H$\alpha$ method is 0.4 dex \citep[][]{Kennicutt+09}.

\begin{figure}[t!]
    \centering
    \includegraphics[width=\linewidth, trim=7mm 1mm 15mm 6mm]{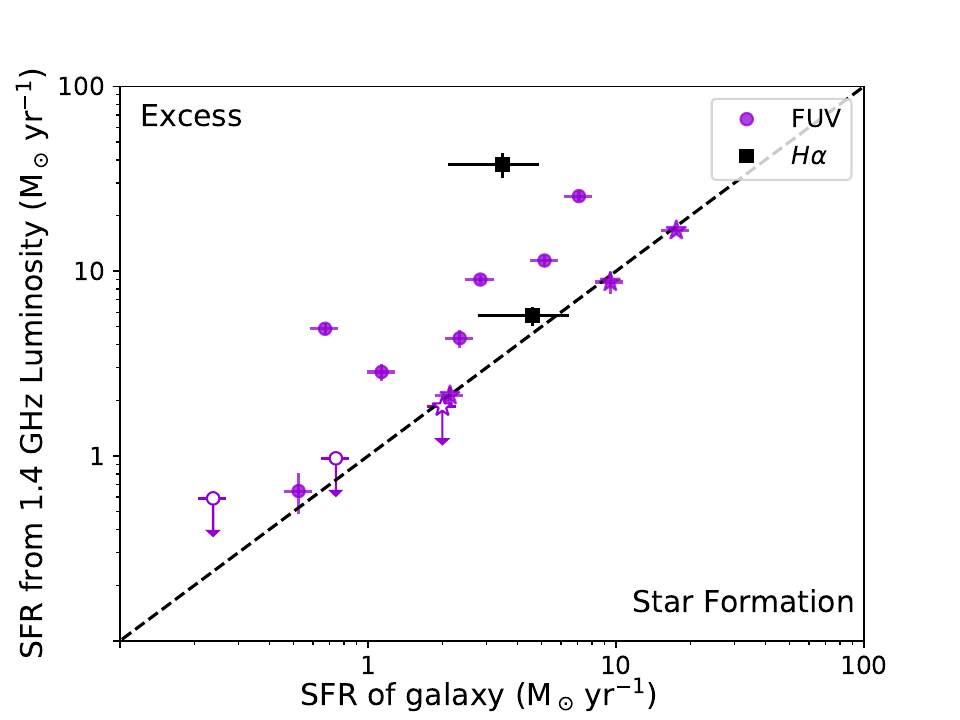}
    \caption{Comparison of the 1.4 GHz, radio-based SFRs to the host galaxy SFRs for 15 of the 10 GHz-detected radio sources in our sample of SPMs. J0206$-$0017, J0843+3549, and J1018+3613 were removed from this analysis due to the presence of an IR AGN. Host galaxy SFRs were determined using either the IR-corrected FUV (purple circles) or H$\alpha$ luminosity (black squares). Unfilled data points are non-detections at 1.4 GHz, and their radio-based SFRs are upper limits determined using a $3\sigma$ detection threshold from FIRST. Points above the dashed line exhibit radio excess, while those below do not. 
    }
    \label{fig:radio_origins}
\end{figure}

The radio-based SFR is plotted against the galaxy-based SFR for each of our 10 GHz-detected SPM sources in Figure~\ref{fig:radio_origins}. The host galaxy SFRs are in the range $0.2\, \mathrm{M}_\odot\, \mathrm{yr}^{-1} \leq$ SFR $\leq 17\, \mathrm{M}_\odot\, \mathrm{yr}^{-1}$. We find 4 radio sources that do not have excess radio emission, indicating that their radio emission could be explained by star-formation processes alone. These are: J1015+3914, \object[VIII Zw 101]{J1041+1105}, J1445+5134, and J1617+2512. We note that J0851+4050, J1041+1105 and \object[UGC 09804 NED01]{J1517+0409} are non-detections at 1.4 GHz, and thus their radio-based SFRs are upper limits. 9 of our sources fall on either side of the radio-excess line within their 1$\sigma$ errors, including all of those sources that do not show excess radio emission. Only 4 sources have excess radio emission above a factor of $3\sigma$ from what is expected by star-formation process. These are: J0759+2750, J1304+6520, J1433+3444, and J1655+2639. Now that we have identified which radio sources do or do not show excess radio emission, we seek to answer what physical process is the dominant means of radio emission production for each. 

\subsection{Thermal Emission from Star Formation}
\label{sec:thermal}
The first consideration for the origin of the radio emission is that of thermal bremsstrahlung (free-free) emission produced by ionized hydrogen in active star-forming regions. In the optically-thin regime, radio emission that is dominated by a free-free component is characterized by a flat spectral index of $\alpha = -0.1$ \citep[][]{Condon_92,Murphy+11,Klein+18}. Each of the 4 non-radio-excess sources are in the optically-thin regime at GHz frequencies, as indicated by their broadband radio SEDs (see Figure~\ref{fig:seds_alldetect} for J1015+3914, J1445+5134, and J1617+2512, and Figure~\ref{fig:seds_somedetect} for J1041+1105). However, their optically-thin spectral index values, $\alpha_3^{10}$, range from -1.46 to -0.64, with J1041+1105 having a lower limit of -1.43. Although there may be a contribution from free-free emission in each of these radio sources, it is clear from their spectral index values that free-free emission is not the dominant radio emission mechanism for any.
This is not unexpected, as free-free emission from HII regions does not usually dominate the radio SED for $\nu<10$ GHz \citep{Condon_92, Murphy_13}.

\subsection{Radio Supernova and Supernova Remnants}
\label{sec:rsn}
Our second consideration for the origin of the nuclear radio emission in our SPM sources is from non-thermal emission produced by either an individual radio supernova (RSN) or a population of supernova remnants (SNRs). We first consider an individual RSN as the progenitor for the radio emission. RSN are morphologically compact radio sources that span a range of radio luminosity \citep[][]{Weiler+02} and spectral index values \citep[][]{Bendo+16, Klein+18, Galvin+18, Emig+20}. The radio emission associated with a RSN is powered by synchrotron processes. Generally, for star-forming galaxies, this synchrotron emission is diffuse, tracing the host galaxy's morphology. In the optically thin regime, RSN associated with a Type Ib/c event have $\alpha < -1$ ($S_\nu \propto \nu^\alpha$), while those associated with a Type II event have a shallower spectral index $\alpha > -1$ \citep[][]{Weiler+02}. The radio luminosity of an individual RSN will peak a few 100s of days after the initial SN explosion, reaching a maximum 5 GHz luminosity of $L_\mathrm{6cm\, peak} \approx 1.3 \times 10^{27}$ erg s$^{-1}$ Hz$^{-1}$ \citep[][]{Weiler+02}. However, two of the most luminous RSN, SN1998bw \citep[][]{Kulkarni+98} and PTF11qcj \citep[][]{Corsi+14, Palliyaguru+19}, have a peak luminosity value as high as $10^{29}$ erg s$^{-1}$ Hz$^{-1}$ at 5 GHz. We take this luminosity to be the upper limit to what a RSN can achieve and compare this to the extrapolated 5 GHz luminosity for each of our SPM detections.

To extrapolate to the 5 GHz luminosity, we use $\alpha_3^{10}$ for each source (Table \ref{table:spec_fit}, column 4) determined by performing a linear fit to the log of the 3 and 10 GHz flux density values. After extrapolation, 2 of the 15 detections have a 5 GHz luminosity greater than $10^{29}$ erg s$^{-1}$ Hz$^{-1}$: J0759+2750, and J1304+6520. All of the sources with a lower limit to $\alpha_3^{10}$ are below this luminosity. The remaining 13 have a median luminosity value of $4.3\times 10^{28}$ erg s$^{-1}$ Hz$^{-1}$, which is only a factor of 2.3 lower than this RSN luminosity limit. While the spectral index and luminosity values do not rule out the individual RSN origin for these 15 sources, it is extremely unlikely that each radio source is associated with an individual, extremely luminous, nuclear RSN. Nonetheless, we pursue a more robust argument to rule out this scenario.

\citet[][]{Chomiuk&Wilcots_09} determined an expression that relates the maximum 1.4 GHz luminosity of a RSN to the SFR of its host galaxy:
\begin{equation}
    L_\mathrm{1.4\, GHz}^\mathrm{max} = \left(95^{+32}_{-23} \right) \mathrm{SFR}^{0.98 \pm 0.12}\, ,
\end{equation}
where the 1.4 GHz luminosity is in units of 10$^{24}$ erg s$^{-1}$ Hz$^{-1}$ and the SFR is measured in M$_\odot$ yr$^{-1}$. 

For a given SFR, we first use this relation to determine the maximum 1.4 GHz luminosity of a RSN, then extrapolate this to a 10 GHz luminosity to compare to our VLA sources. There is some freedom here in which value to choose for the spectral index. \citet[][]{Chomiuk&Wilcots_09} use $\alpha = -0.5$ when deriving the synchrotron emission from a RSN. This comes from the assumption that the cosmic ray (CR) energy spectrum is a power law of the form E$^{-2}$, which gives a synchrotron spectral index of $\alpha_\mathrm{syn}=-0.5$. However, focusing on the most luminous RSNs, SN1998bw has a steeper spectral index of $\alpha_\mathrm{syn}=-0.75$ \citep[][]{Chevalier&Fransson}, and PTF11qcj has a varying late-time spectral index $\alpha_\mathrm{syn} \gtrsim -1$ \citep[][]{Corsi+14}. \citet[][]{Bjornsson_13} note that the spectral index should approach a value of $\alpha_\mathrm{syn}=-1$ in the optically-thin regime, and, indeed, this is in agreement with those values listed in Table 1 of \citet[][]{Chevalier&Fransson}. For our analysis, we have used a spectral index value of $\alpha_\mathrm{syn}=-0.5$, as is done in \citet[][]{Chomiuk&Wilcots_09}. We chose this spectral index value because it is the shallowest among those discussed. Thus, if any of our sources lie above the extrapolated RSN luminosity using a spectral index value of $\alpha_\mathrm{syn}=-0.5$, they will certainly do the same for a steeper spectral index value. Using this method, we find that the observed 10 GHz luminosity of each radio source is greater than the expected luminosity of an individual RSN by at least a factor of 80. It is evident that an individual, luminous RSN is not responsible for the radio emission in these SPM galaxies.

For the 4 radio sources that do not show excess radio emission, it is likely, then, that their radio emission is produced by a population of SNRs. Only one of these radio sources (J1015+3914) is hosted by a star-forming galaxy as determined by its optical emission-line ratios. Two are identified as SF-AGN composite galaxies (J1445+5134, J1617+2512), and the remaining is a LINER (J1041+1105).

The median spectral luminosity at 10 GHz of these 4 radio sources is $1.4\times10^{28}$ erg s$^{-1}$ Hz$^{-1}$. For comparison, the nuclear starbursts identified by \citet{Song+22} in sample of 63 local (U)LIRGS with SFRs in the range of $0.14-13\, \mathrm{M_\odot}\, \mathrm{yr}^{-1}$ have a median spectral luminosity of $5.8\times10^{27}$ erg s$^{-1}$ Hz$^{-1}$, or about a factor of 2 lower than what we have found for our radio SF sources. However, higher luminosity radio SF sources do exist: NGC 4945, a powerful, local starburst with a SFR of 1.5 $\mathrm{M_\odot}\, \mathrm{yr}^{-1}$, has a higher spectral luminosity at 10 GHz of $2.4\times10^{28}$ erg s$^{-1}$ Hz$^{-1}$ \citep{Lenc&Tingay_09}. To emphasize, only J1015+3914 may be powered by SF processes alone, as determined by its emission-line ratios. The other three may have contributions from another ionization process, e.\,g., an AGN. Interestingly, this source has the highest 10 GHz luminosity ($4.3\times10^{28}$ erg s$^{-1}$ Hz$^{-1}$) of any of the non-excess sources.

Among these sources with a detection in the FIRST catalog, none are resolved at the spatial scales probed by the $5\arcsec$ FIRST beam. That is, they do not display the diffuse emission morphology that is characteristic of synchrotron emission from SNRs. This morphology does become more apparent at low frequency (LoTSS or RACS), and at higher spatial resolution for a few sources, e.\,g., J1015+3914 (Figure~\ref{fig:1015}), though remains elusive at GHz frequency in others, e.\,g., J1041+1105 (Figure~\ref{fig:1041}) and J1617+2512 (Figure~\ref{fig:1617}). This is discussed in more detail for each individual source in Appendix~\ref{sec:app_a}. When examining the ratio of peak flux to integrated flux density however, all of these sources become resolved at 10 GHz (see Figure~\ref{fig:ratios}). Deeper observations at 1.4 and 3 GHz may reveal lower surface brightness emission indicative of this characteristically diffuse nature.

\subsection{AGN}
\label{sec:agn}
For the 11 radio-excess sources, it is likely that their radio emission is dominated by an AGN component. We did not find a one-to-one match between the radio AGN classification and that derived from our BPT analysis (Section~\ref{sec:em_line}). Instead, the radio AGN are found to occupy all categories of the BPT classifications: \object[WISEA J111340.54+271428.3]{J1113+2714} and \object[UGC 06557]{J1135+2913} are SF; J0759+2750, \object[UGC 04892]{J0916+4542} are SF-AGN composites; J1304+6520, J1433+3444 are Seyfert AGN; J0851+4050, J1511+0417 are LINERs; and J1511+2309, J1517+0409, J1655+2639 are hosted by quiescent systems. In addition to these 11, we have already concluded that the radio emission in J0206$-$0017, J0843+3549, and J1018+3613 is each associated with an AGN. From the emission-line perspective alone, this indicates that radio AGN activity may be present during an ongoing or recent stage of star formation activity in the host SPM. In total, we find that the nuclear radio emission for 14 of the 18 sources at 10 GHz, or 78\%, is dominated by a radio AGN.

We begin by discussing the three radio sources hosted in quiescent emission-line galaxies: J1511+2309, J1517+0409, and J1655+2639. As noted before in Section~\ref{sec:em_line}, the spectra of J1511+2309 and J1655+2639 contain H$\beta$ absorption that is most likely from a stellar origin due to its absorption trough being centered on the rest-frame H$\beta$ wavelength. Because there is not a significant detection of any H$\beta$ emission, as reported by OSSY \citep{Oh+11}, in these two spectra, we have classified them as quiescent, even though the [OIII]$\lambda5007$, H$\alpha$, and [NII]$\lambda6583$ lines are detected with a S/N ratio $>3$. The radio emission properties, however, indicate the clear presence of radio AGN activity, as we discuss further in this section. J1517+0409 is unique among these quiescent emission-line SPMs and radio AGN. The optical spectrum for this SPM is largely featureless: only the [NII]$\lambda6583$ line is detected with a S/N ratio $>3$. Additionally, it was a non-detection by each of the archival radio surveys that observed it (RACS, FIRST, VLASS), making it the only 10 GHz-detected source with no additional radio detection(s). The radio-based SFR of J1517+0409 is within the radio-excess area shown in Figure~\ref{fig:radio_origins}, though this was determined by using a 3$\sigma$ detection limit to the FIRST image of this source. With more sensitive 1.4 GHz observations, it is likely that J1517+0409 would not show radio-excess, and the origin of its radio emission would need to be re-examined. However, considering the data available during this analysis, we count J1517+0409 to be among the likely radio AGN discovered by our observations. 

\subsubsection{Radio AGN Morphologies} \label{sec:agn_morph}
For these radio AGN, almost all of them display only a single-component morphology except for J0843+3549, J1433+3444, and J1511+0417. The LoTSS map of J1433+3444 alone displays the extended morphology characteristic of AGN (Figure~\ref{fig:1433}.) The majority of these radio AGN do not have collimated jets and/or extended lobe emission that is easily and clearly identifiable in any of their intensity maps (see Appendix~\ref{sec:app_a} for all intensity maps). This is perhaps unsurprising given that almost all of our 10 GHz sources are radio-quiet, whereas radio-loud AGN are nearly ubiquitously associated with highly relativistic emission arising from radio jets. Yet, the majority of our sources are also not dominated by flat-spectrum radio cores. These objects are identified via their unresolved radio emission, flat spectral index ($\alpha \geq -0.5$), and high brightness temperature, indicative of non-thermal emission, and are almost always associated with an AGN. J0206$-$0017, J1433+3444 and J1511+0417 are the most likely sources to contain a dominant radio core when considering their unresolved morphology and flat spectral index values (Table~\ref{table:spec_fit}). Indeed, this is the case for J0206$-$0017, as VLBI observations by \citet{Walsh+23} revealed that the radio emission remains compact down to pc scale, retains its flat spectral index, and shows chromatic variation in its position, confirming its radio-core nature. J0851+4050, J0916+4542, J1113+2714 and J1517+0409 only have lower limits to their spectral index value so the true nature of their radio AGN emission is ambiguous.

It is evident from the ratio of peak-to-integrated flux density at 10 GHz that some of our single-component sources are moderately resolved (Figure~\ref{fig:ratios}). Since there is little evidence for kpc-scale radio jets or lobes at GHz frequencies, aside from J0843+3549 and J1511+0417, this moderately resolved nature of the nuclear emission could be an indication of young or frustrated radio jets, which are confined to only the central, sub-kpc region of their host galaxy. J1655+2639 is the most identifiable example of this scenario. The radio emission is moderately resolved along a linear feature (VLA X panel of Figure~\ref{fig:1655}) that has a steep spectral index of $\alpha_3^{10}=-0.97 \pm 0.03$. 

Alternatively, the moderately resolved nature of some of these radio AGN sources can be explained by the presence of non-thermal emission associated with star formation. As noted, 2 of our radio AGN sources are hosted by SF emission-line galaxies, and 2 more are classified as SF-AGN composite. J0759+2750, hosted by a SF-AGN composite galaxy, is perhaps an archetypal source for SF and AGN activity because its 1.4 GHz radio luminosity is $>3\sigma$ higher than what is expected from star formation alone, but the 10 GHz morphology shows both a diffuse, non-linear component and an unresolved component (VLA X panel of Figure~\ref{fig:0759}). J0916+4542, also hosted by a SF-AGN composite galaxy, is similarly resolved by its peak-to-integrated flux density ratio at 10 GHz, although the point-like morphology makes the nature of the diffuse emission unclear (Figure~\ref{fig:0916}).

Like J0759+2750, J1135+2913, hosted by a SF galaxy, shows low surface brightness emission in its 10 GHz map (VLA X panel of Figure~\ref{fig:1135}), though its radio emission is only a factor of 2.2$\sigma$ higher than expected from star formation. However, the diffuse emission in J1135+2913 forms a linear feature. This is particularly evident once this source is imaged with a \textit{uv}-taper to create a lower resolution map at 10 GHz. It is likely that this feature is a radio jet associated with the radio AGN, like J1655+2639, though we cannot rule out that there is no contribution to the diffuse radio emission from star-formation processes. 

The diffuse emission of J1304+6520 is more difficult to interpret (Figure~\ref{fig:1304}). Like J0759+2750, J1304+6520's 1.4 GHz radio luminosity is $>3\sigma$ higher than what is expected from star formation alone. However, unlike J0759+2750, J1304+6520 is hosted by a Seyfert AGN emission-line galaxy. So, if this diffuse emission is from SF processes, it is not evident that such would be the case from its optical emission-lines. To test if this diffuse emission would resolve into a jet-like feature, we created 10 GHz maps using different weighting schemes. However, these maps did not provide clear evidence of a radio jet. Further observations are required to ascertain the nature of this radio emission. For this analysis, we cannot conclusively determine the emission is from a jet, and do not count it as such as a result.

Lastly, we discover a potential precessing radio jet in J0843+3549. The emission of the compact feature is moderately resolved in all of its intensity maps, and the position angle (PA) of this moderately resolved feature changes from a range of 131$\degr$-151$\degr$ at kpc scales to $-18\degr$ in our 10 GHz intensity map, nearly aligning with the second radio component (Figure~\ref{fig:0843}). The different observing frequencies probe different periods in the AGN's evolution because of their differing resolution elements. Thus, the change in PA from the largest spatial scales ($2.5\arcsec-6\arcsec$ resolution) to the smallest ($\sim0.25\arcsec$) is indicative of time evolution in the PA of the radio jet. The near-alignment of the second radio component with the moderately-resolved structure of the primary points towards a common origin for both emission features. To confirm this precessing jet, further observations are needed over a wide frequency and spatial resolution coverage.

Then, 4 of the 14 radio AGN in our sample (29\%) show evidence for a compact radio jet from their morphology.

\subsubsection{Dual AGN Candidates}\label{sec:dagn}
J0843+3549 and J1511+0417 are both radio doubles; that is, they show two morphologically distinct radio components (see Figure~\ref{fig:0843} and Figure~\ref{fig:1511+0417}). For both, the radio AGN is likely hosted by the southern, dominant component in each source. The two point spectral index value $\alpha_3^{10}$ for these dominant components is $-1.37 \pm 0.05$ and $0.19 \pm 0.03$ for J0843+3549 and J1511+0417, respectively. J0843+3549 is thus a candidate compact steep spectrum (CSS) object, which are compact radio sources less than 20 kpc in linear size and have $\alpha \leq -0.5$ \citep{O'Dea&Saikia_21}. J1511+0417 is host to a radio core, as is evident by the inverted spectral index and unresolved morphology of the southern component. The two components are separated by 1.6 kpc for J0843+3549, and 2.1 kpc for J1511+0417. For both systems, there is no clear radio emission which connects the dominant source to the weaker one. J1511+0417 is particularly noteworthy because of the co-location of the northern radio component with a second optical nucleus. Although this galaxy merger is classified as a post-merger system, which are defined as containing only a single optical nucleus, it's clear through the visual identification of distinct optical nuclei that this system is in an earlier stage of galaxy merger evolution, prior to the merging of the stellar nuclei. Indeed, this is corroborated by the \textit{GAIA} photometric catalog containing an additional source identification at the location of the northern optical nucleus. The northern radio component has a two point spectral index of $-0.60 \pm 0.04$, also making it a candidate CSS object. These characteristics make J1511+0417 a candidate dual AGN.

J0843+3549 has also been identified as a dual AGN candidate by previous work. Using deep NIR imaging, \citet{Koss+18} revealed a population of hidden nuclear mergers in a sample of heavily obscured, hard X-ray selected AGN. They identified a second IR source in the central kpcs of this optically selected post-merger galaxy that was blended into the optical nucleus in its low-resolution SDSS image. We identified two radio components in J0843+3549, the dominant of which is co-spatial with the central IR/optical component and the second component is found to the north of this. However, \citet{Koss+18} identified the second IR nucleus located 2.9 kpc to the east of the dominant component. If indeed the second IR nucleus of \citet{Koss+18} is associated with this galaxy merger and not a chance projection, we find no evidence of radio emission associated with it in our 10 GHz map, down to a $3\sigma$ luminosity limit of $2.7\times10^{20}$ W Hz$^{-1}$. We require further observations of J0843+3549 to confirm the nature of the northern radio source, since the two components are blended in the VLASS Quick Look image and we do not have lower frequency observations of higher resolution such as for J1511+0417.

\section{Discussion}\label{sec:discussion}
\subsection{Prevalence of RQ Emission}
Much attention has been given in recent years to studying the connection between merging galaxy systems and the triggering of a radio-loud or radio-quiet AGN \citep{RA+11,Bessiere+12,RA+12,Chiaberge+15,KW+17_apj,Pierce+22}. The general finding is that radio-loud AGN are associated with merging galaxy systems, while the fraction of radio-quiet AGN hosted by merging galaxies is often indistinguishable from non-merging systems \citep{Chiaberge+15}, suggesting that the radio-quiet phase is ubiquitous in the formation of all early-type galaxies \citep{RA+13}. This is particularly striking given that 89\% (16/18) of the radio emission we detected is radio-quiet, whether its origin be from an AGN, SF, or a combination of both; and 86\% (12/14) of the radio AGN are radio-quiet. From the overall sample, this translates into 57\% (16/28) of our post-mergers hosting radio-quiet emission, 43\% (12/28) hosting a radio-quiet AGN, and 7\% hosting a radio-loud AGN.

We emphasize important distinctions between our study and those focusing on the incidence of galaxy mergers in radio-loud/radio-quiet systems. These studies \citep[e\,g.,][]{Chiaberge+15,Breiding+23}{}{} placed constraints on the merger fraction of AGN host galaxies; they began by selecting for a sample of luminous quasars and examined the host galaxy morphology for signs of ongoing or recent gravitational interaction. Our sample, however, is of known merging galaxies that were selected only because of their host galaxy morphology and the presence of a single optical nucleus \citep{Carpineti+12}. Then, the merger fraction for our radio-quiet and radio-loud AGN is unity by definition; that is, 100\% of the radio-AGN are hosted by mergers. We cannot make direct comparisons to similar, though distinct, AGN-merger studies because of the contrasting selection criterion used for the different samples. Additionally, this means that the C12 sample is unbiased towards AGN activity, whereas previous studies either favored, or outright required, the identification of a radio AGN. The nature of our study is to holistically examine the radio properties of these post-merger galaxies and constrain the progenitor of their radio emission, including SF-related activity. This is not to say that previous works have found that all merging galaxies will produce a radio-loud AGN; such a case is clearly unrealistic by the existence of inactive merging galaxy systems, as is the case for a number of the SPMs in both their radio emission and optical emission-line activity presented in this work.

The deep nature of our observations may also play a key role in the high fraction of radio-quiet AGN in our sample. The sensitivity limits from ongoing and past GHz-frequency radio surveys are close to, if not more than, an order of magnitude less sensitive than what we achieved with our 10 GHz observations. Without the high-significance 10 GHz detection revealed by our observations, a number of these radio sources would be classified as non-detections by standard survey selection criterion, which are often $\geq5\sigma$, and thus would be excluded from such a study examining the link between merging galaxies and the incidence of radio-loud/radio-quiet emission. However, because of the 10 GHz detection that is co-spatial with the low-significance radio emission revealed by these surveys, we are confident that this emission is of astrophysical origin and should be examined as such. Again, the nature of our study makes it impractical to compare our detection rates to the merger fraction of AGN hosts from previous studies. Nonetheless, it is expected that with deeper, more sensitive observations, the number of radio-quiet AGNs hosted by galaxy mergers will increase. Such observations are required to attain a full representation of the radio-quiet AGN population, as is evidenced by the high fraction of radio-quiet AGN in our post-merger sample.

\subsubsection{The Role of Mergers in RQ AGN}
The predominance of radio-quiet AGN among our radio AGN sample points to a scenario in which the majority of the SMBHs powering these AGN have low black hole spin values. This comes from the framework first proposed by \citet{B&Z_77}, in which the energy extracted from a spinning BH via a highly magnetized accretion disk results in the launching of a jet. The radio-quiet/radio-loud dichotomy can then be explained by the SMBH possessing either a low or high spin \citep{W&C_95}.

Because these radio AGN are hosted exclusively by post-merger galaxies, we can explore this SMBH spin paradigm from an interesting perspective: namely, the coalescence of a supermassive black hole binary (SMBHB). \citet{W&C_95} proposed that the mass ratio of the SMBHB imposes significant evolutionary effects on the radio-loud nature of the coalesced SMBH. Only for SMBHBs with a mass ratio of order unity, with each SMBH of high mass ($\geq 10^8 M_\odot$), will the resultant coalesced SMBH be highly spinning and thus able to form a radio-loud AGN. This event is intrinsically rare, since the mass function of SMBHs declines for high mass values \citep{McLure+04,Hopkins+06,Gultekin+09}, making the formation of such a SMBHB rare as well. Leaving out J1511+0417, which is not a post-merger system and likely would not yet have formed a SMBHB, the radio-loud AGN are outnumbered by the radio-quiet AGN in our sample by a factor of 5.5, which is consistent with the broad expectations of the \citet{W&C_95} framework.

There is observational evidence to support the merger-spin framework to explain the radio-loud/radio-quiet dichotomy. \citet{deRuiter+05} and \citet{C&B_06} have found that radio-loud AGN are ubiquitously hosted by cored galaxies, i.\,e., those that show a flattening of their brightness distribution towards the optical nucleus. Cored galaxies themselves are products of SMBHB evolution. Once the SMBHB reaches pc separation, it will scatter stars whose orbits form close encounters with itself, creating the cored brightness distribution profile of the post-merger galaxy \citep{M&M_05}. Then, at least a subset of cored post-merger galaxies will host a coalesced SMBH, some of which may form a radio-quiet AGN and a rare few a radio-loud AGN. This gives a self-consistent model for the triggering and high relative fraction of radio-quiet AGN emission we have discovered in our sample of post-merger galaxies. Follow-up radio and optical observations would be needed to test this hypothesis for our sample of post-mergers. Very Long Baseline Interferometry (VLBI) is needed to probe the state of any SMBHB in these radio AGN. In Section~\ref{sec:ngvla}, we present the results of simulated observations using the current and next-generation VLBI instruments to ascertain the feasibility of these studies. Space-based or adaptive optics optical observations would be needed to model the brightness distribution profiles of these post-mergers to assess if they are truly cored galaxies.

Alternatively, accretion can lead to a spinning up of the SMBH. As discussed in \citet{Volonter+13} and \citet{Chiaberge+15}, coherent accretion, i.\,e., the accreted material has a constant angular momentum axis, will lead to a spinning up of the single SMBH. However, this naturally requires the accretion to be constant over a significant time evolution, which, in turn, requires a large gaseous reservoir for the SMBH to reside in. For gas-rich galaxy mergers, it is well established that such a reservoir can be created via tidal torquing of the gas, which drives it towards the nucleus where it forms a circumbinary disk \citep{DiMatteo+05}. By examining the colors of the SPM sample, C12 concluded that at least 55\% of these SPMs are the product of a galaxy merger consisting of one or more late-type, e.\,g., gas-rich, galaxies. This accretion-driven spin up of the SMBH may be a plausible origin for the radio emission in a subsample of these SPMs. However, it should be noted that this framework can only produce radio-loud AGN; the radio-quiet AGN require an alternative explanation. 

\subsection{Impact of AGN Feedback}
Important to the overall discussion of post-merger evolution is an AGN's ability to either trigger or cut-off star formation in post-mergers through AGN feedback. The centralization of gas that occurs during a gas-rich galaxy merger can both fuel accretion to power an AGN and also act as a catalyst for triggering starburst activity. The chronology of these two processes is crucial: if the AGN activity is not prompt during the starburst period, AGN feedback will have little to no effect on the SFR of the host galaxy.

\citet{Kaviraj+15} found that the host galaxies for a sample of VLBI-detected, merger-triggered radio AGN were a factor of 3 more likely than stellar mass- and redshift-matched inactive early-type galaxies to lie on the UV-optical red sequence. Using this and timescale arguments, the authors argue that these merger-triggered radio AGN are inefficient at regulating the SF in the host galaxy. \citet{Shabala+17} found a similar result for VLBI-detected radio AGN hosted by gas-rich minor mergers. By reconstructing the SF history of the host galaxies, they found that none of the radio AGN were triggered within 400 Myr of the onset of the starburst activity in the host, limiting their ability to impact the overall SFR. \citet{Carpineti+12} also found that the fraction of star-forming galaxies peaks in ongoing mergers, but the fraction of optical AGN peaks in post-merger systems, indicating different triggering times for these two processes during the merger evolution.  

Our radio-detected sources span a broad expanse of radio luminosity and dominant emission mechanisms. As shown in Section~\ref{sec:rsn}, at least 4 of these sources are likely dominated by emission from SNRs associated with past or ongoing SF. 3 of these 4 radio sources are hosted by LINER or SF-AGN composite emission-line galaxies. These 3 may represent the very earliest stages of AGN feedback occurring in the post-mergers, as there is evidence for both an AGN and recent SF activity, or at least ambiguously for the LINER. On the other hand, 4 of the likely radio AGN are hosted by either SF or SF-AGN emission-line galaxies. These may represent the next stage of AGN feedback, as the AGN is now the dominant radio emission mechanism, outshining in the radio band the total emission from the supernova remnants associated with the SF. There is multi-wavelength evidence for potential feedback processes occurring in at least these 8 post-mergers based on their radio and emission-line activity, though, as shown in Section~\ref{sec:radio_excess}, almost all of the post-merger galaxies appear to be forming stars at a rate $\geq1\,\mathrm{M_\odot}\,\mathrm{yr}^{-1}$. Although for the SF-dominated radio sources we, naturally, cannot estimate the effect of an AGN-powered jet on feedback, we can discuss how our population of radio AGN fits within the realm of AGN feedback.  

Establishing the total AGN jet power from its radio luminosity is not straightforward (see \citealt{Godfrey&Shabala_16}). However, the capability of the jet to substantially impact the surrounding interstellar medium (ISM), creating feedback, can be broadly interpreted even from an order of magnitude estimation. We used the relation of \citet{Cavagnolo+10} to estimate the jet power for each of the radio AGN in our sample. We found a range of jet powers spanning $10^{41}-10^{43}$ erg s$^{-1}$, making these low power radio jets. These jet powers are actually favorable for feedback process: low power jets, confined to the central kpcs of the host, are more effective at large, sustained disruption of the surrounding ISM compared to high power jets that easily puncture through the dense circumnuclear ISM and remain collimated at large distances from the launching region \citep{Mukherjee+16}. Indeed, the compact morphology of the radio sources favors this scenario, as none of the radio AGN show collimated jets beyond the nuclear region of their host galaxy. In conjunction with the substantial SFRs of the host post-merger galaxies, this makes these sources good candidates to study the impact of AGN feedback in post-merger galaxies by searching for AGN-triggered bursts of SF activity (positive feedback) and/or multiphase outflows (negative feedback).

\subsection{Spectral Index of SF-Related Emission}
In the GHz regime, radio emission related to the shocks propagated by SNRs is optically thin, with a canonical spectral index value $\alpha\sim-0.8$ \citep{Condon_92}. By comparing the radio-based SFR to that derived from host galaxy properties, we have identified 4 compact, nuclear, 10 GHz radio sources that are likely dominated by SF-related processes. Among these, we have placed strict constraints on the optically-thin spectral index for 3 of these sources using their 3 and 10 GHz flux densities. These 3 radio sources have a median optically-thin spectral index $\alpha_3^{10} = -1.14\pm0.08$, significantly steeper than the canonical value of $\alpha\sim-0.8$. Such a difference may be cause for concern. However, recent work by \citet{Klein+18} has provided a more in-depth analysis of the broadband radio spectra of integrated SNR populations. By using nearly 2 decades of frequency coverage, these authors found that for 14 SF galaxies, the synchrotron spectral index at low frequency, i.\,e., $\nu<$ 1 GHz, was similar to the canonical value. However, in the 1-12 GHz frequency range, the spectra required either a break or an exponential decline, indicating a steepening of the radio spectra. These features would be caused either by significant synchrotron or inverse-Compton loses to the high-energy electrons produced by the SN shocks, although \citet{Klein+18} explain that a cutoff in the synchrotron spectrum is difficult to explain without inducing a single-injection scenario, which is unlikely when considering the integrated properties of a SNR population.

This situation is alleviated somewhat by the emission-line activity for each of the SF-dominated radio sources. Only J1015+3614 is identified as a purely SF galaxy from its emission-line ratios, and its radio source has the flattest optically-thin spectral index of these sources of interest with $\alpha_3^{10}=-0.64\pm0.10$. This is not dissimilar to the spectral index values found for other SF galaxies \citep{Chyzy+18,An+21}, and while slightly flatter than the canonical $\alpha\sim-0.8$, this can be explained by the ratio of CR electrons favoring a younger, more energetic population, e.\,g., due to recent supernova activity. J1445+5134 and J1617+2512 have been identified as SF-AGN composites by their emission-line ratios, and have much steeper optically-thin spectral index values, $-1.46\pm0.04$ and $-1.31\pm0.11$, respectively. These steep spectral index values for SF galaxies have been observed before: using a sample of 41 6 GHz-detected submillimeter galaxies (SMGs), \citet{Thomson+19} found that the median 1.4-6 GHz spectral index for a sub-sample of bright SMGs to be $-1.35\pm0.24$. Additionally, the radio spectra of the bright SMGs showed a steepening at these GHz frequencies when compared to the same spectra at MHz frequencies. \citet{Thomson+19} discussed that such a phenomenon may be caused by the mixing of distinct electron populations accelerated by decoupled processes that dominate at different frequencies and at different spatial scales. We have presented this hypothesis to explain the spectral breaking (Section~\ref{sec:radio_spec}) and radio morphology (Section~\ref{sec:agn_morph}) for a number of different radio sources in our sample. We further extend this idea to the two SF-dominated radio sources hosted by SF-AGN composite emission-line galaxies. If there is a contribution from both a steep-spectrum radio AGN component and SNR population to the observed radio emission, such a mixing may cause a steepening of the radio spectral index. This would most likely arise from resolution effects, i.\,e., the diffuse emission from SNRs is resolved out at higher frequency, leaving only the steep-spectrum, compact AGN component. A scenario in which the diffuse emission from SNRs is cutoff at higher frequency is unlikely, as discussed previously \citep{Klein+18}. Multi-band SED modeling of these sources is needed to better understand the fractional contribution of the AGN \citep{Dietrich+18,RP+20}.

The remaining SF-dominated radio source is J1041+1105. Unlike for the other SF-dominated radio sources, we could only place a lower limit on the optically-thin spectral index of J1041+1105 ($\alpha_3^{10}>-1.43$) due to the non-detection of any radio emission at 3 GHz. Like J1445+5134 and J1617+2512, $\alpha_3^{10}$ for this source may also be steeper than expected, given its SF-dominated nature. However, when considering the 1.4 GHz flux density limit, which is more sensitive than the 3 GHz limit, the lower limit to the spectral index becomes -0.81. We know, then, that J1041+1105 does not show the same type of highly steep spectral index at GHz frequencies that J1445+5134 and J1617+2512 do. This may be indicative of the more canonical radio emission associated with SF, like J1015+3614. However, unlike J1015+3614, the host galaxy of J1041+1105 is a LINER. Although the exact nature of LINER emission is ambiguous, the surface brightness profiles of the low-ionization emission lines found in LINER hosts though integral field spectroscopy favor a scenario in which the ionization is powered by post-asymptotic giant branch stars \citep{Yan+12,Singh+13}. Such a scenario is also favorable due to the ubiquity of these stars in all galaxies, especially those with little active star formation. Radio observations, however, seemingly favor the presence of an AGN over pure SF-related processes to produce the observed radio emission \citep{Filho+04,Singh+15}. Although J1041+1105 may possibly have an optically-thin spectral index close to the canonical SF value of -0.8, more sensitive 1.4 and 3 GHz observations are required to better understand the association of this nuclear radio source to its LINER emission-line host galaxy.

\section{SMBHBs in SPM Systems}\label{sec:ngvla}
\subsection{Do Our SPMs Host SMBHBs?}
Radio observations are a powerful tool to probe both the kpc- and pc-scale environment to search for and confirm SMBHB candidates. At kpc scales, the morphology of the extended jets or lobes may hold signatures of SMBHB evolution: an X-shaped morphology due to a coalesced SMBHB \citep{Begelman+80,merritt_spinflip_02}, or an S-shaped (helical) morphology due to jet precession caused by a SMBHB (e.g., \citealt{Rubinur+17}). These radio structures have steep spectral index measurements ($\alpha\leq-0.5$), making low frequency observations particularly advantageous towards their identification. After visual inspection, we find no evidence for any of these kpc-scale, SMBHB evolutionary signatures in the radio morphology of these SPM galaxies at any observing frequency (see Appendix~\ref{sec:app_a} for multi-frequency radio maps for each SPM). However, the absence of these S- or X-shaped morphologies does not dismiss that these SPMs may harbor a SMBHB.

Likewise, only 2 of these SPMs (J0841+3549 and J1511+0417), show any evidence of DAGN behavior, as discussed in Section~\ref{sec:dagn}. Our 10 GHz observations would be the most adept at identifying DAGN candidates because of their sensitivity (nominal image RMS $\sim15$ $\mu$Jy) and high angular resolution, which would be able to resolve potential blended radio cores in lower resolution images. Then, we find no evidence for secondary radio emission in the remaining 12 radio AGN down to a range of limiting $3\sigma$ luminosities $L_\nu=8\times10^{19}-7\times10^{20}$ W Hz$^{-1}$. A second AGN in these systems would be extremely radio-faint and require ultra-deep sensitivities to detect. This is also true for the SF-dominated sources: we find no evidence of multi-component radio emission in any of these systems above a 10 GHz luminosity of $L_\nu=3.7\times10^{20}$ W Hz$^{-1}$.

\textit{Gaia}'s superb astrometric precision has enabled a number of searches for DAGN and SMBHBs candidates that utilize astrometric variability induced by photocenter pseudo-motion of the unresolved SMBH pair, or varstrometry \citep{Hwang+20,Shen+21,Chen+22,Schwartzman+23}. This technique is particularly powerful for systems with $z>0.5$, as \textit{Gaia}'s astrometry is optimized for compact sources \citep{M&S_22}. Below this redshift, extended features in the host galaxy, e.\,g., tidal tails, will induce false astrometic noise \citep{Souchay+22}. Because our sample of SPMs were selected to have extended, tidal features and $z<0.1$, any analysis using the photometric center from \textit{Gaia} is unreliable, including searching for offsets in the radio and optical photometric centers. 

Then, for the majority of the 10 GHz-detected SPM galaxies in our sample (16/18; 89\%), we find no evidence for SMBHB evolution at the (sub-)kpc scale. We emphasize, however, that the lack of evidence does not preclude that any of these SPMs may host a SMBHB system.

\subsection{Searches with Very Long Baseline Interferometry}
Very Long Baseline Interferometry (VLBI) offers a plethora of direct and indirect methods to identify SMBHB candidates. Among these, the identification of dual, flat-spectrum radio cores at pc-scale separation provides the most compelling evidence of any SMBHB, a technique that is only feasible because of VLBI's unique milliarcsecond angular resolution. Indeed, this technique has so far provided the best evidence for a SMBHB, hosted in the elliptical galaxy 0402+379, through imaging \citep{Rodriguez+06} and proper motion constraints \citep{Bansal+17}. VLBI observations can also provide corroboratory evidence of SMBHB candidates through indirect methods. Significant position angle differences between the pc- and kpc-scale radio jet may be indicative of binary evolution (e.\,g., \citealt{Mooley+18}), as the jet opening angle widens as the binary loses energy due to the emission of gravitational waves \citep{Kulkarni&Loeb16}. Periodic variability observed in both the radio luminosity and radio core position may be indicative of SMBHB-induced precession of the radio jet of sub-pc SMBHB candidates (e.\,g., \citealt{Sudou+03,Stirling+03,Kun+14}).

We are interested in establishing the feasibility of performing VLBI observations to search for a SMBHB in each of the 14 radio AGN we have discovered with our 10 GHz VLA observations. We did this by simulating two different VLBI observatories: the Very Long Baseline Array (VLBA), and the Next Generation Very Large Array (ngVLA). For each frequency band of each array, we calculated the expected continuum RMS image sensitivity of a 1-hour long integration and a 10-hour long integration. Here, we are using integration to represent the total time on-source for each target of interest. Each observation, encompassing this on-source time, is then necessarily longer than 1 hour and 10 hours to account for overheads. In actual targeted observations for searches of SMBHBs, Walsh et al. (in prep) found that roughly 75\% of their total observation time was on-source, with the remaining 25\% dedicated to scans of amplitude and phase reference calibrators. So, we can expect an increase of approximately 25\% for each integration time, or observations which are approximately 1.3 and 13.3 hours for the 1-hour and 10-hour integration, respectively. For the VLBA, we have assumed an efficiency factor $\eta_s$ of 0.8, and that all 10 antennas, thus 90 baselines, are included for each simulated integration. Our simulated integrations use the L-band (21 cm), S-band (13 cm), C-band (6 cm and 5 cm), X-band (4 cm), Ku-band (2 cm), K-band (1.4 cm), Ka-band (1.25 cm), and Q-band (0.7 cm) receivers for the VLBA. For each frequency band, we have assumed the maximum possible data rate (2048 Mbps for L- and S-band, 4096 Mbps for all others) and used the SEFD provided for the VLBA\footnote{\url{https://science.nrao.edu/facilities/vlba/docs/manuals/oss/bands-perf}}. For the simulated ngVLA integrations, we used the ngVLA sensitivity calculator Python script\footnote{\url{https://gitlab.nrao.edu/vrosero/ngvla-sensitivity-calculator}} to calculate the expected continuum RMS image sensitivity. We did this for each of the central frequencies listed for the VLBA, since the larger bandwidths of the ngVLA receivers encapsulate multiple VLBA receiver frequency ranges. We simulated the 1-hour and 10-hour long integrations at the first 5 bands of the ngVLA for this analysis, with central frequencies at 2.4, 8, 16, 27, and 41 GHz. 
We have not taken into account the RFI environment for any of these simulated integrations. This is especially prevalent at lower frequencies, where up to half of the bandwidth may be unusable due to the persistent, dominating presence of RFI. Each of these simulated integrations is an ideal case and represents a lower limit to what is achievable in actual observations.

\begin{figure*}[t]
    \centering
    \includegraphics[width=\textwidth, trim=20mm 5mm 20mm 5mm]{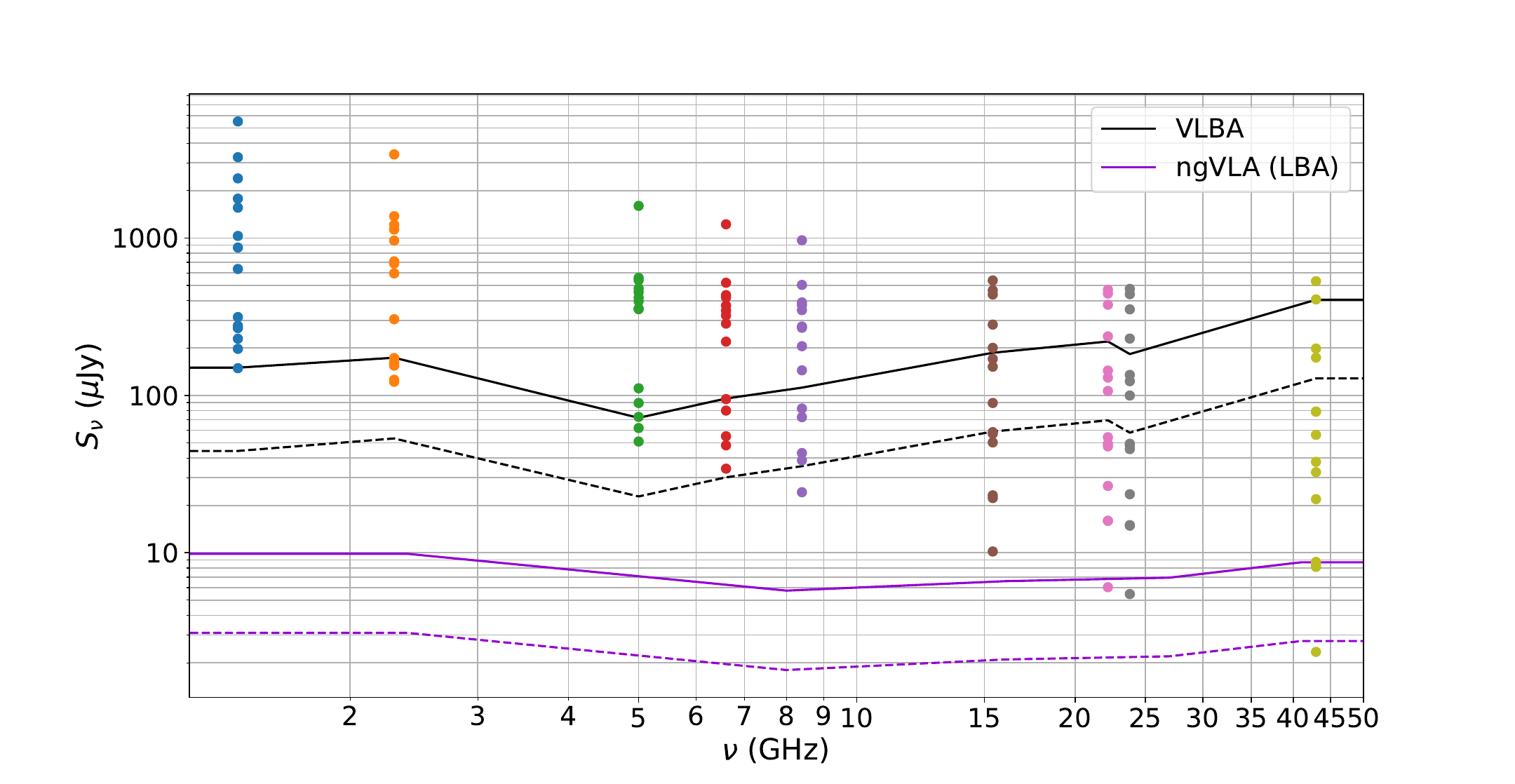}
    \caption{VLBI-scale flux density estimates of the 14 newly-discovered radio AGN from our sample of SPM galaxies plotted with the simulated, $5\sigma$ sensitivity curves of a 1-hour and 10-hour integration with the Very Long Baseline Array (VLBA) and the Long Baseline Array (LBA) of the Next Generation Very Large Array (ngVLA). These data points represent the potential flux density of a supermassive black hole binary (SMBHB), which can only be resolved with VLBI. Points above a $5\sigma$ sensitivity curve indicate a detection. At lower frequency ($\nu<10$ GHz), the VLBA observations may already reach the desired sensitivity to achieve a significant detection of milliarcsecond-scale radio emission. The notable difference is at higher frequency ($\nu>10$ GHz), where a 1-hour integration with the ngVLA vastly improves the probability of detection compared to a 10-hour VLBA integration.}
    \label{fig:ng_sens}
\end{figure*}

To calculate the expected milliarcsecond-scale radio flux density at each frequency, we used the 10 GHz flux density value (Table~\ref{table:flux_values}, column 6) and the $\alpha_3^{10}$ spectral index value (Table~\ref{table:spec_fit}, column 4) for each of the radio AGN. For these radio AGN with more than one resolved component (J0843+3549 and J1511+0417), we used the 10 GHz flux density value of the dominant radio component. We began by extrapolating the 10 GHz flux density to each of the central frequencies listed above. Here, we have assumed that the broader radio SED for each of these radio components follows a simple power law $S_\nu \propto \nu^\alpha$, where $\alpha$ is $\alpha_3^{10}$. We have used the 10 GHz flux density values because this represents the best approximation to the radio flux density we would expect from a dominant radio core. All other flux density measurements were taken at lower frequency and angular resolution. Because of this, those flux density measurements are more likely to have contributions from non-core-related phenomena, such as steep-spectrum features, e.\,g., a radio jet, or star formation, especially for the case of the 144 MHz LoTSS data. Indeed, only J0206$-$0017 and J1511+0417 have a flat spectral index ($\alpha>-0.5$) in the optically-thin regime, which is expected if the dominant contributor to the radio flux density were from a radio core. We do note that J1433+3444 may also be a flat spectrum object since $\alpha_3^{10} = -0.47 \pm 0.07$. The same could also be true for all source for which we could only place a lower limit on $\alpha_3^{10}$ (J0851+4050, J0916+4542, J1113+2714, and J1517+0409). This is critical to establishing the expected population of detected radio sources for each of our simulated integrations. If we systematically overestimate the expected milliarcsecond-scale flux density, we will also overestimate the number of significant detections achievable in each of our simulated integrations. We have also assumed that $\alpha_3^{10}$ also represents the dominant VLBI component. This certainly does not need to be the case, as even unresolved features at sub-arcsecond scale may be resolved out at the mas-scale probed by VLBI observations, possibly revealing a dominant radio core at mas scales. However, we are using these values since they best represent the physical situation as we can currently determine. Once again, we only wish to estimate what the VLBI-scale emission properties are; only through actual observation could these flux density values be determined. 

After extrapolating the 10 GHz flux density to the designated frequency values, we apply a factor of 0.3 in converting from sub-arcsecond-scale flux density to mas-scale flux density. This value was chosen from the analysis of \citet[][]{Deller&Middelberg_14}. In their analysis, these authors determined the ratio of peak VLBI flux density at 1.4 GHz to peak FIRST flux density for a large sample of VLBI sources detected in the mJy Imaging VLBA Exploration at 20 cm (mJIVE-20) survey. Overall, they found that 30\%-35\% of all sources have compact VLBI emission in which the majority of the FIRST flux is recovered, with this trend increasing towards lower FIRST flux density. Indeed, for FIRST sources with a flux density measuring from 1-2 mJy that are detected with VLBI, all recover at least 32\% of the FIRST flux density value at VLBI scales, with about 25\% of sources having greater than 64\% of this value recovered. We acknowledge that this extrapolation was determined only for 1.4 GHz observations, whereas our analysis uses the flux density determined at 10 GHz. Our factor of 0.3, then, may even underestimate the VLBI-scale flux density for each source. Though, for the sake of this analysis, this is preferred to an overestimation.

We now have estimations for the VLBI-scale flux density for each of the SPM sources detected with our 10 GHz observations. 
Figure~\ref{fig:ng_sens} plots these estimated VLBI-scale flux density values with the simulated continuum image RMS sensitivities of a 1-hour long and 10-hour long integration with the VLBA and ngVLA. For each simulated integration, the sensitivity curve represents a 5$\sigma$ detection threshold. Points that fall above these curves represent a detection, while those below are not expected to be detected. Notably, lower frequency integrations ($\nu<10$ GHz) with the VLBA may already achieve a sensitivity to reach a significant detection of VLBI-scale radio emission. While these frequencies may not be optimal for isolating the core emission, the flux density information they provide is nonetheless critical to establishing the spectral index value of the potential radio core associated with each binary constituent. The significant improvement in detection threshold appears in the higher frequency integrations ($\nu>10$ GHz). At these frequencies, we expect that the vast majority of sources would not be detected even with a 10-hour integration time using the VLBA. However, with the ngVLA, we find that a 1-hour integration time is sufficient to detect all of the sources at 15 GHz, and the majority of sources at 22, 23, and 43 GHz. These higher frequencies observations are best at isolating the radio core by resolving out larger-scale emission and provide high astrometric precision due to their high angular resolution.

\section{Summary}
\label{sec:summary}
In this paper, we have analyzed the emission properties of a sample of 30 local post-merger galaxies from \citet{Carpineti+12} to search for star formation- and AGN-related activity. Our main results are as follows:

\begin{enumerate}
    \item \textbf{Diverse Emission-Line Activity:} Using the optical emission-line flux ratios derived from the OSSY catalog \citep{Oh+11}, and standard BPT diagram analyses, 43\% of the post-mergers are optically quiescent, 10\% are dominated by SF, 13\% by a combination of SF and AGN, 13\% by Seyfert AGN, and 20\% by LINER activity.
    
    \item \textbf{Low-luminosity Radio Emission:} Of those with detectable radio emission, through both archival radio surveys and new, high resolution, 10 GHz observations with the VLA, the vast majority are radio-quiet, with only 2 reaching the spectral luminosity threshold ($\nu L_\nu>10^{32}$ W) to be classified as radio-loud. We discovered a number of nuclear radio sources at high significance ($\geq 5\sigma$) with our 10 GHz observations that were otherwise non-detections by archival radio surveys, emphasizing the importance of deep observations to reveal the full population of radio-quiet systems.

    \item \textbf{Prevalence of Compact Radio Emission:} At the largest spatial extents, sampled by 144 MHz LoTSS observations, all of the detected radio sources have a diffuse or extended emission component. Only J0843+3549 and J1433+3444 display an AGN-like morphology among these sources. The nuclear emission is unresolved in all but one of the 15 sources at 1.4 GHz ($5\arcsec$ resolution), indicating that compact, nuclear emission is prevalent in these post-mergers. At the most compact scales, sampled by our 10 GHz observations ($\approx 0.2\arcsec$) the sources show a variety of AGN- and SF-related morphologies.

    \item \textbf{Radio Spectra and Spectral Index Measurements:} Of the 12 radio sources with well-sampled (4 or more detections) spectra, we found 3 that showed evidence of significant curvature ($|q|\geq0.2$) in their broadband spectrum, with one being a Gigahertz peaked spectrum (GPS) source (J1617+2512). Two spectra were found to be inverted ($q\geq0.2$), though we believe this is likely due to an overall flattening of the spectrum at high frequency and not a true inversion. The spectra of J0843+3549, J1015+3914, J1135+2953, and 1304+6520 are poorly fit by both a simple and curved power law, as indicated by the $\chi_{red}^2$ value for each fit. This is either due to variability at one or more flux values, or a blending of two distinct electron populations at low angular resolution. We also calculated the spectral index value $\alpha_3^{10}$ to the compact radio emission using the 3 and 10 GHz flux values, or its lower limit for sources without a 3 GHz detection. These $\alpha_3^{10}$ values range from $0.19 \geq \alpha \geq -1.74$, though the majority have a steep spectral index $\alpha < -0.5$. 

    \item \textbf{SF Activity in Post-Mergers:} The 1.4 GHz luminosity of 4 of the radio sources (J1015+3914, J1041+1105, J1445+5134, and J1617+2512) can be explained by SF processes alone. For each, we have determined that their emission is most likely due to a population of supernova remnants, as the expected luminosity from an individual, luminous radio supernova is at least a factor of 80 dimmer than each of their observed 10 GHz luminosity. Further, these 4 radio sources are hosted by either SF, SF-AGN composite, or LINER emission-line galaxies, providing corroboratory evidence of ongoing or recent SF activity in each of these post-mergers. It is notable that the spectral index value for 3 of these 4 is significantly steeper than the canonical value $\alpha \sim -0.8$ for shock-dominated sources \citep{Condon_92}. However, these spectral index values are not dissimilar to those of other shock-dominated sources \citep[e.\,g.,]{Chyzy+18,Thomson+19,An+21}, and may be indicative of an older CR electron population due to evolved supernova activity. Alternatively, as two of these sources are hosted by SF-AGN composite emission-line galaxies, the steep spectral index may be due to resolution effects in which the diffuse synchrotron is resolved out at higher angular resolution, revealing compact, steep-spectrum AGN emission.

    \item \textbf{Discovery of Radio AGN:} We have discovered 14 likely radio AGN in these post-mergers: 3 because of their association with a known AGN and 11 for which we found excess radio emission compared to SF predictors. 86\% (12/14) of these radio AGN are radio-quiet, with only 14\% (2/14) being radio-loud. The post-merger hosts are found to occupy all regions of the BPT diagrams, indicating that radio AGN activity may be present even during stages of star-forming activity of the post-merger evolution. We also report on the discovery of a precessing jet in the dual AGN candidate J0843+3549 \citep{Koss+18}, and discover a new dual AGN candidate, J1511+0417.

    \item \textbf{The Origin of Radio AGN Activity in Mergers:} The prevalence of radio-quiet AGN among our radio AGN population lends itself to a scenario in which radio-quiet AGN in ongoing or recent galaxy mergers may be more populous than previously believed. Because our sample is comprised of late-stage and post-merger systems, the high fraction of radio-quiet AGN can be explained by SMBH spin up due to the coalescence of a supermassive black hole binary (SMBHB; \citet{W&C_95}). In this framework, radio-loud AGN are only produced for the most massive binary systems, which are intrinsically rare due to the sharp decline at high mass of the SMBH mass function \citep{McLure+04, Hopkins+06, Gultekin+09}. Indeed, we have found that the radio-quiet AGN outnumber the radio-loud AGN by a factor of 5.5. Alternatively, gas-rich mergers may produce radio-loud AGN if the SMBH sustains coherent accretion for an extended period of time. Both scenarios need further observations to test rigorously. 

    \item \textbf{Jet-ISM Feedback:} We estimated the total power of the jets for our sample of radio AGN. The jet powers span a range of $10^{41}-10^{43}$ erg s$^{-1}$, making them low power. The majority of the post-mergers have a SFR $\geq1$ M$_\odot$ yr$^{-1}$, indicating that the AGN may play an important role in providing positive or negative feedback. Importantly, these low power jets, confined to the central kpcs of the host, are more effective at large, sustained disruption of the surrounding ISM compared to high power jets that easily puncture through the dense circumnuclear ISM and remain collimated at large distances from the launching region \citep{Mukherjee+16}. Indeed, the compact morphology of these radio sources agrees with this scenario. These radio AGN are then good candidates to study the impact of AGN feedback in post-merger systems by searching for signatures of multi-phase gas outflows.

    \item \textbf{Next-generation Searches of SMBHBs:} Lastly, we simulated 1- and 10-hour integrations at multiple frequencies with the Very Long Baseline Array (VLBA) and the VLBI capabilities of the Next Generation VLA (ngVLA). These simulations present the necessary time commitment for each instrument to reach the deep sensitivities required to perform robust searches for a SMBHB in each of these radio AGN hosted by a post-merger galaxy. We estimated the milliarcsecond-scale flux density of the radio source using the spectral index value $\alpha_3^{10}$ for each radio AGN and additional factors from the literature (e.\,g., \citet{Deller&Middelberg_14}). We found that at low frequency ($\nu<10$ GHz), the VLBA can already perform these robust searches, though the low frequency, milliarcsecond environment of radio AGN will often be dominated by extended, steep spectrum emission, making radio core identification difficult. The ngVLA will be a particularly powerful instrument for searches of SMBHBs at high frequency ($\nu>10$ GHz), where the dual, flat-spectrum cores of the SMBHB are the dominant emission signature. These high frequency, high angular resolution observations also offer significantly better astrometric precision than low frequency observations, which will be important for constraining proper motion measurements of the SMBHB constituents. 
\end{enumerate}

Our study of the multi-wavelength emission properties of 30 post-merger galaxies has discovered a number of exciting phenomena and individual sources. Future work on this topic will expand the sample population of post-merger galaxies, include better frequency coverage, and examine the radio emission of galaxy mergers at various stages of their evolution. This work will further our understanding of the astrophysical processes occurring during the merger sequence, the impact of AGN feedback, and establish new radio sources for which to follow-up with VLBI with the hope of detecting individual SMBHBs.

\section*{Acknowledgments}
\begin{acknowledgments}
GW and SBS were supported in this work by NSF award grant \#1815664. We thank Amy Reines for the discussions on multi-wavelength analyses to determine the origin of radio emission and Julie Comerford for recommendation of the OSSY catalog to perform our optical spectral analysis. %Some of the data presented in this paper were obtained from the Mikulski Archive for Space Telescopes (MAST) at the Space Telescope Science Institute. The specific observations analyzed can be accessed via \dataset[https://doi.org/10.17909/T9H59D]{https://doi.org/10.17909/T9H59D}. STScI is operated by the Association of Universities for Research in Astronomy, Inc., under NASA contract NAS5–26555. Support to MAST for these data is provided by the NASA Office of Space Science via grant NAG5–7584 and by other grants and contracts. This publication makes use of data products from the Wide-field Infrared Survey Explorer, which is a joint project of the University of California, Los Angeles, and the Jet Propulsion Laboratory/California Institute of Technology, funded by the National Aeronautics and Space Administration. 
The National Radio Astronomy Observatory is a facility of the National Science Foundation operated under cooperative agreement by Associated Universities, Inc. The NANOGrav collaboration, which funded some components associated with this research, receives support from National Science Foundation (NSF) Physics Frontiers Center award \#1430284 and \#2020265. This research has made use of NASA's Astrophysics Data System Bibliographic Services.
\end{acknowledgments}

\facilities{VLA, LOFAR, ASKAP, \textit{Galex}, \textit{WISE}}
\software{
CASA \citep{CASA_22},
Numpy \citep{numpy+11},
Scipy \citep{scipy+20},
Matplotlib \citep{Hunter_07},
Astropy \citep{astropy_18}
}

\bibliography{bib}{}
\bibliographystyle{aasjournal}

\appendix
\restartappendixnumbering

\section{Individual Sources}\label{sec:app_a}
The criterion for the morphology to be unresolved, resolved, marginally resolved, or multi-component are described in Section~\ref{sec:radio_morph}.
The radio contour maps for LoTSS, RACS, FIRST, and VLASS are presented using the nominal image RMS of each survey (see Section~\ref{sec:surveys}). This does not reflect what the local RMS may be for each image, especially for SPMs located near strong sources, which is particularly an issue for the MHz-frequency surveys. If there is a discrepancy between the nominal survey and local image RMS for any of our sources that directly affects the morphology or detection level of the source, we note this in the appropriate subsections. The LoTSS, RACS, FIRST, VLASS/VLA S, VLA X, and Pan-STARRs panels are nominally $1\arcmin\times1\arcmin$, $1.5\arcmin\times1.5\arcmin$, $1\arcmin\times1\arcmin$, $15\arcsec\times15\arcsec$, $5\arcsec\times5\arcsec$, and $1\arcmin\times1\arcmin$ in size, respectively, unless otherwise noted in the Figure caption for each source. The NVSS panel for J1304+6520 is $5\arcmin\times5\arcmin$ in size.

\subsection{J0206$-$0017}\label{sec:J0206-0017}
\begin{figure}
    \centering
    \includegraphics[width=\textwidth,trim=20mm 15mm 15mm 0mm,clip]{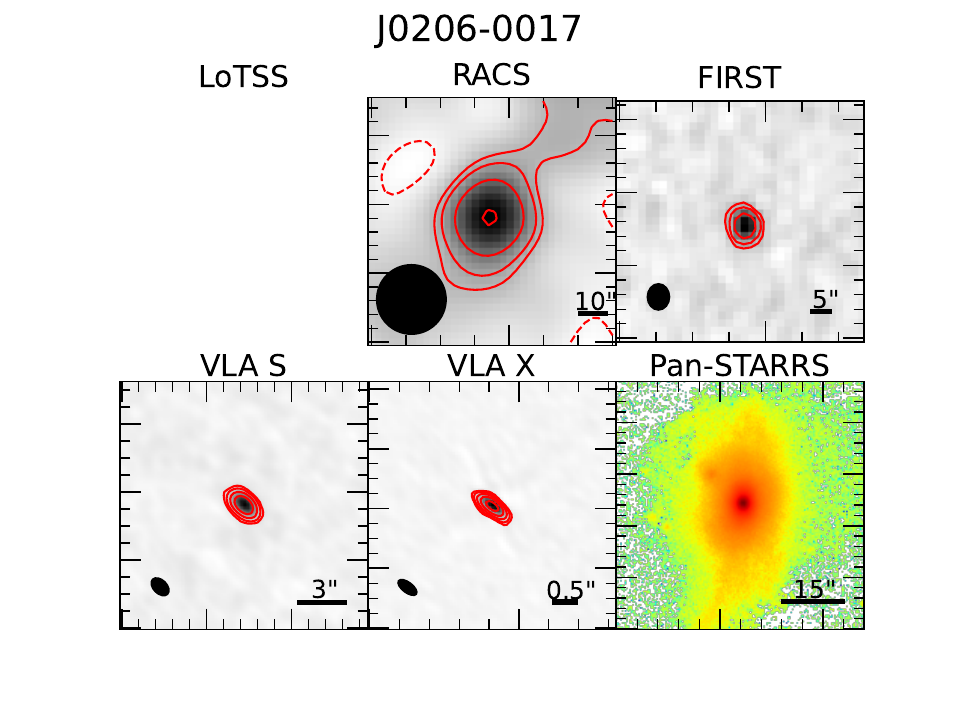}
    \caption{Intensity maps and contour overlays, and Pan-STARRS $r$-band image of the host galaxy of J0206-0017. The synthesized beam is shown in the bottom left corner for the radio surveys, and a scale bar is shown in the bottom right. Contours are shown at -3, 3, 5 and increase by a factor of 2 thereafter times the nominal image RMS for LoTSS, RACS and FIRST, and the off-source RMS for the VLA S (35 $\mu$Jy) and VLA X (30 $\mu$Jy) panels. The Pan-STARRS image has been resampled and falsely colored to emphasize the tidal features of the host galaxy. The local image RMS of this RACS field is greater than the nominal image RMS of RACS. Thus, the compact nature of J0206$-$0017's 888 MHz emission is not represented accurately here.}
    \label{fig:0206}
\end{figure}

J0206$-$0017 is a radio AGN associated with the changing-look AGN Mrk 1018 \citep{McElroy+16,Walsh+23}. The radio emission is unresolved by RACS, FIRST, and at 3 GHz and 10 GHz (Figure~\ref{fig:0206}). We note that the local image RMS of this RACS field is 450 $\mu$Jy, and thus the compact nature of J0206$-$0017's 888 MHz radio emission is not truly reflected in Figure~\ref{fig:0206}. The radio spectral index reflects the compact nature of this source, as it is flat with $\alpha_3^{10}=-0.13\pm0.04$. The broadband radio SED is well fit by a curved power law and appears to be inverted ($q=0.29\pm0.01$). This is most likely due to the flattening of the radio spectrum at GHz frequencies and not truly indicative of an inverted spectrum.

\subsection{J0759+2750}\label{sec:J0759+2750}
\begin{figure}
    \centering
    \includegraphics[width=\textwidth, trim=15mm 15mm 15mm 0mm,clip]{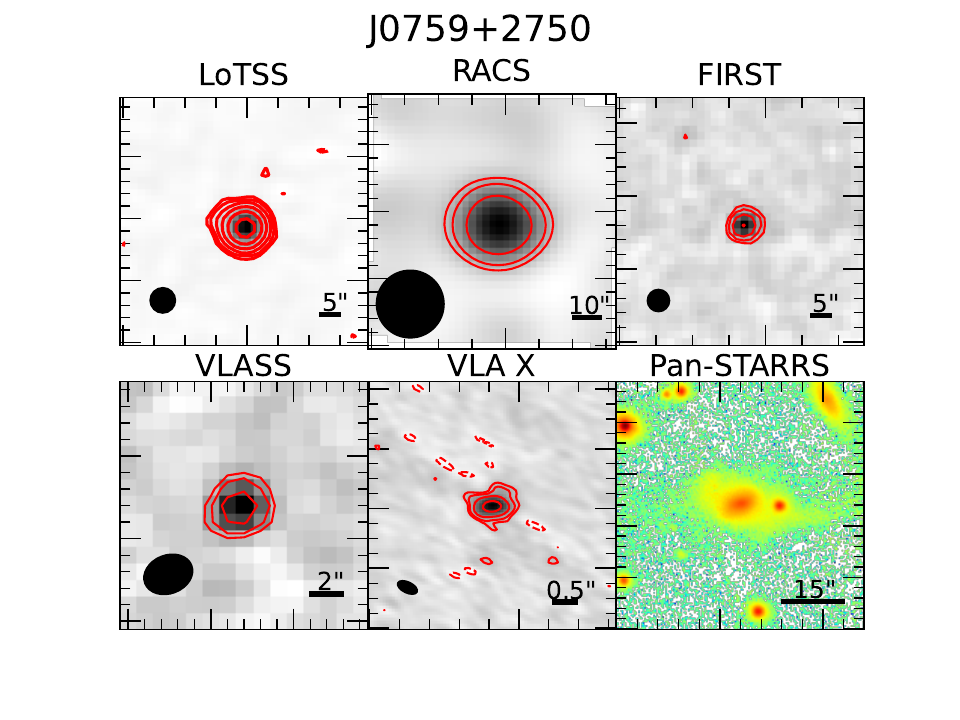}
    \caption{Intensity maps and contour overlays, and Pan-STARRS $r$-band image of the host galaxy of J0759+2750. The synthesized beam is shown in the bottom left corner for the radio surveys, and a scale bar is shown in the bottom right. Contours are shown at -3, 3, 5 and increase by a factor of 2 thereafter times the nominal image RMS for LoTSS, RACS, FIRST and VLASS, and the off-source RMS for the VLA X (13 $\mu$Jy) panel. The Pan-STARRS image has been resampled and falsely colored to emphasize the tidal features of the host galaxy.}
    \label{fig:0759}
\end{figure}

J0759+2750 is a radio AGN hosted by a SF-AGN composite emission-line galaxy. The radio emission is resolved by both LoTSS and at 10 GHz, but unresolved by RACS, FIRST, and VLASS (Figure~\ref{fig:0759}). The eastern extension to the 144 MHz emission is similar to the tidal feature visible in the host galaxy in that direction but does not overlap perfectly with that feature. The 144 MHz extension is slightly broader in width than the tidal feature. The 10 GHz morphology shows both a diffuse, non-linear component and an unresolved component, perhaps indicative of both nuclear SF and AGN activity. The spectral index is steep with $\alpha_3^{10}=-0.75\pm0.07$. The broadband radio spectrum is well fit by a curved power law, but does not exhibit significant curvature ($q=-0.04\pm0.02$).

\subsection{J0843+3549}\label{sec:J0843+3549}
\begin{figure}
    \centering
    \includegraphics[width=\textwidth, trim=15mm 15mm 15mm 0mm,clip]{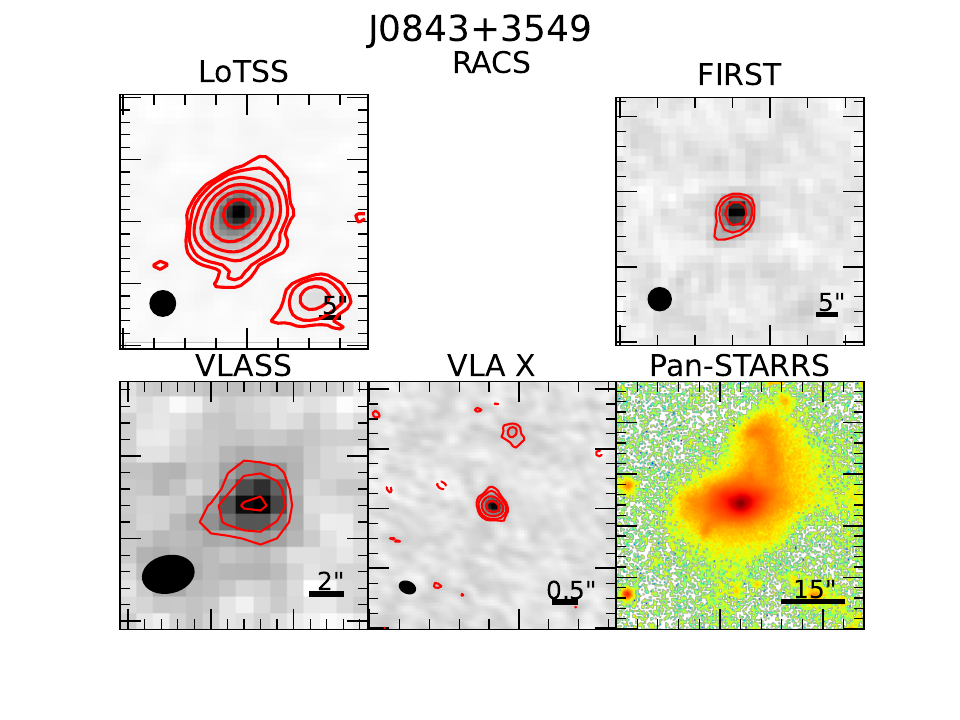}
    \caption{Intensity maps and contour overlays, and Pan-STARRS $r$-band image of the host galaxy of J0843+3549. The synthesized beam is shown in the bottom left corner for the radio surveys, and a scale bar is shown in the bottom right. Contours are shown at -3, 3, 5 and increase by a factor of 2 thereafter times the nominal image RMS for LoTSS, FIRST and VLASS, and the off-source RMS for the VLA X (13 $\mu$Jy) panel. The Pan-STARRS image has been resampled and falsely colored to emphasize the tidal features of the host galaxy.}
    \label{fig:0843}
\end{figure}

J0843+3549 is a radio AGN hosted by a Seyfert AGN emission-line galaxy. The radio emission is resolved by LoTSS, FIRST, and VLASS, and shows multiple components at 10 GHz (Figure~\ref{fig:0843}). The extended feature to the compact emission in each map has a varying deconvolved position angle (PA) from $131.5\degr\pm5.1\degr$ at 144 MHz to $-18\degr\pm15\degr$ at 10 GHz. This extreme PA change from kpc- to sub-kpc-scales shows that J0843+3549 is likely host to a precessing radio jet. A second resolved component is revealed at 10 GHz located 1.55$\arcsec$ (1.6 kpc at $z=0.054$) away from the southern, brighter component. We find no evidence for radio emission associated with the second IR nucleus of \citet{Koss+18}, located to the east of the dominant, central component, down to $L_\nu =2.7\times10^{20}$ W Hz$^{-1}$ at 10 GHz. The 144 MHz component located $27.3\arcsec$ to the southwest is associated with the galaxy cluster GMBCG J130.93151+35.82210 \citep{Hao+10} at a redshift of $z=0.475$ \citep{Rozo+15}. The spectral index is steep with $\alpha_3^{10}=-1.37\pm0.05$. However, higher resolution observations are needed to resolve the independent components and isolate their contributions to the total radio spectrum, as well as measure their separate spectral index values. The broadband radio spectrum is not well fit by either a simple or curved power law. This is likely because we are probing two separate electron populations producing the observed radio emission. The 144 MHz emission probed by LoTSS is most probably associated with supernova remnants, while at GHz frequencies the radio AGN is the dominant emission component.

\subsection{J0851+4050}\label{sec:J0851+4050}
\begin{figure}
    \centering
    \includegraphics[width=\textwidth, trim=15mm 15mm 15mm 0mm,clip]{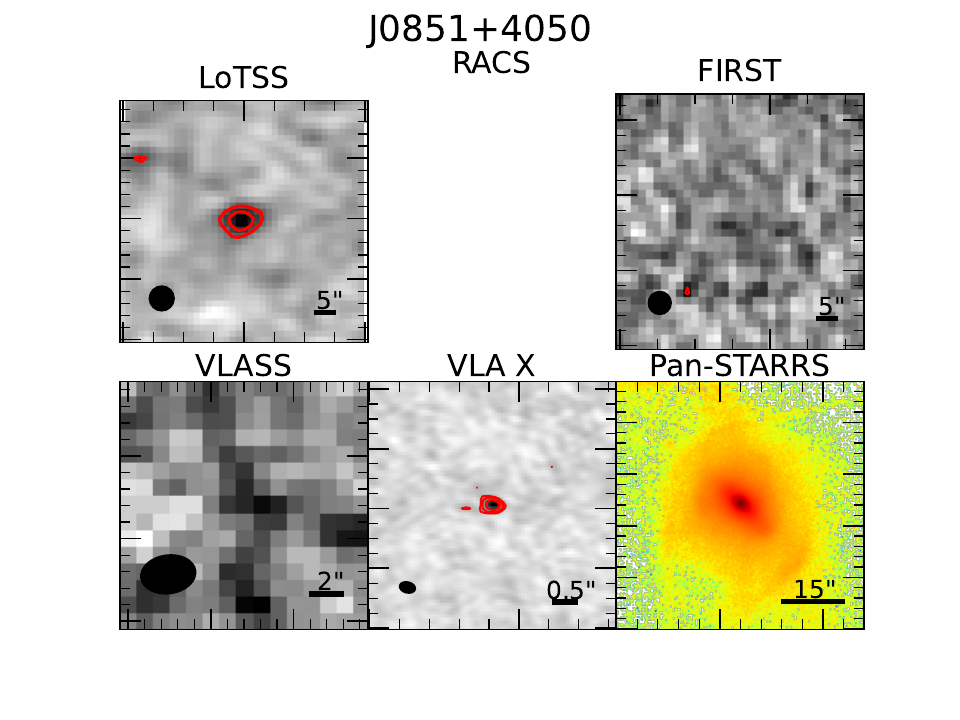}
    \caption{Intensity maps and contour overlays, and Pan-STARRS $r$-band image of the host galaxy of J0851+4050. The synthesized beam is shown in the bottom left corner for the radio surveys, and a scale bar is shown in the bottom right. Contours are shown at -3, 3, 5 and increase by a factor of 2 thereafter times the nominal image RMS for LoTSS, FIRST and VLASS, and the off-source RMS for the VLA X (13 $\mu$Jy) panel. The Pan-STARRS image has been resampled and falsely colored to emphasize the tidal features of the host galaxy.}
    \label{fig:0851}
\end{figure}

J0851+4050 is a possible radio AGN hosted by LINER emission-line galaxy. The radio emission is marginally resolved by LoTSS but unresolved at 10 GHz, and is a non-detection in FIRST and VLASS (Figure~\ref{fig:0851}). The resolved emission at 144 MHz is perpendicular to the tidal features visible in the host galaxy. Because it is a non-detection at 1.4 GHz, the radio-based SFR we calculated in Section~\ref{sec:radio_excess} is an upper limit. Given this constraint, it still falls in the radio excess region of Figure~\ref{fig:radio_origins}. More sensitive observations at 1.4 GHz may change this, but our data cannot determine otherwise, hence our classification as a possible radio AGN. Without a detection at 3 GHz, the lower limit to the spectral index is $\alpha_3^{10}>-0.41$, likely making this a flat-spectrum source. More sensitive observations at 3 GHz are needed to determine if the broadband radio spectrum is better fit by a simple or curved power law and to measure $\alpha_3^{10}$.

\subsection{J0916+4542}\label{sec:J0916+4542}
\begin{figure}
    \centering
    \includegraphics[width=\textwidth, trim=15mm 15mm 15mm 0mm,clip]{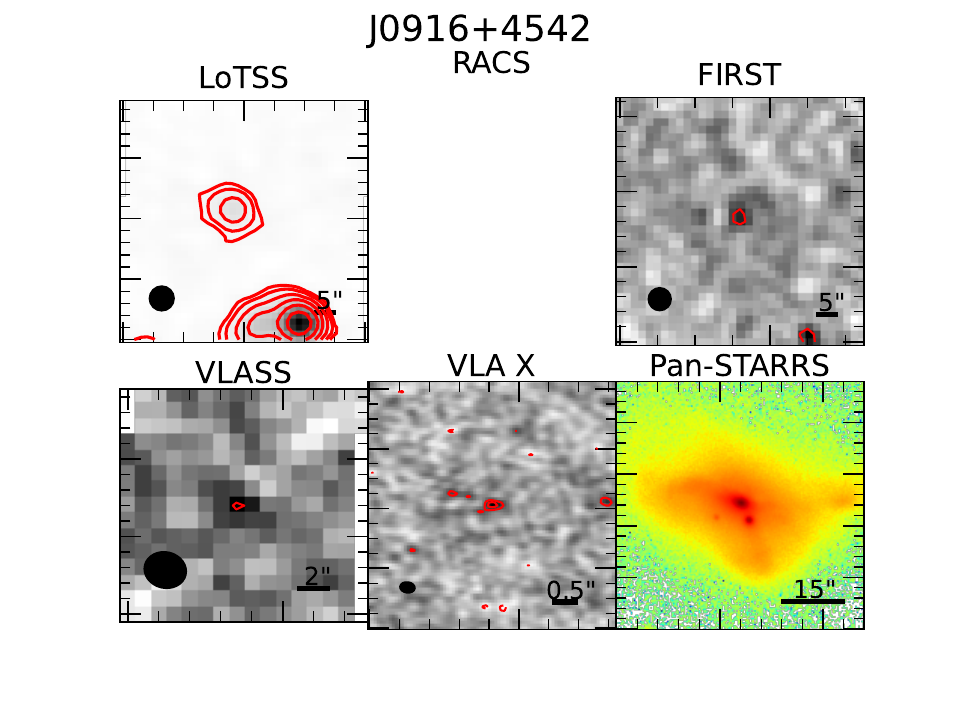}
    \caption{Intensity maps and contour overlays, and Pan-STARRS $r$-band image of the host galaxy of J0916+4542. The synthesized beam is shown in the bottom left corner for the radio surveys, and a scale bar is shown in the bottom right. Contours are shown at -3, 3, 5 and increase by a factor of 2 thereafter times the nominal image RMS for LoTSS, FIRST and VLASS, and the off-source RMS for the VLA X (13 $\mu$Jy) panel. The Pan-STARRS image has been resampled and falsely colored to emphasize the tidal features of the host galaxy.}
    \label{fig:0916}
\end{figure}

J0916+4542 is a radio AGN hosted by a SF-AGN composite emission-line galaxy. The radio emission is resolved by LoTSS and at 10 GHz, and unresolved by FIRST (Figure~\ref{fig:0916}). The diffuse emission at 144 MHz is approximately extended along the east-west tidal features of the host galaxy. The radio source located $32\arcsec$ to the southwest does not appear to be associated with the host SPM, and may either be associated with SDSS J091649.73+454130.4 or may be an FRII-like lobe associated with WISEA J091653.20+454128.7. Without a detection at 3 GHz, the lower limit to the spectral index is $\alpha_3^{10}>-0.91$. More sensitive observations at 3 GHz are needed to determine if its radio spectrum is steep or flat and to determine if the broadband radio spectrum is better fit by a simple or curved power law.

\subsection{J1015+3914}\label{sec:J1015+3914}
\begin{figure}
    \centering
    \includegraphics[width=\textwidth, trim=15mm 15mm 15mm 0mm,clip]{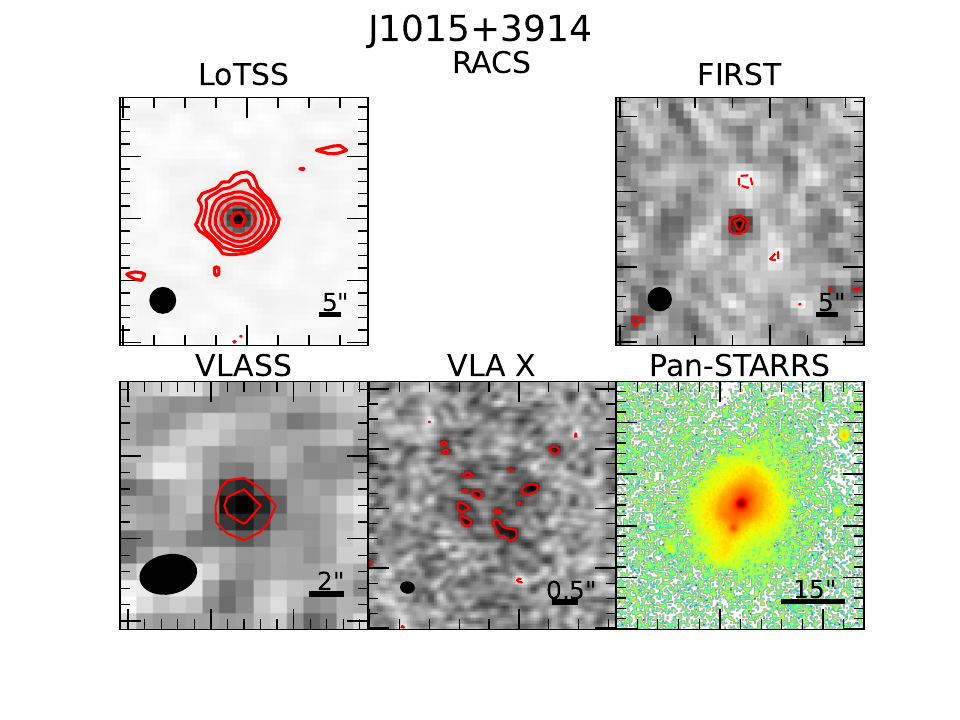}
    \caption{Intensity maps and contour overlays, and Pan-STARRS $r$-band image of the host galaxy of J1015+3914. The synthesized beam is shown in the bottom left corner for the radio surveys, and a scale bar is shown in the bottom right. Contours are shown at -3, 3, 5 and increase by a factor of 2 thereafter times the nominal image RMS for LoTSS, FIRST and VLASS, and the off-source RMS for the VLA X (12 $\mu$Jy) panel. The Pan-STARRS image has been resampled and falsely colored to emphasize the tidal features of the host galaxy.}
    \label{fig:1015}
\end{figure}

J1015+3914 is a SF-dominated radio hosted by a SF emission-line galaxy. The radio emission is resolved by LoTSS and at 10 GHz, marginally resolved by VLASS, and unresolved by FIRST (Figure~\ref{fig:1015}). The resolved features at 144 MHz to the northeast are not spatially coincident with the visible tidal features of the host galaxy. The diffuse morphology of the 10 GHz radio emission is highly indicative of recent SF activity. The radio spectrum is steep with $\alpha_3^{10}=-0.64\pm0.10$. The broadband radio spectrum is poorly fit by either a simple or curved power law. We hypothesize that this results from two distinct electron populations, one indicative of galactic-scale diffuse emission at 144 MHz, and a second population that is more compact and nuclear, perhaps indicative of more recent, nuclear SF activity. Better sampling of the radio SED is needed to confirm this.

\subsection{J1018+3613}\label{sec:J1018+3613}
\begin{figure}
    \centering
    \includegraphics[width=\textwidth, trim=15mm 15mm 15mm 0mm,clip]{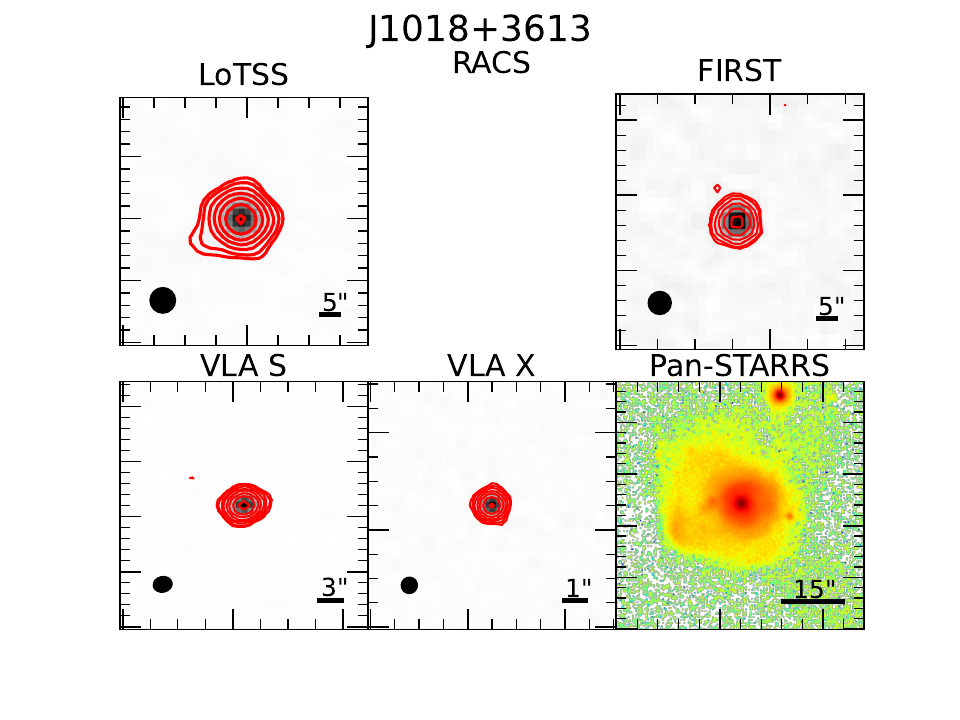}
    \caption{Intensity maps and contour overlays, and Pan-STARRS $r$-band image of the host galaxy of J1018+3613. The synthesized beam is shown in the bottom left corner for the radio surveys, and a scale bar is shown in the bottom right. Contours are shown at -3, 3, 5 and increase by a factor of 2 thereafter times the nominal image RMS for LoTSS and FIRST, and the off-source RMS for the VLA S (15 $\mu$Jy) and VLA X (10 $\mu$Jy) panels. The Pan-STARRS image has been resampled and falsely colored to emphasize the tidal features of the host galaxy. The VLA S and VLA X panels $30\arcsec\times30\arcsec$ and $10\arcsec\times10\arcsec$ in size due to the lower resolution at these frequencies of the VLA in B-configuration.}
    \label{fig:1018}
\end{figure}

J1018+3613 is a radio AGN hosted by a Seyfert AGN emission-line galaxy. The radio emission is resolved only by LoTSS, and is unresolved by FIRST, and at 3 GHz and 10 GHz (Figure~\ref{fig:1018}). The extended feature to the southeast in the LoTSS map is not spatially coincident with the southeast tidal structure visible in the host galaxy. J1018+3613 is one of two radio sources at any frequency that is considered radio-loud ($\nu L_\nu > 10^{32}$ W). The spectral index is steep with $\alpha_3^{10}=-0.97\pm0.03$. The broadband radio spectrum is well fit by a curved power law and shows evidence of curvature ($q=-0.19\pm0.02$), with a peak frequency at 290 MHz.

\subsection{J1041+1105}\label{sec:J1041+1105}
\begin{figure}
    \centering
    \includegraphics[width=\textwidth, trim=15mm 15mm 15mm 0mm,clip]{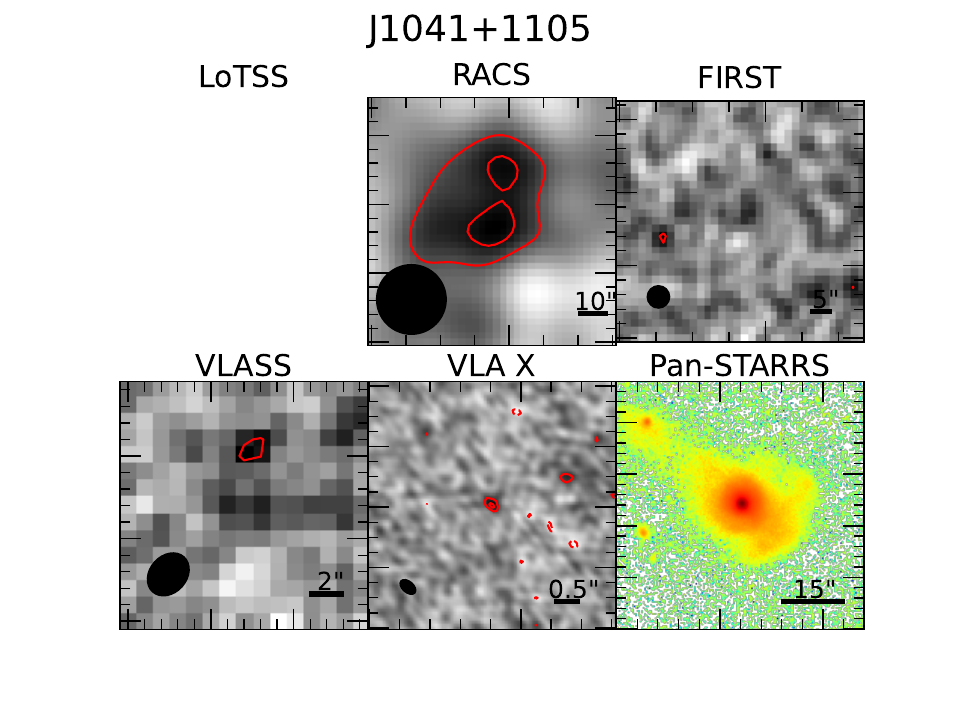}
    \caption{Intensity maps and contour overlays, and Pan-STARRS $r$-band image of the host galaxy of J1041+1105. The synthesized beam is shown in the bottom left corner for the radio surveys, and a scale bar is shown in the bottom right. Contours are shown at -3, 3, 5 and increase by a factor of 2 thereafter times the nominal image RMS for RACS, FIRST and VLASS, and the off-source RMS for the VLA X (12 $\mu$Jy) panel. The Pan-STARRS image has been resampled and falsely colored to emphasize the tidal features of the host galaxy. The local image RMS of this RACS field is greater than the nominal image RMS of RACS. Thus, the morphology of J1041+1105's 888 MHz emission is not represented accurately here.}
    \label{fig:1041}
\end{figure}

J1041+1105 is likely a SF-dominated radio source hosted by a LINER emission-line galaxy. The radio emission is resolved by RACS, unresolved at 10 GHz, and is a non-detection by both FIRST and VLASS (Figure~\ref{fig:1041}). The morphology at 888 MHz is difficult to gauge, as the local image RMS is $\sim$470 $\mu$Jy. When considering this constraint, only the southern radio component shown in the RACS panel of Figure~\ref{fig:1041}, produced using an image RMS equal to the nominal survey sensitivity of 250 $\mu$Jy, remains. Indeed, this southern component is spatially coincident with the host galaxy, while the northern and diffuse components are likely artifacts. Because it is a non-detection at 1.4 GHz, the radio-based SFR we calculated in Section~\ref{sec:radio_excess} is an upper limit. Given this constraint, it still falls in the SF region of Figure~\ref{fig:radio_origins}, hence our classification as a likely SF-dominated radio source. Without a detection at 3 GHz, the lower limit to the spectral index is $\alpha_3^{10}>-1.43$. More sensitive observations are needed to determine if its radio spectrum is steep or flat and to determine if the broadband radio spectrum is better fit by a simple or curved power law.

\subsection{J1113+2714}\label{sec:J1113+2714}
\begin{figure}
    \centering
    \includegraphics[width=\textwidth, trim=15mm 15mm 15mm 0mm,clip]{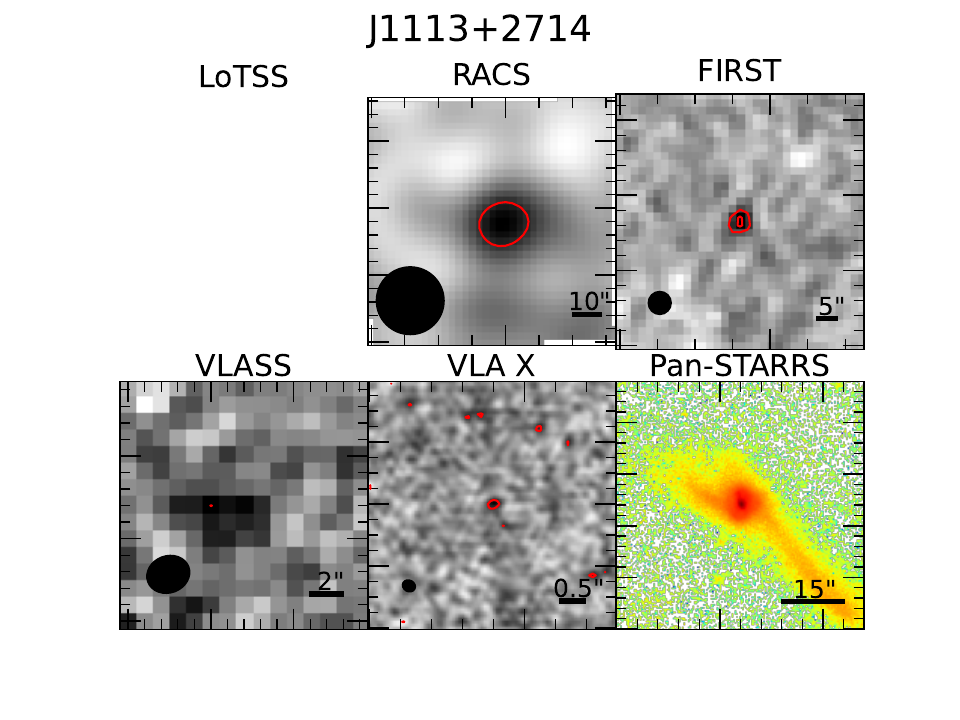}
    \caption{Intensity maps and contour overlays, and Pan-STARRS $r$-band image of the host galaxy of J1113+2714. The synthesized beam is shown in the bottom left corner for the radio surveys, and a scale bar is shown in the bottom right. Contours are shown at -3, 3, 5 and increase by a factor of 2 thereafter times the nominal image RMS for RACS, FIRST and VLASS, and the off-source RMS for the VLA X (12 $\mu$Jy) panel. The Pan-STARRS image has been resampled and falsely colored to emphasize the tidal features of the host galaxy.}
    \label{fig:1113}
\end{figure}

J1113+2714 is a radio AGN hosted by a SF emission-line galaxy. The radio emission is unresolved by RACS, FIRST and at 10 GHz, and is a non-detection by VLASS (Figure~\ref{fig:1113}). The unresolved emission is spatially coincident with the optical nucleus of the host galaxy. Without a detection at 3 GHz, the lower limit to the spectral index is $\alpha_3^{10}>-1.42$. More sensitive observations are needed to determine if its radio spectrum is steep or flat and to determine if the broadband radio spectrum is better fit by a simple or curved power law.

\subsection{J1135+2953}\label{sec:J1135+2953}
\begin{figure}
    \centering
    \includegraphics[width=\textwidth, trim=15mm 15mm 15mm 0mm,clip]{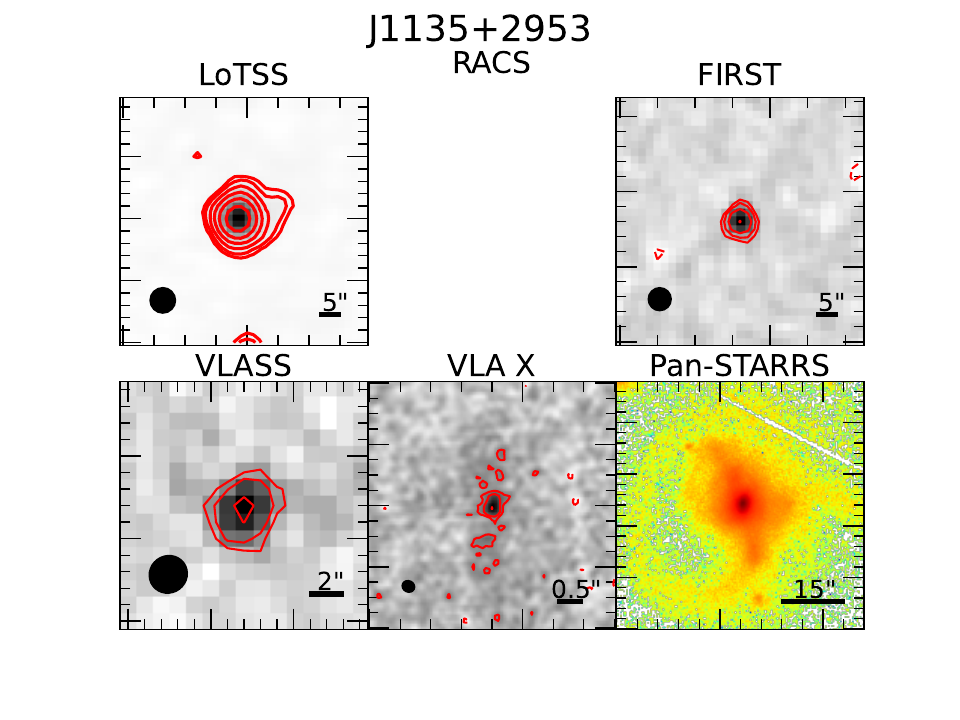}
    \caption{Intensity maps and contour overlays, and Pan-STARRS $r$-band image of the host galaxy of J1041+2953. The synthesized beam is shown in the bottom left corner for the radio surveys, and a scale bar is shown in the bottom right. Contours are shown at -3, 3, 5 and increase by a factor of 2 thereafter times the nominal image RMS for LoTSS, FIRST and VLASS, and the off-source RMS for the VLA X (12 $\mu$Jy) panel. The Pan-STARRS image has been resampled and falsely colored to emphasize the tidal features of the host galaxy.}
    \label{fig:1135}
\end{figure}

J1135+2953 is a radio AGN hosted by a SF emission-line galaxy. The radio emission is resolved by LoTSS, VLASS, and at 10 GHz, and unresolved by FIRST (Figure~\ref{fig:1135}). The extension to the northwest in the LoTSS map is similar in PA to the tidal feature visible in the host galaxy, though the two features do not perfectly align. The emission becomes compact at 1.4 GHz, but clearly has a resolved morphology at 10 GHz. The morphology at 10 GHz is inversion symmetric, and the 1.4 GHz emission shows excess radio luminosity over SF predictors, likely indicating that this is a radio AGN with a bipolar jet. The spectral index is steep with $\alpha_3^{10}=-1.74\pm0.05$. The broadband radio spectrum is not well fit by either a simple or curved power law. This is likely because we are probing two separate electron populations producing the observed radio emission. The 144 MHz emission probed by LoTSS is most probably associated with supernova remnants, while at GHz frequencies the radio AGN is the dominant emission component.  

\subsection{J1304+6520}\label{sec:J1304+6520}
\begin{figure}
    \centering
    \includegraphics[width=\textwidth, trim=15mm 15mm 15mm 0mm,clip]{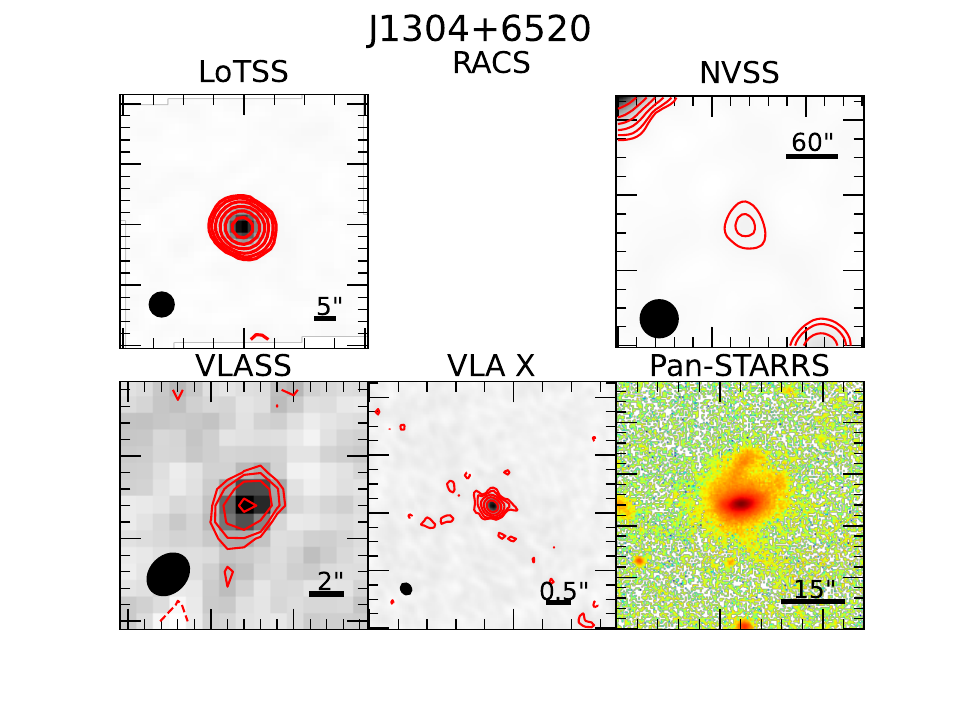}
    \caption{Intensity maps and contour overlays, and Pan-STARRS $r$-band image of the host galaxy of J1304+6520. The synthesized beam is shown in the bottom left corner for the radio surveys, and a scale bar is shown in the bottom right except for NVSS, where it is in the upper right. Contours are shown at -3, 3, 5 and increase by a factor of 2 thereafter times the nominal image RMS for LoTSS, NVSS and VLASS, and the off-source RMS for the VLA X (12 $\mu$Jy) panel. The Pan-STARRS image has been resampled and falsely colored to emphasize the tidal features of the host galaxy.
    }
    \label{fig:1304}
\end{figure}

J1304+6520 is a radio AGN hosted by a Seyfert AGN emission-line galaxy. The radio emission of J1304+2953 is resolved by LoTSS and at 10 GHz, and unresolved by NVSS and VLASS (Figure~\ref{fig:1304}). Although the emission is resolved by LoTSS, we find no evidence for emission coincident with the tidal features of the host galaxy. J1304+6520 is one of two radio sources at any frequency that is considered radio-loud ($\nu L_\nu > 10^{32}$ W). Indeed, the 1.4 GHz luminosity of J1304+6520 is $3\sigma$ greater than what is expected by SF-predictors, indicating this is a radio AGN. However, the morphology at 10 GHz is unexpected for a radio AGN as it does not resolve into the standard bipolar or one-sided jet morphology, or show a single, unresolved core. Different weighting schemes during imaging neither showed this jet-like morphology nor resolved out some of the extended features we observed at 10 GHz. This is discussed further in Section~\ref{sec:agn_morph}. More observations at high angular are needed to study its morphology. The spectral index is steep with $\alpha_3^{10}=-0.91\pm0.07$. The broadband radio spectrum is not well fit by either a simple or curved power law. Although the reduced $\chi^2$ value for the simple power law fit is not extremely poor, visual inspection of this fit to the broadband spectrum clearly shows deviant behavior. As with other sources showing this spectral behavior, this is most likely from two different electron populations being probed at 144 MHz and GHz frequencies, respectively.

\subsection{J1433+3444}\label{sec:J1433+3444}
\begin{figure}
    \centering
    \includegraphics[width=\textwidth, trim=15mm 15mm 15mm 0mm,clip]{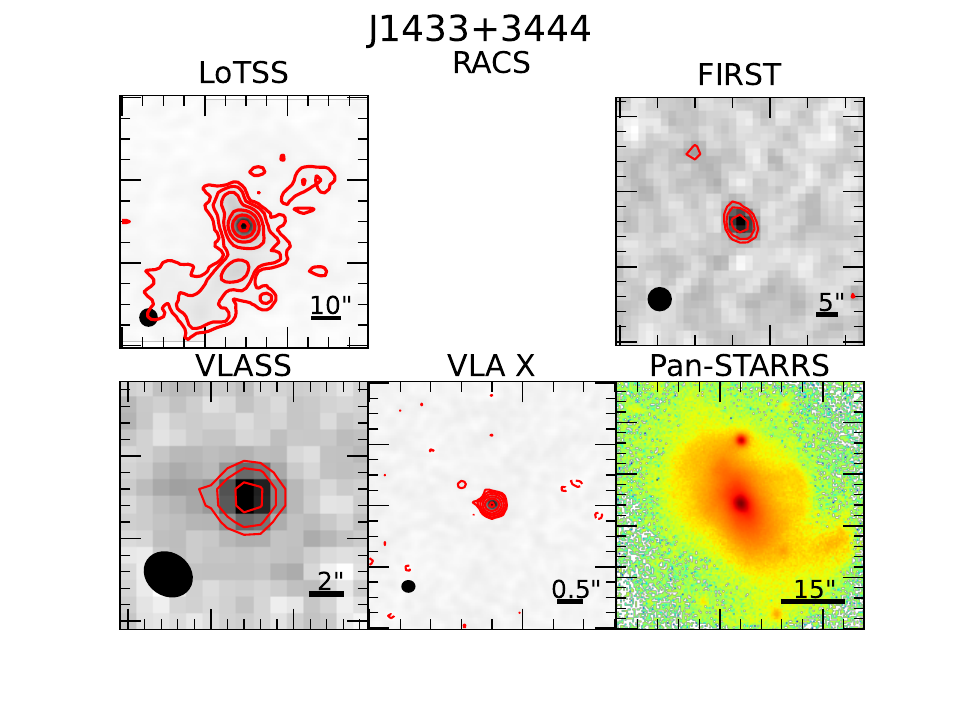}
    \caption{Intensity maps and contour overlays, and Pan-STARRS $r$-band image of the host galaxy of J1433+3444. The synthesized beam is shown in the bottom left corner for the radio surveys, and a scale bar is shown in the bottom right. Contours are shown at -3, 3, 5 and increase by a factor of 2 thereafter times the nominal image RMS for RACS, FIRST and VLASS, and the off-source RMS for the VLA X (12 $\mu$Jy) panel. The Pan-STARRS image has been resampled and falsely colored to emphasize the tidal features of the host galaxy. The LoTSS panel is $1.5\arcmin\times1.5\arcmin$ in size to show the full extent of the FRI-like morphology.
    }
    \label{fig:1433}
\end{figure}

J1433+3444 is a radio AGN hosted by a LINER emission-line galaxy. J1433+3444 displays the most AGN-like radio morphology of all sources detected by LoTSS (Figure~\ref{fig:1433}). The 144 MHz morphology extends approximately 80$\arcsec$ (55 kpc at $z=0.034$) at a PA of -39.3$\degr$ in a one-sided, FRI-like structure, with possible deboosted, counter-jet emission. The base of this FRI-like jet is located approximately 9.1$\arcsec$ (6.2 kpc) to the southwest of the compact emission feature, but there is no evidence for jet bending or curvature that may explain this offset. The core is also marginally extended at an approximate PA of 24$\degr$, parallel to the tidal structure visible in the host galaxy's morphology extending to the northeast. The source is moderately resolved by FIRST and at 10 GHz, and resolved by VLASS. The VLASS contour map shows a marginal extension to the east. This feature also appears in the naturally-weighted 10 GHz map, but appears to resolve into smaller components in a uniformally-weighted map. We do note, however, that the ratio of peak-to-integrated 10 GHz flux density for this source is unity within $3\sigma$, hence our morphological classification as moderately resolved. The spectral index is flat with $\alpha_3^{10}=-0.47\pm0.07$. The broadband radio spectrum is well fit by a curved power law but does not exhibit significant curvature ($q=0.08\pm0.02$).

\subsection{J1445+5134}\label{sec:J1445+5134}
\begin{figure}
    \centering
    \includegraphics[width=\textwidth, trim=15mm 15mm 15mm 0mm,clip]{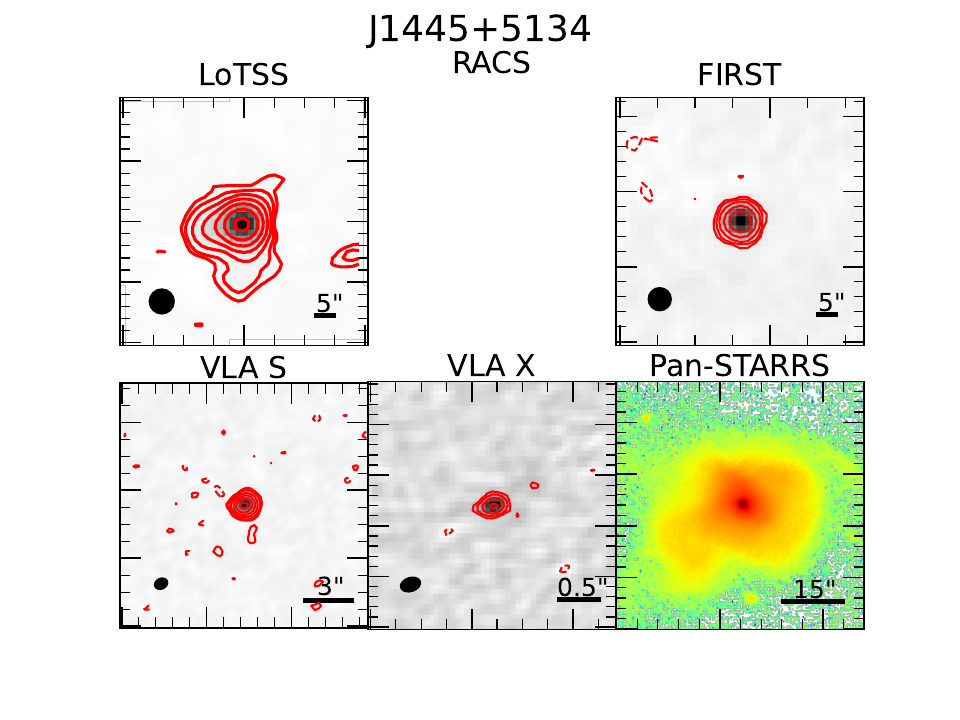}
    \caption{Intensity maps and contour overlays, and Pan-STARRS $r$-band image of the host galaxy of J1445+5134. The synthesized beam is shown in the bottom left corner for the radio surveys, and a scale bar is shown in the bottom right. Contours are shown at -3, 3, 5 and increase by a factor of 2 thereafter times the nominal image RMS for LoTSS and FIRST, and the off-source RMS for the VLA S (45 $\mu$Jy) and VLA X (40 $\mu$Jy) panels. The Pan-STARRS image has been resampled and falsely colored to emphasize the tidal features of the host galaxy. The VLA X panel is $3\arcsec\times3\arcsec$ in size.}
    \label{fig:1445}
\end{figure}

J1445+5134 is a SF-dominated radio source hosted by a SF-AGN composite emission-line galaxy. The radio emission is resolved by LoTSS and at 3 and 10 GHz, and unresolved by FIRST (Figure~\ref{fig:1445}.) The extended components in the LoTSS map are not spatially coincident with any of the tidal features visible in the host galaxy's morphology. Even though the GHz-frequency radio emission is SF-dominated, the diffuse, host-galaxy-like morphology that is standard for such sources is not observed in J1445+5134. The spectral index is steep with $\alpha_3^{10}=-1.46\pm0.04$. The broadband radio spectrum is well fit by a curved power law, and shows evidence of curvature ($q=-0.18\pm0.02$) with a peak frequency at 155 MHz. 

\subsection{J1511+0417}\label{sec:J1511+0417}
\begin{figure}
    \centering
    \includegraphics[width=\textwidth, trim=15mm 15mm 15mm 0mm,clip]{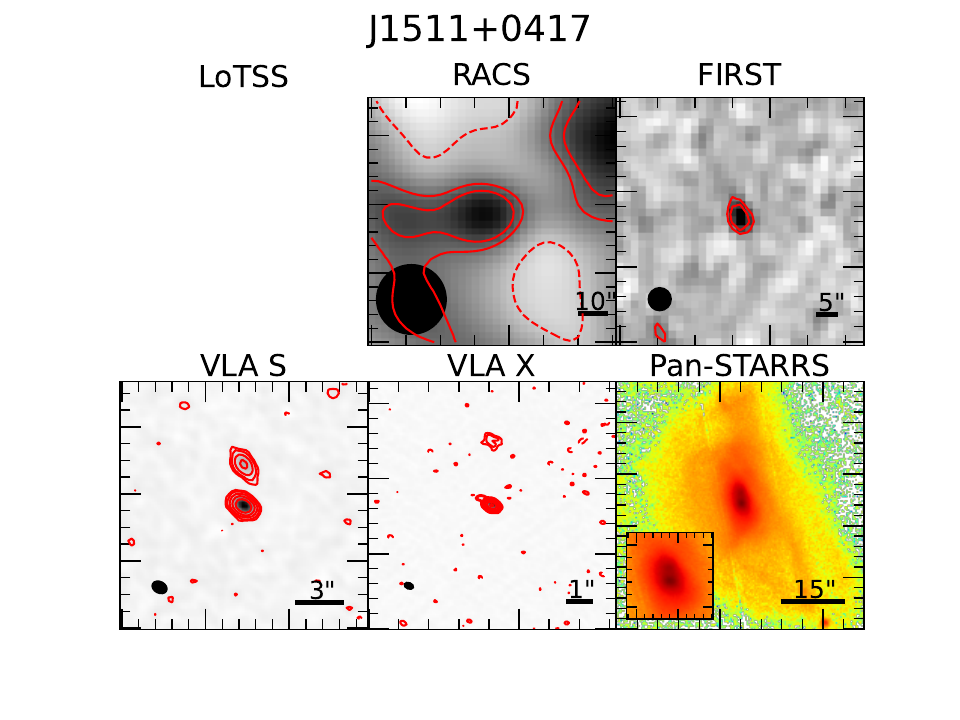}
    \caption{Intensity maps and contour overlays, and Pan-STARRS $r$-band image of the host galaxy of J1511+0417. The synthesized beam is shown in the bottom left corner for the radio surveys, and a scale bar is shown in the bottom right. Contours are shown at -3, 3, 5 and increase by a factor of 2 thereafter times the nominal image RMS for RACS and FIRST, and the off-source RMS for the VLA S (12 $\mu$Jy) and VLA X (7 $\mu$Jy) panels. The Pan-STARRS image has been resampled and falsely colored to emphasize the tidal features of the host galaxy. The inset panel shows the dual nuclei of the host galaxy. The VLA X panel is $10\arcsec\times10\arcsec$ in size to show the second radio component. The local image RMS of this RACS field is greater than the nominal image RMS of RACS. Thus, the morphology of J1511+0417's 888 MHz emission is not represented accurately here.}
    \label{fig:1511+0417}
\end{figure}

J1511+0417 is a radio AGN hosted by a LINER emission-line galaxy. The radio emission is unresolved by FIRST, has multiple components at 3 GHz and 10 GHz, and is a non-detection by RACS (Figure~\ref{fig:1511+0417}). Like J0206$-$0017 and J1041+1105, the local image RMS in RACS ($\approx$770 $\mu$Jy) for J1511+0417 is significantly higher than the nominal survey RMS (250 $\mu$Jy), leading to a misinformed, apparent $5\sigma$ detection in Figure~\ref{fig:1511+0417}. We have not included this as a detection, as this emission is below a $3\sigma$ detection threshold using the local image RMS, nor is it listed in the RACS catalog. The maps at 3 and 10 GHz show that this source has two components that are separated by 2.5$\arcsec$ (2.1 kpc at $z=0.042$). The southern, dominant component is unresolved at both frequencies and has an inverted spectral index value of $\alpha_3^{10}=0.19\pm0.03$. The northern component is marginally resolved at 3 GHz and is resolved at 10 GHz and has a steep spectral index value of $\alpha_3^{10}=-0.60$. It is particularly noteworthy that both radio sources are co-spatial with distinct optical nuclei of the host galaxy (see inset of the Pan-STARRS panel in Figure~\ref{fig:1511+0417}). This makes J1511+0417 a candidate DAGN.

\subsection{J1511+2309}\label{sec:J1511+2309}
\begin{figure}
    \centering
    \includegraphics[width=\textwidth, trim=15mm 15mm 15mm 0mm,clip]{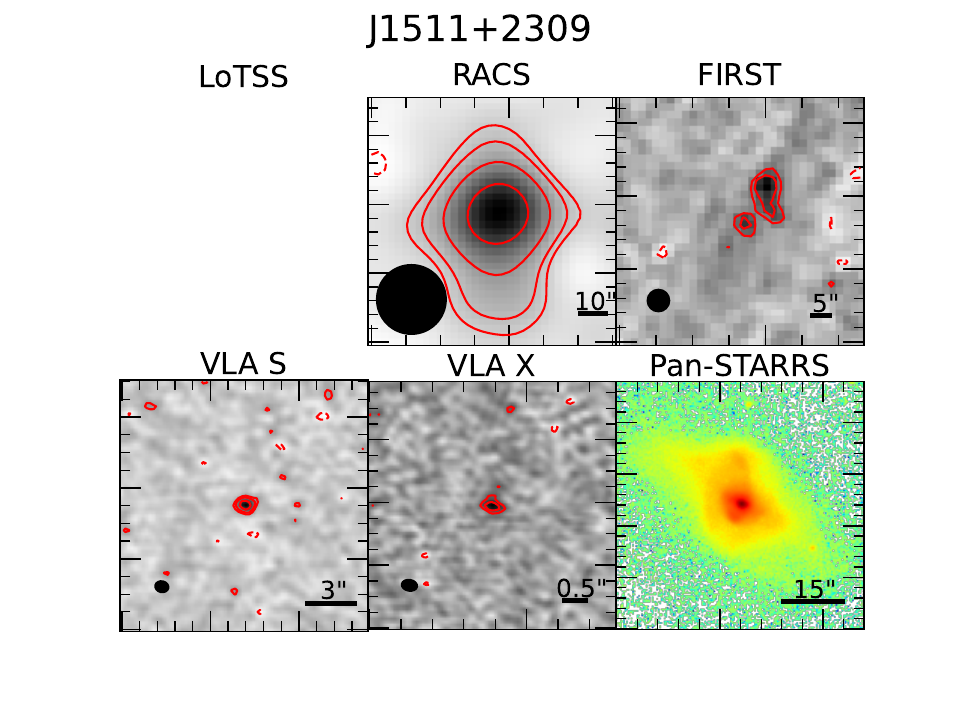}
    \caption{Intensity maps and contour overlays, and Pan-STARRS $r$-band image of the host galaxy of J1511+2309. The synthesized beam is shown in the bottom left corner for the radio surveys, and a scale bar is shown in the bottom right. Contours are shown at -3, 3, 5 and increase by a factor of 2 thereafter times the nominal image RMS for RACS and FIRST, and the off-source RMS for the VLA S (28 $\mu$Jy) and VLA X (17 $\mu$Jy) panels. The Pan-STARRS image has been resampled and falsely colored to emphasize the tidal features of the host galaxy.}
    \label{fig:1511+2309}
\end{figure}

J1511+2309 is a radio AGN hosted by an optically-quiescent emission-line galaxy. The radio emission is resolved by RACS and at 10 GHz, and unresolved by FIRST and at 3 GHz (Figure~\ref{fig:1511+2309}). We discuss the choice of morphological classification at 1.4 GHz in Section~\ref{sec:radio_morph}. At 888 MHz (RACS), the source is extended in a north-south direction, with a marginal extension in the east-west direction. The synthesized beam of RACS is too large (25$\arcsec$) to study if these extended features are associated with any tidal structures visible in the host galaxy. The emission then becomes compact at higher frequency until it displays a similar northern extension at 10 GHz. The spectral index is steep with $\alpha_3^{10}=-0.57\pm0.08$. The broadband radio spectrum is well fit by a curved power law and shows evidence of inversion ($q=1.61\pm0.36$). However, we believe that the spectrum likely isn't truly inverted but is flattening at higher frequency, similar to J0206$-$0017. Higher frequency observations are be needed to confirm this.

\subsection{J1517+0409}\label{sec:J1517+0409}
\begin{figure}
    \centering
    \includegraphics[width=\textwidth, trim=15mm 15mm 15mm 0mm,clip]{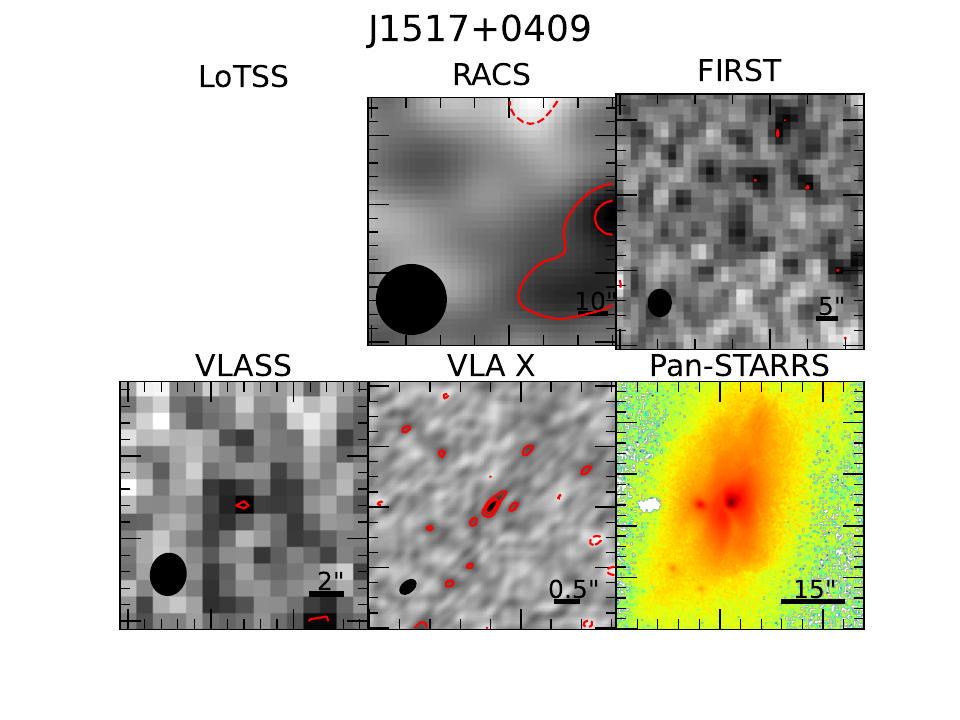}
    \caption{Intensity maps and contour overlays, and Pan-STARRS $r$-band image of the host galaxy of J1517+0409. The synthesized beam is shown in the bottom left corner for the radio surveys, and a scale bar is shown in the bottom right. Contours are shown at -3, 3, 5 and increase by a factor of 2 thereafter times the nominal image RMS for RACS, FIRST and VLASS, and the off-source RMS for the VLA X (12 $\mu$Jy) panel. The Pan-STARRS image has been resampled and falsely colored to emphasize the tidal features of the host galaxy.}
    \label{fig:1517}
\end{figure}

J1517+0409 is a possible radio AGN hosted by an optically-quiescent emission-line galaxy. The radio emission is only detected at 10 GHz, unresolved, and co-spatial with the optical nucleus of the host galaxy (Figure~\ref{fig:1517}). We found no evidence for emission at the $3\sigma$ level or greater in any of the RACS, FIRST, or VLASS maps centered on the 10 GHz position of the radio source. Because it is a non-detection at 1.4 GHz, the radio-based SFR we calculated in Section~\ref{sec:radio_excess} is an upper limit. Given this constraint, it still falls in the radio excess region of Figure~\ref{fig:radio_origins}. More sensitive observations at 1.4 GHz may change this, but our data cannot determine otherwise, hence our classification as a possible radio AGN. Without a detection at 3 GHz, the lower limit to the spectral index is $\alpha_3^{10}>-1.02$. More sensitive observations are needed to determine if its radio spectrum is steep or flat and to determine if the broadband radio spectrum is better fit by a simple or curved power law.

\subsection{J1617+2512}\label{sec:J1617+2512}
\begin{figure}
    \centering
    \includegraphics[width=\textwidth, trim=15mm 15mm 15mm 0mm,clip]{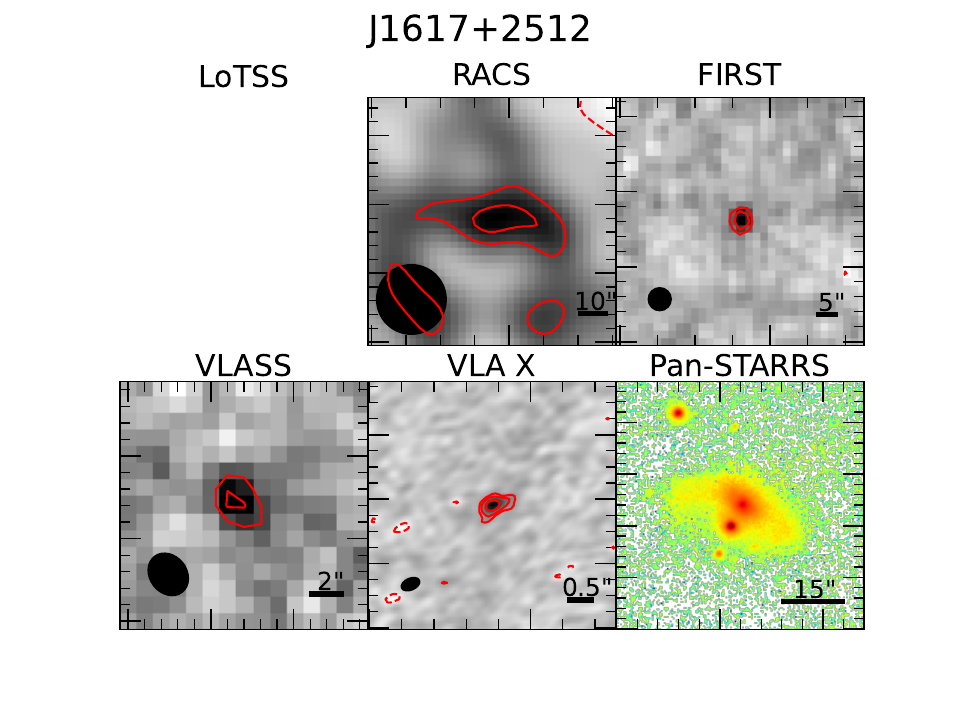}
    \caption{Intensity maps and contour overlays, and Pan-STARRS $r$-band image of the host galaxy of J1617+2512. The synthesized beam is shown in the bottom left corner for the radio surveys, and a scale bar is shown in the bottom right. Contours are shown at -3, 3, 5 and increase by a factor of 2 thereafter times the nominal image RMS for RACS, FIRST and VLASS, and the off-source RMS for the VLA X (11 $\mu$Jy) panel. The Pan-STARRS image has been resampled and falsely colored to emphasize the tidal features of the host galaxy.
    }
    \label{fig:1617}
\end{figure}

J1617+2512 is a SF-dominated radio source hosted by a SF-AGN composite emission-line galaxy. The source is marginally resolved by RACS, unresolved by FIRST, and resolved by VLASS and at 10 GHz (Figure~\ref{fig:1617}). The RACS emission is generally diffuse but the synthesized beam is too large (25$\arcsec$) to confidently assess if this diffuse nature is spatially coincident with the tidal features visible in the host galaxy. The source is extended to the southwest in the VLASS map, and marginally extended along a roughly northwest-southeast direction at 10 GHz. Although it is a SF-dominated radio source, we do not see the typical diffuse emission morphology indicative of this at 1.4 GHz. The spectral index is steep with $\alpha_3^{10}=-1.31\pm0.11$. The broadband radio spectrum is well fit by a curved power law shows evidence of curvature ($q=-0.37\pm0.08$), and has a peak frequency at 1.11 GHz, making this the only Gigahertz peaked spectrum (GPS) source in our sample.

\subsection{J1655+2639}\label{sec:J1655+2639}
\begin{figure}
    \centering
    \includegraphics[width=\textwidth, trim=15mm 15mm 15mm 0mm,clip]{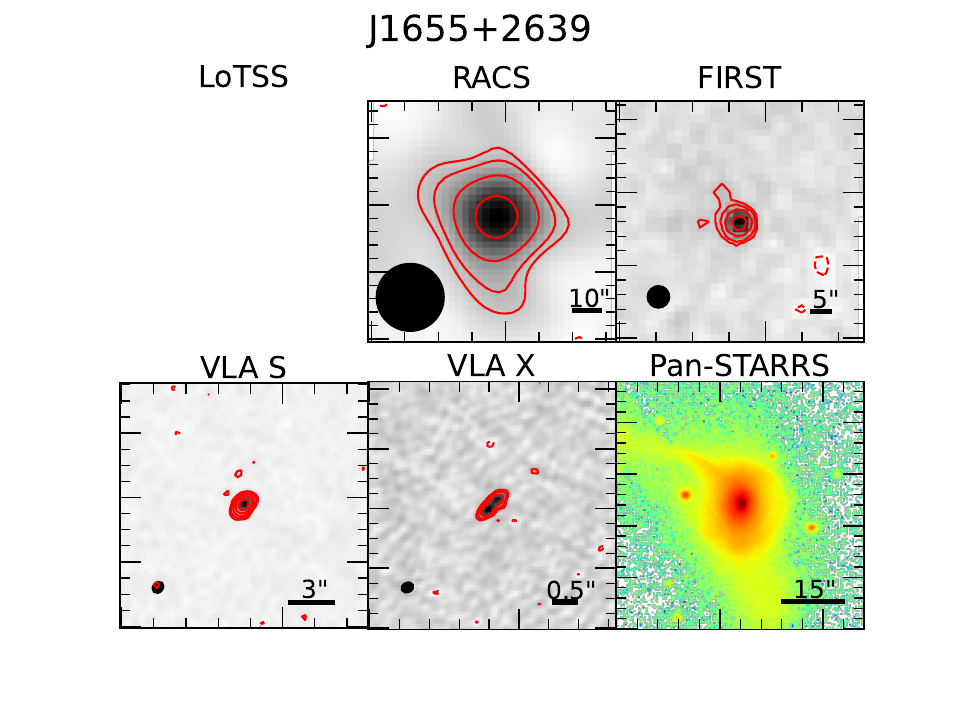}
    \caption{Intensity maps and contour overlays, and Pan-STARRS $r$-band image of the host galaxy of J1655+2639. The synthesized beam is shown in the bottom left corner for the radio surveys, and a scale bar is shown in the bottom right. Contours are shown at -3, 3, 5 and increase by a factor of 2 thereafter times the nominal image RMS for RACS and FIRST, and the off-source RMS for the VLA S (34 $\mu$Jy) and VLA X (19 $\mu$Jy) panels. The Pan-STARRS image has been resampled and falsely colored to emphasize the tidal features of the host galaxy.}
    \label{fig:1655}
\end{figure}

J1655+2639 is a radio AGN hosted by an optically-quiescent emission-line galaxy. The source is marginally resolved by RACS and FIRST, and resolved at 3 and 10 GHz (Figure~\ref{fig:1655}). The 888 MHz emission is extended along the eastern and southern tidal features visible in the host galaxy. These 888 MHz features are slightly exaggerated in Figure~\ref{fig:1655}, as the image was produced assuming the nominal image RMS for RACS of $250\mu$Jy, but the local image RMS is actually $\approx500\mu$Jy The 1.4 GHz emission is also extended along the eastern tidal feature, although the dominant component to the radio emission is compact. Our observations at 3 and 10 GHz show the jet-like morphology of this radio AGN at a PA of $\approx 140\degr$. The radio excess ($>3\sigma$) and morphology of this source make it among the clearest examples of a radio AGN in our sample. The spectral index is steep with $\alpha_3^{10}=-0.98\pm0.03$. The broadband radio spectrum is well fit by a simple power law.

\end{document}